\documentclass{article}     

\usepackage{amsmath, amsfonts,amsthm,amssymb}
\usepackage{dsfont}
\usepackage[english]{babel}
\usepackage[left=2.0cm, top=2.7cm, bottom=2.50cm,right=2.0cm]{geometry}
\usepackage{verbatim}
\usepackage{mathrsfs}
\usepackage{bm}
\usepackage{color}
\usepackage{graphicx}
\usepackage{url}
\usepackage{bm}
\usepackage{colortbl}
\RequirePackage{natbib}

\newtheorem{thm}{Theorem}[section]
\newtheorem{cor}[thm]{Corollary}
\newtheorem{lem}[thm]{Lemma}

\theoremstyle{definition}

\newtheorem{assum}[thm]{Assumption}
\newtheorem{rem}[thm]{Remark}
\numberwithin{equation}{section}

\newcommand{\be}{\begin{equation}}
\newcommand{\ee}{\end{equation}}
\newcommand{\bq}{\begin{eqnarray}}
\newcommand{\eq}{\end{eqnarray}}

\def\ed{{\,\stackrel{\frak {D}}{=}\,}}
\def\ld{\;{\stackrel{\frak {D}}{\longrightarrow}}\;}

\def\ld{{\stackrel{\frak {D}}{\longrightarrow}}}

\def\calF{{\mathcal F}}

\def\calN{{\mathcal N}}

\def\bbr{{\mathbb R}}
\def\bbe{{\mathbb E}}
\def\bbp{{\mathbb P}}

\def\ed{\;{\stackrel{\frak {D}}{=}}\;}

\def\ld{\;{\stackrel{\frak {D}}{\longrightarrow}}\;}

\definecolor{Red}{rgb}{0,0,0}

\definecolor{DRed}{rgb}{0,0,0}

\definecolor{Green}{rgb}{0,0,0}

\definecolor{Blue}{rgb}{0.00, 0.00, 0.00}
\newcommand{\Blue}{\color{Blue}}
\definecolor{PaleGrey}{rgb}{0,0,0}

\title{High-order short-time expansions for ATM option prices of exponential L\'{e}vy models}
\author{Jos\'{e} E. Figueroa-L\'{o}pez\thanks{Department of Statistics, Purdue University, West Lafayette, IN, 47907,  USA ({\tt figueroa@purdue.edu}). Research supported in part by the NSF Grants DMS-1149692 and DMS-0906919.}
\and Ruoting Gong\thanks{Department of Mathematics, Rutgers, The State University of New Jersey, Piscataway, NJ, 08904, USA ({\tt rgong@math.rutgers.edu}).}
\and Christian Houdr\'{e}\thanks{School of Mathematics, Georgia Institute of Technology, Atlanta, GA, 30332, USA ({\tt houdre@math.gatech.edu}). Research supported in part by the the Simons Foundation grant \#246283.
\newline The authors gratefully acknowledge the constructive and insightful comments provided by two anonymous referees, which significantly contributed to improve the quality of the manuscript.  The authors {also thank Victor Rivero for bringing our attention to the formula (\ref{PstStb})} as well as Sveinn \'Olafsson for pointing out some {mistakes in a previous proof of a lemma} and other helpful comments.}}

\date{April 5, 2014}

\begin{document}

\maketitle

\begin{abstract}

The short-time asymptotic behavior of option prices for a variety of models with jumps has received much attention in recent years. In the present work, a novel second-order approximation for ATM option prices is derived for a large class of exponential L\'{e}vy models with or without Brownian component. The results hereafter shed new light on the connection between both the volatility of the continuous component and the jump parameters and the behavior of ATM option prices near expiration. In the presence of a Brownian component, the second-order term, in time-$t$, is of the form $d_{2}\,t^{(3-Y)/2}$, with {$d_{2}$ only depending} on $Y$, the degree of jump activity, on $\sigma$, the volatility of the continuous component, and on an additional parameter controlling the intensity of the ``small" jumps (regardless of their signs). This extends the well known result that the leading first-order term is $\sigma t^{1/2}/\sqrt{2\pi}$. In contrast, under a pure-jump model, the dependence on $Y$ and on the separate intensities of negative and positive small jumps are already reflected in the leading term, which is of the form $d_{1}t^{1/Y}$. The second-order term is shown to be of the form {$\tilde{d}_{2} t$} and, therefore, its order of decay turns out to be independent of $Y$. The asymptotic behavior  of the corresponding Black-Scholes implied volatilities is also addressed. Our method of proof is based on an integral representation of the option price involving the tail probability of the log-return process under the share measure and a suitable change of probability measure under which the pure-jump component of the log-return process becomes a $Y$-stable process. Our approach is sufficiently general to cover a wide class of L\'{e}vy processes which satisfy the latter property and whose L\'{e}vy densitiy can be closely approximated by a stable density near the origin. Our numerical results show that {the} first-order term typically exhibits rather poor performance and that the second-order term {{can} significantly improve the approximation's accuracy, {particularly} in the absence of a Brownian component}.

\vspace{0.3 cm}

\noindent{\textbf{AMS 2000 subject classifications}: 60G51, 60F99, 91G20, 91G60.}

\vspace{0.3 cm}

\noindent{\textbf{Key words}: exponential L\'{e}vy models; CGMY and tempered stable models; short-time asymptotics; at-the-money option pricing; implied volatility.}

\end{abstract}

\section{Introduction}

It is generally recognized that the standard Black-Scholes option pricing model is inconsistent with options data, while still being used in practice because of its simplicity and the existence of tractable solutions. Exponential L\'{e}vy models generalize the classical Black-Scholes setup by allowing jumps in stock prices while preserving the independence and the stationarity of returns. There are several reasons for introducing jumps in financial modeling. First of all, sudden sharp shifts in the price level of financial assets often occur in practice, and these ``jumps" are hard to handled within continuous-paths models. Second, the empirical returns of financial assets typically exhibit distributions with heavy tails and high kurtosis, which again are hard to replicate within purely-continuous frameworks. Finally, market prices of vanilla options display skewed implied volatilities (relative to changes in the strikes), in contrast to the classical Black-Scholes model which predicts a flat implied volatility smile. Moreover, the fact that the implied volatility smile and skewness {phenomena} become much more pronounced for short maturities is a clear indication of the need for jumps in the underlying models used to price options. {The reader is referred} to \cite{CT04} for further motivations on the use of jump processes in financial modeling.

One of the first applications of jump processes in financial modeling originates with~\cite{Mandelbrot}, who suggested a pure-jump stable L\'{e}vy process to model power-like tails and self-similar behavior in cotton price returns. \cite{Merton:76} and \cite{Press} subsequently considered option pricing and hedging problems under an exponential compound Poisson process with Gaussian jumps and an additive independent non-zero Brownian component. A similar exponential compound Poisson jump-diffusion model was more recently studied in \cite{Kou:2002}, where the jump sizes are distributed according to an asymmetric Laplace law. For infinite activity exponential L\'{e}vy models, \cite{BNS:1998} introduced the normal inverse Gaussian (NIG) model, while the extension to the generalized hyperbolic class was studied by \cite{EKP:1998}. \cite{MadSen:1990} introduced the symmetric variance gamma (VG) model while its asymmetric extension was later analyzed in \cite{MadMil:1991} and \cite{MCC:1998}. Both models are built on Brownian subordination; the main difference being that the log-return process in the NIG model is an infinite variation process with Cauchy-like behavior of small jumps, while in the VG model the log-price is of finite variation with infinite but relatively low activity of small jumps. The class of ``truncated stable" processes was first introduced by \cite{Koponen:1995}. It was advocated for financial modeling by \cite{Cont:1997} and \cite{Matacz}, and {was} further developed in the context of asset price modeling by \cite{CGMY:2002}, who introduced the terminology CGMY. Nowadays, the CGMY model is considered to be a prototype of the general class of models with jumps and enjoys widespread applicability.

The CGMY model is a particular case of the more general KoBoL class of \cite{BL02}, which in turn is a subclass of the semi-parametric class of tempered stable processes (TSPs) introduced by \cite{Rosinski:2007}\footnote{The terminology Tempered Stable is widely used in the literature, but unfortunately there is no uniform usage. It is worth noting that Rosi\'{n}ski's class (\cite{Rosinski:2007}) is much more general than the ``tempered stable" class introduced in several classical references of financial modeling with jumps (e.g. \cite{Applebaum2004}, \cite{CT04}, \cite{Kyprianou}).}. TSPs form a rich class with appealing features for financial modeling. One of their most interesting features lies in their short-time and long-time behaviors. Indeed, denoting by $Y\in (0,2)$ the index of a TSP, $L:=(L_{t})_{t\geq{}0}$, it follows that
\begin{equation}\label{SRBTSP}
\left(h^{-1/Y}L_{ht}\right)_{t\geq 0}\ld\left(Z_{t}\right)_{t\geq 0},\quad h\to 0,
\end{equation}
where $Z:=\left(Z_{t}\right)_{t\geq 0}$ is a strictly $Y$-stable L\'{e}vy process and $\ld$ denotes convergence in distribution. In fact, (\ref{SRBTSP}) holds when $Y\in(0,1)$ provided $L$ is driftless, while it holds when $Y\in(1,2)$ regardless of the mean of $L$. Roughly, (\ref{SRBTSP}) suggests that the short-time behavior of TSP is akin to that of a stable process, with its heavy-tailed distribution and self-similarity property which are desirable for financial modeling as emphasized, for example, by \cite{Mandelbrot}. In contrast,
\begin{equation}\label{LRBTSP}
\left(h^{-1/2}L_{ht}\right)_{t\geq 0}\ld\left(W_{t}\right)_{t\geq 0},\quad h\to\infty,
\end{equation}
where $\left(W_{t}\right)_{t\geq 0}$ is a Brownian Motion, which suggests that in the long-horizon, the process is Brownian-like. In terms of increments, if we were to consider the consecutive ``high-frequency" increments of the process $L$, say $\{L_{\underline{h} i}-L_{\underline{h} (i-1)}\}_{i\geq 1}$ with $\underline{h}\ll 1$, these will exhibit statistical features consistent with those of a {self-similar} stable time series. But, low-frequency increments, say $\{L_{\bar{h}i}-L_{\underline{h}(i-1)}\}_{i\geq 1}$ with $\bar{h}\gg\underline{h}$, will have Gaussian like distributions.

Stemming in part from their importance for model calibration and testing, small-time asymptotics of option prices have received a lot of attention in recent years (see, e.g., \cite{BBF:2002}, \cite{BBF2:2004}, \cite{FFF:2010}, \cite{FFK:2012}, \cite{FigForde:2012}, \cite{FordeJac:2009}, \cite{FordeJac:2010}, \cite{FordeJacLee:2010}, \cite{Gatheral:2009}, \cite{Henry:2009}, \cite{Paulot:2009}, \cite{Rop10}, \cite{Tankov}). {Reviewed here are} only the studies most closely related to ours, focusing in particular on the at-the-money (ATM) case. \cite{CW03} first analyzed, partially via heuristic arguments, the first-order asymptotic behavior of an It\^{o} semimartingale with jumps. Concretely, \cite{CW03} {argued} that ATM option prices of pure-jump models of bounded variation decrease at the rate $O(t)$, while they are just $O(t^{1/2})$ in the presence of a Brownian component. By analyzing the particular case of a stable pure-jump component, \cite{CW03} also {argued} that, for other cases, the rate could be $O(t^{\beta})$ for some $\beta\in(0,1)$. \cite{MuhNut:2009} formally showed that, in the presence of a continuous-time component, the leading term of ATM option prices is of order $\sqrt{t}$, for a relatively broad class of It\^{o} models, while for a type of It\^{o} processes with $\alpha$-stable-like small jumps and $\alpha>1$, the leading term is $O(t^{1/\alpha})$ (see also~\cite[Proposition 4.2]{FigForde:2012}, \cite[Theorem 3.7]{FigGH:2011}, and~\cite[Proposition 5]{Tankov} for related results in exponential L\'{e}vy models). However, none of these papers {gives} higher order asymptotics for the ATM option prices.

{The present paper studies} the small-time behavior for at-the-money call (or equivalently, put) option prices
\begin{equation}\label{CallPriceDfn}
{\bbe\left[\left(S_{t}-S_{0}\right)^{+}\right]=S_{0}\bbe\left[\left(e^{X_{t}}-1\right)^{+}\right],}
\end{equation}
under the exponential L\'{e}vy model
\begin{equation}\label{ExpLvMdl}
S_{t}:=S_{0}e^{X_{t}},\qquad {t\geq 0},
\end{equation}
where $X:=(X_{t})_{t\geq 0}$ is the superposition of a tempered-stable-like process $L:=(L_{t})_{t\geq 0}$ and of an independent Brownian motion $(\sigma W_{t})_{t\geq 0}$, i.e.,
\begin{equation}\label{CGMY1stDfn}
X_{t}:=\sigma W_{t}+L_{t},\qquad {t\geq 0},
\end{equation}
where $W:=(W_{t})_{t\geq 0}$ is a standard Brownian motion independent of $L$. The term ``tempered stable" is understood here in a much more general sense than in several classical sources of financial mathematics (e.g., \cite{Applebaum2004}, \cite{CT04}, \cite{Kyprianou}) and even more general than in \cite{Rosinski:2007}. Roughly (see Section \ref{Sec:TmpStble} below for the explicit conditions), {consider} {a L\'{e}vy process} whose L\'{e}vy measure $\nu$ admits a density $s:\bbr\backslash\{0\}\to (0,\infty)$ such that, for some $Y\in(1,2)$, the function $q(x):=s(x)|x|^{Y+1}$ satisfies the following fundamental property:
\begin{align}\label{KyCondValAllRs}
\lim_{x\searrow 0}\frac{1}{x}\left(C_{+}-q(x)\right)=\beta_{+},\quad\lim_{x\nearrow 0}\frac{1}{x}\left(C_{-}-q(x)\right)=-\beta_{-},
\end{align}
for some positive constants $C_{+}$, $C_{-}$, $\beta_{+}$, $\beta_{-}>0$. Intuitively, (\ref{KyCondValAllRs}) indicates that the small jumps of the process behave like those of a $Y$-stable process. As it turns out (see~\cite[Proposition 1]{rosenbaum.tankov.10}), this class still exhibits the appealing short-time and long-time behaviors of Rosi\'{n}ski's original tempered stable processes as described in (\ref{SRBTSP})-(\ref{LRBTSP}) above. It is worth noting that the previously defined parameter $Y$ coincides with the Blumenthal-Getoor index {of the process $L$,}
\begin{align*}
{BG:=\inf\left\{r\geq 0:\,\int_{|x|\leq 1}|x|^{r}\,\nu(dx)<\infty\right\},}
\end{align*}
which measures the ``degree of jump activity" of the process in that {(see, e.g., \cite{Sato:1999}):}
\begin{align*}
{\sum_{u\leq t}\left|\Delta L_{u}\right|^{r}<\infty\,\,\,\text{a.s.}\quad\text{if and only if}\quad r>Y.}
\end{align*}
As it turns out {(see~\cite{Jacod05}), for any $r>Y>1$,}
\begin{align*}
{\lim_{t\to 0}\frac{1}{t}\bbe\left(\psi(L_{t})\right)=\int_{\mathbb{R}\setminus\{0\}}\psi(x)\,\nu(dx),}
\end{align*}
for any bounded continuous function $\psi$ such that $\psi(x)=O(|x|^{r})$ as $x\to 0$.

Under the standing assumption (\ref{KyCondValAllRs}) {(and other conditions)}, we show that the first-order asymptotic behavior of (\ref{CallPriceDfn}) in short-time takes the form
\begin{equation}\label{1stOAp}
{\lim_{t\to 0}t^{-\frac{1}{Y}}\mathbb{E}\left[\left(S_{t}-S_{0}\right)^{+}\right]=S_{0}\mathbb{E}\left(Z^{+}\right),}
\end{equation}
where $Z$ is {a certain} centered {$\alpha$-stable} random variable under $\mathbb{P}$. When $\sigma\neq 0$, $Z\sim\calN(0,\sigma^{2})$ (i.e., {$\alpha=2$}) {and, as already shown in~\cite{Tankov} and~\cite{Rop10}, (\ref{1stOAp}) holds true with} $\mathbb{E}(Z^{+})=\sigma/\sqrt{2\pi}$. If $\sigma=0$ and $L$ is symmetric, {then $\alpha=Y$ and} the characteristic function of $Z$ is explicitly given by
\begin{align*}
\bbe\left(e^{iuZ}\right)=\exp\left\{-2C\Gamma(-Y)\left|\cos\left(\frac{1}{2}Y\pi\right)\right|\,|u|^{Y}\right\},
\end{align*}
where $C:=C_{+}=C_{-}$ can be interpreted as a measure of the ``intensity of small jumps" of the process. In that case, (see~\cite[(25.6)]{Sato:1999}),
\begin{equation}\label{1stOrdFA}
d_{1}:=\mathbb{E}\left(Z^{+}\right)=\frac{1}{\pi}\Gamma\left(1-\frac{1}{Y}\right)\left(2C\Gamma(-Y)\left|\cos\left(\frac{\pi Y}{2}\right)\right|\right)^{\frac{1}{Y}}.
\end{equation}
We refer the reader to Remark \ref{Rem:LedingTermPJ} below for the explicit expression of the leading term in the general case. Interestingly enough, {in} the presence of a continuous component (i.e., $\sigma\neq 0$ {in (\ref{CGMY1stDfn})}), the first-order asymptotic term only reflects information {about} the {volatility $\sigma$}, in sharp contrast with the pure-jump case where the leading term depends on the parameter $C$ {and} the index $Y$, which in turn respectively control the intensity of small jumps and the degree of jump activity of the process, as already stated above. The intuition behind (\ref{1stOAp}) is actually easy to explain. Indeed, since
\begin{align*}
{\left(S_{t}-S_{0}\right)^{+}=S_{0}\left(e^{X_{t}}-1\right)^{+}=S_{0}\left(e^{X_{t}^{+}}-1\right)\sim S_{0}(X_{t})^{+},}\qquad {(t\to{}0)}
\end{align*}
it is expected that, {for $\sigma=0$,}
\begin{equation}\label{IntuitFirsOrder}
	\lim_{t\to 0}t^{-\frac{1}{Y}}\mathbb{E}\left[\left(S_{t}-S_{0}\right)^{+}\right]={S_{0}}\lim_{t\to 0}\mathbb{E}\left(t^{-\frac{1}{Y}}X_{t}^{+}\right)={S_{0}\lim_{t\to 0}\mathbb{E}\left(t^{-\frac{1}{Y}}L_{t}^{+}\right)}={S_{0}}\mathbb{E}\left(Z_{1}^{+}\right),
\end{equation}
in light of (\ref{SRBTSP}). {In} the presence of {a Brownian} component {(i.e., $\sigma\neq{}0$), it is} known that $t^{-1/2}X_{t}\ld \sigma W_{1}$ as $t\to{}0$ (see \cite[pp. 40]{Sato:1999}).

The asymptotic result (\ref{1stOAp}) is in agreement with the result of {\cite[Theorem 5]{Tankov}}, which showed that for a pure-jump L\'{e}vy process $(X_{t})_{t\geq 0}$ (i.e., $\sigma=0$ in (\ref{CGMY1stDfn})), whose characteristic function is of the form
\begin{equation}\label{TnkvChFSpec}
\bbe\left(e^{iuX_{t}}\right)={\exp\left\{t\left(iu\gamma-|u|^{Y}f(u)\right)\right\}},
\end{equation}
with $Y\in (1,2)$ and a function $f$ satisfying $\lim_{u\nearrow\infty}f(u)=\hat{c}_{+}\in (0,\infty)$ and $\lim_{u\searrow-\infty}f(u)=\hat{c}_{-}\in (0,\infty)$,
\begin{equation}\label{1stOApTankov}
{\lim_{t\to 0}t^{-\frac{1}{Y}}\mathbb{E}\left[\left(S_{t}-S_{0}\right)^{+}\right]=\frac{S_{0}}{2\pi}\left(\hat{c}_{+}^{1/Y}+\hat{c}_{-}^{1/Y}\right).}
\end{equation}
An expression similar to (\ref{1stOAp}) {was} also obtained in~\cite[Theorem 4.4]{MuhNut:2009} for a more general class of pure-jump martingales. Nevertheless, as previously mentioned, none of the these papers obtained second or higher order terms for {ATM option} prices.

The main result of the present paper establishes, for the exponential L\'{e}vy model {(\ref{ExpLvMdl})-(\ref{CGMY1stDfn})}, a second-order correction term to (\ref{1stOAp}). To the best of our knowledge, this is the first second-order result in the literature of exponential L\'{e}vy models. {To wit,} the second-order asymptotic behavior of the ATM call option price (\ref{CallPriceDfn}) in short-time is of the form
\begin{align}\label{Asymp1}
{\frac{1}{S_{0}}\mathbb{E}\left[\left(S_{t}-S_{0}\right)^{+}\right]}=d_{1}t^{\frac{1}{Y}}+d_{2}t+o(t),\quad t\to 0,
\end{align}
in the pure-jump case (i.e., $\sigma=0$ in (\ref{CGMY1stDfn})), while in the presence of {an independent} Brownian component (i.e., $\sigma\neq 0$ {in (\ref{CGMY1stDfn})}),	
\begin{align}\label{Asymp2}
{\frac{1}{S_{0}}\mathbb{E}\left[\left(S_{t}-S_{0}\right)^{+}\right]}=d_{1}t^{\frac{1}{2}}+d_{2}t^{\frac{3-Y}{2}}+o\left(t^{\frac{3-Y}{2}}\right),\quad t\to 0,
\end{align}
for different constants $d_{1}$ and $d_{2}$ {determined} explicitly. {In} the presence of {an independent Brownian} component, the second-order term $d_{2}$ {only depends on} the degree of jump activity $Y$ {and} {on} the parameter $C:=(C_{+}+C_{-})/2$, which measures the net intensity of the small jumps of $X$ (regardless of their sign). In particular, it is impossible to discern the difference between $C_{+}$ and $C_{-}$ at this second-order {approximation}. In contrast, in {the} pure-jump case (i.e., $\sigma=0$ in (\ref{CGMY1stDfn})), the degree of jump activity of the L\'{e}vy process is already present in the first-order term. As a byproduct of our asymptotic results for {ATM} option prices, we also give the asymptotic behavior of the corresponding {ATM} Black-Scholes implied volatility, denoted by $\hat{\sigma}(t)$. Concretely, {in} the presence of {an independent} Brownian component (i.e., $\sigma\neq 0$ in (\ref{CGMY1stDfn})),
\begin{align}\label{AsyIVGerCGMY0}
\hat{\sigma}(t)=\sigma+\frac{(C_{+}+C_{-})2^{-\frac{Y}{2}}}{Y(Y-1)}\Gamma\left(1-\frac{Y}{2}\right)\sigma^{1-Y}t^{1-\frac{Y}{2}}+o\left(t^{1-\frac{Y}{2}}\right),\quad t\rightarrow 0,
\end{align}
while, in the absence of the continuous component,
\begin{align}\label{AsyIVPureCGMY0}
\hat{\sigma}(t)=\sigma_{1}t^{\frac{1}{Y}-\frac{1}{2}}+\sigma_{2}t^{{\frac{1}{2}}}+o(t^{{\frac{1}{2}}}),\quad t\rightarrow 0,
\end{align}
for some constants $\sigma_{1},\sigma_{2}$ that we will determine explicitly.
	
The results (\ref{Asymp1})-(\ref{AsyIVPureCGMY0}) {are novel} and, {moreover,} not as easily guessed as the first-order term. Indeed, for instance, the intuition behind (\ref{IntuitFirsOrder}) does not seem to be directly transferable to (\ref{Asymp1})-(\ref{Asymp2}) since, from the second-order Taylor expansion of the exponential,
\begin{align*}
	{\frac{1}{S_{0}}\left(S_{t}-S_{0}\right)^{+}=e^{X_{t}^{+}}-1\approx X_{t}^{+}+\frac{1}{2} X_{t}^{2}\;{{\,\stackrel{\frak {D}}{\approx}\,}}\;t^{\frac{1}{Y}}Z_{1}^{+}+\frac{1}{2}t^{\frac{2}{Y}}Z_{1}^{2},}
\end{align*}
where for the third approximation we again used that $t^{-1/Y}X_{t}\;{{\,\stackrel{\frak {D}}{\rightarrow}\,}}\; Z_{1}$ as $t\to{}0$. However, $\mathbb{E} Z_{1}^{2}=\infty$ and the argument fails after taking expectations. Our derivation is fully probabilistic and builds on two facts. First, {make} use of the following model-free representation due to \cite{CM09}:
\begin{align}\label{CMR}
{\frac{1}{S_{0}}\mathbb{E}\left[\left(S_{t}-S_{0}\right)^{+}\right]}=\mathbb{P}^{*}\left(X_{t}>E\right)=\int_{0}^{\infty}e^{-x}\mathbb{P}^{*}\left(X_{t}>x\right)dx,
\end{align}
where $\mathbb{P}^{*}$ is the martingale probability measure obtained when one takes the stock as the num\'{e}raire (i.e., $\mathbb{P}^{*}(A):=\mathbb{E}\left(S_{t}{\bf 1}_{A}\right)$ {for every $A\in\mathcal{F}_{t}$, $t\geq 0$}) and $E$ is an independent mean-one exponential random variable under $\mathbb{P}^{*}$. The measure $\mathbb{P}^{*}$ is also called the share measure (see~\cite{CM09}). Second, {further} change the probability measure $\bbp^{*}$ to another probability measure, say $\widetilde{\bbp}$, under which the pure jump-component of $X$ becomes a stable L\'{e}vy process, independent of the continuous component of $X$. This change, in turn, {enables} to exploit some key features of stable processes such as self-similarity and the asymptotic behavior of their marginal {densities}.

Let us finish this introductory section with a brief digression on the relevance as well as {further} extensions and applications of our results:
\begin{itemize}
\item L\'{e}vy models are often criticized for their partial ability to fit volatility surfaces across maturities and to account for some stylized features of stock prices such as volatility clustering. A natural question is then whether the results hereafter can be extended to more complex models. The short answer is ``yes". Even though a complete exposition of this point is {beyond} the scope of the present manuscript, {the reason can be broadly justified here. This} has its origins in the work of \cite{MuhNut:2009}, where first-order asymptotics of ATM option prices for a relatively general martingale process with jumps and stochastic volatility {are} obtained by first showing the analog for a suitable class of L\'{e}vy models and then proving that the option prices under the general model can be closely approximated by those of a suitably chosen L\'{e}vy model in the class. It is therefore expected that our second-order expansions will be valid for a much more general class of It\^{o} models satisfying certain regularity conditions. Thus, for instance, it is expected that (\ref{AsyIVGerCGMY0}) will hold with $\sigma$ replaced with the spot volatility at time $0$, say $\sigma_{0}$, which itself can be random depending on an additional risky factor such as in the Heston model. {We refer the reader to  \cite{FigOlaf} for more details about the latter type of results.}
\item From a qualitative point of view, the short-time expansions stated here {allow} to connect information of the model's parameters to key features of the option prices and the implied volatility smile {and}, at a more basic level, to identify (and rank) the parameters (or numerical features of functional parameters) that {most} influence the behavior of the option prices and implied volatilities in short-time. Hence, for instance, (\ref{AsyIVGerCGMY0}) {asserts that}, under a continuous component and an infinite stable-like jump activity component, the most important feature that determines the short-time behavior of the implied volatility, below the spot volatility $\sigma$, is the intensity of small-jumps as measured by $C:=(C_{+}+C_{-})/2$. {Moreover, this} influence is ``felt at the rate'' $t^{1-Y/2}$, {which is} much stronger than how the jump features start to be ``felt" in the case of a finite jump activity, where the rate is of order $t^{1/2}$ (see Remark \ref{CompMedSca} below).
\item From a quantitative point of view, as {already} mentioned, short-time asymptotics of option prices are relevant in model testing and calibration. The first type of application {is} already present in the seminal work of \cite{CW03}. {For} the second type of application, high-order expansions, as the ones obtained here, can facilitate numerical calibration {by} suggesting, {for instance,} proper functional forms for extrapolation purposes or {for} setting starting points of numerical calibration methods. We refer to Remark \ref{MoreOnCalibration} below for more information about {this} latter type of application.
\end{itemize}

The present article is organized as follows. Section \ref{Sec:TmpStble} introduces the class of L\'{e}vy processes studied thereafter and some probability measures transformations {needed} throughout the paper. Section 3 establishes the second-order asymptotics of the call option price under the pure-jump model ({i.e., $\sigma=0$ in (\ref{CGMY1stDfn})}). Section 4 establishes the second-order asymptotics of the call option price under an additional {independent} Brownian component ({i.e., $\sigma\neq 0$ in (\ref{CGMY1stDfn})}). Section \ref{Sec:CGMY} illustrates our second-order asymptotics for the important particular class of CGMY models, recovering our preliminary results first presented in~\cite{FigGH:2012b}. Section \ref{Numerics} assesses the performance of the asymptotic expansions through a detailed numerical analysis for the CGMY model. The proofs of our main results are deferred to the appendices.

\section{A {Tempered-Stable-Like Model} With Brownian {Component}}\label{Sec:TmpStble}

Let $L:=(L_{t})_{t\geq 0}$ be a pure-jump L\'{e}vy process with triplet $(0,b,\nu)$ and let $W:=(W_{t})_{t\geq 0}$ be a Wiener process, independent of $L$, defined on a complete filtered probability space $(\Omega,\calF,(\calF_{t})_{t\geq 0},\bbp)$. {Assume} zero interest rate and that $\bbp$ is a martingale measure for the exponential L\'{e}vy model
\begin{align*}
S_{t}=S_{0}e^{X_{t}},\quad\text{with }\,X_{t}:=\sigma W_{t}+L_{t},\quad {t\geq 0},
\end{align*}
where $(S_{t})_{t\geq 0}$ represents the price process of a non-dividend paying risky asset. Equivalently, the L\'{e}vy triplet $(\sigma^{2},b,\nu)$ of $X:=(X_{t})_{t\geq 0}$ is such that
\begin{equation}\label{NdCndTSMrt}
{\rm (i)}\,\,\int_{1}^{\infty}e^{x}\nu(dx)<\infty,\quad\text{and}\quad {\rm (ii)}\,\,\bbe\left(e^{X_{1}}\right)=\exp\left(b+\frac{\sigma^{2}}{2}+\int_{\bbr_{0}}\left(e^{x}-1-x{\bf 1}_{\{|x|\leq 1\}}\right)\nu(dx)\right)=1,
\end{equation}
where hereafter $\bbr_{0}:=\bbr\backslash\{0\}$ and the L\'{e}vy triplets are given relative to the truncation function ${\bf 1}_{\{|x|\leq 1\}}$ (see~\cite[Section 8]{Sato:1999}). Without loss of generality, {also} assume that $(X_{t})_{t\geq 0}$ is the canonical process $X_{t}(\omega):=\omega(t)$ defined on the canonical space $\Omega=\mathbb{D}([0,\infty),\mathbb{R})$ (the space of c\`{a}dl\`{a}g functions $\omega:[0,\infty)\to\mathbb{R}$) equipped with the $\sigma$-field $\mathcal{F}:=\sigma(X_{s}:s\geq{}0)$ and the right-continuous filtration $\mathcal{F}_{t}:=\cap_{s>t}\sigma(X_{u}:u\leq s)$).

As explained in the introduction, we consider a tempered-stable-like L\'{e}vy processes $L$ where the L\'{e}vy measure {$\nu$} admits a density $s:\bbr_{0}\to [0,\infty)$ of the form
\begin{align*}
s(x)=|x|^{-Y-1}q(x),
\end{align*}
for $Y\in(1,2)$ and a bounded measurable function $q:\bbr_{0}\to [0,\infty)$ such that
\begin{align}\label{KyCondValAllRsIntro0}
\lim_{x\searrow 0}\frac{1}{x}\left(C_{+}-q(x)\right)=\beta_{+},\qquad\lim_{x\nearrow 0}\frac{1}{x}\left(C_{-}-q(x)\right)=-\beta_{-},
\end{align}
for some positive constants $C_{+}$, $C_{-}$, $\beta_{+}$ and $\beta_{-}$. Throughout, {the following are the standing} assumptions:
\begin{assum}\label{StnAsm1}
The function $q:\bbr_{0}\to [0,\infty)$ is such that
\begin{align}\label{Eq:StndCnd1a}
&{\rm (i)}\,\,\,q(x)\leq C_{-},\quad\text{for all }\,x<0;\quad\quad {\rm (ii)}\,\,\,q(x)\leq C_{+}e^{-x},\quad\text{for all }\,x>0;\\
\label{Eq:StndCnd1b} &{\rm (iii)}\,\,\,\limsup_{|x|\to\infty}\frac{|\ln q(x)|}{|x|}<\infty;\qquad\quad\,\,{\rm (iv)}\,\,\,\inf_{|x|<\delta}q(x)>0,\quad\text{for any }\,\delta>0.
\end{align}
\end{assum}
A prototypical tempered-stable-like process, as understood here, is the CGMY process of widespread use in mathematical finance. For the CGMY process,
\begin{align}\label{LevyMeasCGMY}
\nu(dx)=\left(\frac{Ce^{-Mx}}{x^{1+Y}}\,{\bf 1}_{\{x>0\}}+\frac{Ce^{Gx}}{|x|^{1+Y}}\,{\bf 1}_{\{x<0\}}\right)dx,
\end{align}
so that $C_{+}=C_{-}=C$, $\beta_{+}=C_{+}M$, and $\beta_{-}=C_{-}G$. From this analogy, {let us} choose to re-express the model in terms of the parameters $M:=\beta_{+}/C_{+}$ and $G:=\beta_{-}/C_{-}$, and the function
\begin{equation}\label{DfnBarq0}
\bar{q}(x):=\frac{q(x)}{C_{+}}{\bf 1}_{\{x>0\}}+\frac{q(x)}{C_{-}}{\bf 1}_{\{x<0\}}.
\end{equation}
In that case, (\ref{KyCondValAllRsIntro0}) can equivalently be written as
\begin{align}\label{KyCondValAllRsIntro}
\lim_{x\searrow 0}\frac{1}{x}\left(1-\bar{q}(x)\right)=M,\qquad\lim_{x\nearrow 0}\frac{1}{x}\left(1-\bar{q}(x)\right)=-G.
\end{align}
The following relations are direct consequences of (\ref{KyCondValAllRsIntro}) and are only stated here for future reference:
\begin{align}\label{UsflLmt}
{\rm (i)}\,\,\,\lim_{\delta\searrow 0}\bar{q}\left(\delta y\right)^{\frac{1}{\delta}}=e^{-yM}{\bf 1}_{\{y>0\}}+e^{-|y|G}{\bf 1}_{\{y<0\}},\quad
{\rm (ii)}\,\,\,\lim_{\delta\searrow 0}\frac{1}{\delta}\ln\bar{q}\left(\delta y\right)=-yM{\bf 1}_{\{y>0\}}-|y|G{\bf 1}_{\{y<0\}},
\end{align}
for any $y\in\bbr\backslash\{0\}$. Let us also note that (\ref{Eq:StndCnd1a}-ii) implies the martingale condition (\ref{NdCndTSMrt}-i) and also that $M>1$.
\begin{rem}
The class of processes considered here is similar to the unifying class of Regular L\'{e}vy Processes of Exponential type (RLPE) as introduced in~\cite{BL02}. It also covers the class of proper tempered stable processes as defined in~\cite{Rosinski:2007} as well as several parametric models typically used in mathematical finance, including the CGMY processes and the more general class of normal tempered stable processes  (cf.~\cite[Section 4.4.3]{CT04}). Let us also remark that the range of $Y$ considered here (namely, $Y\in(1,2)$) may arguably be the most relevant for financial applications in {that} several recent econometric studies of high-frequency financial data {suggest} that the Blumenthal-Getoor index is larger than $1$ (cf.~\cite{AitJacod09}, \cite{Belomestny:2010}, and references therein). Nevertheless, admittedly the Blumenthal-Getoor index is in general relatively hard to estimate and different, {less recent}, studies have indicated values of $Y<1$ for some financial data set (see, e.g., \cite{CGMY:2002}).
\end{rem}

Following a density transformation construction as given in \cite[Definition 33.4 and Example 33.14]{Sato:1999} and using the martingale condition {$\bbe\left(e^{X_{t}}\right)=1$}, {define} a probability measure $\mathbb{P}^{*}$ on $(\Omega,\mathcal{F})$ via
\begin{equation}\label{DSM}
{\left.\frac{d\bbp^{*}}{d\bbp}\right|_{\calF_{t}}}=e^{X_{t}},\quad t\geq 0,
\end{equation}
i.e., $\mathbb{P}^{*}(B)=\mathbb{E}\left(e^{X_{t}}{\bf 1}_{B}\right)$, for any $B\in\mathcal{F}_{t}$ and $t\geq 0$. The measure $\mathbb{P}^{*}$, sometimes {called} the share measure, can be interpreted as the martingale measure when using the stock price as the num\'{e}raire. Under this probability measure, $(X_{t})_{t\geq 0}$ has the representation $(\sigma W^{*}_{t}+L^{*}_{t})_{t\geq 0}$ where, under $\mathbb{P}^{*}$, $W^{*}:=(W^{*}_{t})_{t\geq 0}$ is a Wiener process and $L^{*}:=(L^{*}_{t})_{t\geq 0}$ is a L\'{e}vy process with triplet $(0,b^{*},\nu^{*})$ given by
\begin{align*}
\nu^{*}(dx):=e^{x}\nu(dx)=e^{x}s(x)dx,\qquad b^{*}:=b+\int_{|x|\leq 1}x\left(e^{x}-1\right)s(x)dx+\sigma^{2},
\end{align*}
which is moreover independent of $W^{*}$. Hereafter, {set}
\begin{align*}
q^{*}(x):=q(x)e^{x},\qquad s^{*}(x):=|x|^{-Y-1}q^{*}(x),\qquad\bar{q}^{*}(x):=\bar{q}(x)e^{x}.
\end{align*}
The next relations {are direct consequences of} (\ref{UsflLmt}):
\begin{align}\label{UsflLmtq*}
{\rm (i)}\,\,\,\lim_{\delta\searrow 0}\bar{q}^{*}\left(\delta y\right)^{\frac{1}{\delta}}=e^{-yM^{*}}{\bf 1}_{\{y>0\}}+e^{-|y|G^{*}}{\bf 1}_{\{y<0\}},\quad
{\rm (ii)}\,\,\,\lim_{\delta\searrow 0}\frac{1}{\delta}\ln\bar{q}^{*}\left(\delta y\right)=-yM^{*}{\bf 1}_{\{y>0\}}-|y|G^{*}{\bf 1}_{\{y<0\}},
\end{align}
where {$M^{*}:=M-1$ and $G^{*}:=G+1$.}

An important tool thereafter is to change probability measures from $\bbp^{*}$ to another probability measure, {denoted} by $\widetilde{\bbp}$, under which $(L^{*}_{t})_{t\geq 0}$ is a stable L\'{e}vy process and $(W^{*}_{t})_{t\geq 0}$ is a Wiener process independent of $L^{*}$. Concretely, let
\begin{align}\label{DfnTripletTildeZ}
\tilde{q}(x):=C_{+}{\bf 1}_{\{x>0\}}+C_{-}{\bf 1}_{\{x<0\}},\quad\tilde{\nu}(dx):=|x|^{-Y-1}\tilde{q}(x)dx,\quad\tilde{b}:=b^{*}+\int_{|x|\leq 1}x(\tilde{\nu}-\nu^{*})(dx).
\end{align}
Note that $\tilde{\nu}$ is the L\'{e}vy measure of a $Y$-stable L\'{e}vy process. Moreover, $\tilde{\nu}$ is equivalent to $\nu^{*}$ since clearly $\nu^{*}(dx)=e^{x}q(x)\tilde{q}(x)^{-1}\tilde{\nu}(dx)$ and $q$ is strictly positive in {view} of (\ref{Eq:StndCnd1b}-iv). For future reference, it is convenient to write $\tilde{\nu}$ as
\begin{align*}
\tilde{\nu}(dx)=e^{\varphi(x)}\nu^{*}(dx),\quad\text{with}\quad\varphi(x):=-\ln\bar{q}^{*}(x)=-\ln\bar{q}(x)-x.
\end{align*}
By virtue of~\cite[Theorem 33.1]{Sato:1999}, there exists a probability measure $\widetilde{\bbp}$ locally equivalent\footnote{Equivalently, there exists a process $(U_{t})_{t\geq 0}$ such that ${\widetilde{\bbp}(B)}=\bbe^{*}(e^{U_{t}}{\bf 1}_{B})$, for {any} $B\in\mathcal{F}_{t}$ and $t\geq 0$.} to $\bbp^{*}$ such that $(X_{t})_{t\geq 0}$ is a L\'{e}vy process with L\'{e}vy triplet {$(\sigma^{2},\tilde{b},\tilde{\nu})$} under $\widetilde{\bbp}$, provided that the following condition is satisfied:
\begin{align*}
\int_{\bbr_{0}}\left(e^{\varphi(x)/2}-1\right)^{2}\nu^{*}(dx)=\int_{\bbr_{0}}\left(1-e^{-\varphi(x)/2}\right)^{2}\tilde{\nu}(dx)<\infty.
\end{align*}
To see that the previous condition holds under our assumptions, note that the integral therein can be expressed as
\begin{equation}\label{NExpTIn}
C_{+}\int_{0}^{\infty}\left(1-e^{\frac{1}{2}\ln\bar{q}^{*}(x)}\right)^{2}|x|^{-Y-1}dx+C_{-}\int_{-\infty}^{0}\left(1-e^{\frac{1}{2}\ln\bar{q}^{*}(x)}\right)^{2}|x|^{-Y-1}dx,
\end{equation}
and {that}, by (\ref{UsflLmtq*}-ii), the integrand in the first integral is such that
\begin{align*}
\left(e^{\frac{1}{2}\ln\bar{q}^{*}(x)}-1\right)^{2}\sim\frac{1}{4}\left(\ln\bar{q}^{*}(x)\right)^{2}\sim\frac{(M^{*})^{2}}{4}x^{2},\quad\text{as }\,x\searrow 0.
\end{align*}
This shows that the first integral is finite on any interval $(0,\varepsilon)$. Outside any neighborhood of the origin, this integral is finite in view of (\ref{NdCndTSMrt}-i). The second integral in (\ref{NExpTIn}) can be {similarly} handled using (\ref{UsflLmtq*}) and the fact that
\begin{equation}\label{ImpSignCond}
\varphi(x)=-\ln\bar{q}^{*}(x)\geq 0,\quad\text{for any }\,x<0,
\end{equation}
as seen from the assumption (\ref{Eq:StndCnd1a}-i).

Now that {the existence of the probability measure $\widetilde{\bbp}$ is established}, let us state some {of its} properties and introduce some related terminology. Throughout, $\widetilde{\bbe}$ denotes {the} expectation under $\widetilde{\bbp}$. Letting {$\widetilde{\gamma}:=\widetilde{\bbe}\left(X_{1}\right)=\widetilde{\bbe}\left(L_{1}^{*}\right)$}, {recall} that the centered process $(Z_{t})_{t\geq 0}$, defined by
\begin{equation}\label{SSSP}
Z_{t}:=L_{t}^{*}-t\tilde{\gamma},\quad {t\geq 0},
\end{equation}
is a strictly $Y$-stable under $\widetilde{\bbp}$ {and, thus}, also self-similar, i.e.,
\begin{equation}\label{SSCN}
\left(t^{-1/Y}Z_{ut}\right)_{u\geq 0}\ed\left(Z_{u}\right)_{u\geq 0},\quad\text{for any }\,t>0.
\end{equation}
Let $p_{Z}$ denote the marginal density function of $Z_{1}$ under $\widetilde{\bbp}$. It is well known (see, e.g., \cite[(14.37)]{Sato:1999} and references therein) that
\begin{equation}\label{Asydenpz00}
p_{Z}(v)\sim C_{\pm}|v|^{-Y-1},\quad\text{as }\,v\to\pm\infty,\,\,\,\text{respectively}.
\end{equation}
In particular, for any $v>0$, as $t\to 0$,
\begin{align}\label{AsytailZ100}
\widetilde{\bbp}\left(Z_{t}\geq v\right)&=\widetilde{\bbp}\left(Z_{1}\geq t^{-1/Y}v\right)\sim t\tilde{\nu}([v,\infty))=t\,\frac{C_+}{Y}v^{-Y},\\
\label{AsytailZ100b} \widetilde{\bbp}\left(Z_{t}\leq -v\right)&=\widetilde{\bbp}\left(Z_{1}\leq -t^{-1/Y}v\right)\sim t\tilde{\nu}((-\infty,-v])=t\,\frac{C_{-}}{Y}v^{-Y}.
\end{align}
The following tail estimate can also be deduced from (\ref{Asydenpz00}) (see Appendix \ref{AddtnProofs} for its proof):
\begin{equation}\label{KIn}
\widetilde{\bbp}\left(\left|Z_{1}\right|\geq t^{-1/Y}v\right)=\widetilde{\bbp}\left(|Z_{t}|\geq v\right)\leq\kappa t|v|^{-Y},
\end{equation}
for all $v>0$ and $0<t\leq 1$ and some absolute constant $0<\kappa<\infty$.

Also of use,  below, is the following representation of the log density process
\begin{equation}\label{EMMN}
U_{t}:={\left.\ln\frac{d\widetilde{\bbp}}{d\bbp^{*}}\right|_{\calF_{t}}}=\lim_{\epsilon\to 0}\left(\sum_{\underset{|\Delta X_{s}|>\varepsilon}{s\leq t:}}\varphi\left(\Delta X_{s}\right)-t\int_{|x|>\varepsilon}\left(e^{\varphi(x)}-1\right)\nu^{*}(dx)\right),\quad {t\geq 0},
\end{equation}
(cf.~\cite[Theorem 33.2]{Sato:1999}). The process $(U_{t})_{t\geq 0}$ can be expressed in terms of the jump-measure $N(dt,dx):=\#\{(s,\Delta X_{s})\in dt\times dx\}$ of the process $(X_{t})_{t\geq 0}$ and its compensated measure ${\bar{N}}(dt,dx):=N(dt,dx)-\tilde{\nu}(dx)dt$ (under $\widetilde{\bbp}$). Indeed, {for any $t\geq 0$,}
\begin{align*}
U_{t}&=\lim_{\epsilon\to 0}\left(\int_{0}^{t}\int_{|x|>\varepsilon}\varphi(x)N(ds,dx)-t\int_{|x|>\varepsilon}\left(e^{\varphi(x)}-1\right)e^{-\varphi(x)}\tilde{\nu}(dx)\right)\\
&=\lim_{\epsilon\to 0}\left(\int_{0}^{t}\int_{|x|>\varepsilon}\varphi(x){\bar{N}}(ds,dx)+t\int_{|x|>\varepsilon}\left(e^{-\varphi(x)}-1+\varphi(x)\right)\tilde{\nu}(dx)\right).
\end{align*}
Thus, from the definition of the Poisson integral under a compensated Poisson measure as in~\cite[Theorem 33.2]{Kallenberg},
\begin{equation}\label{DcmLL}
U_{t}=\widetilde{U}_{t}+\eta t,\quad {t\geq 0},
\end{equation}
with
\begin{align}\label{Uplusminuseta}
\widetilde{U}_{t}:=\int_{0}^{t}\int_{\bbr_{0}}\varphi(x){\bar{N}}(ds,dx),\quad\eta:=\int_{\bbr_{0}}\left(e^{-\varphi(x)}-1+\varphi(x)\right)\tilde{\nu}(dx).
\end{align}
In particular, from the definition of $(U_{t})_{t\geq 0}$ in (\ref{EMMN}), the process $e^{-U_{t}}$ is a $\widetilde{\bbp}$-martingale and, {so,}
\begin{equation}\label{ExpMmntTildeU}
\widetilde{\bbe}\left(e^{-\widetilde{U}_{t}}\right)=e^{\eta t},\quad {t\geq 0}.
\end{equation}
\begin{rem}
The constant $\eta$ above is well defined {and, therefore, so is $\widetilde{U}_{t}$}. Indeed, from (\ref{Eq:StndCnd1b}-iv) and (\ref{UsflLmtq*}-ii),
\begin{align*}
\int_{|x|\leq 1}\left|e^{-\varphi(x)}-1+\varphi(x)\right||x|^{-Y-1}dx\leq C_{1}\int_{|x|\leq 1}\varphi^{2}(x)|x|^{-Y-1}dx\leq C_{2}\int_{|x|\leq 1}|x|^{1-Y}dx<\infty,
\end{align*}
for some constants $0<C_{1},C_{2}<\infty$. Similarly, from (\ref{ImpSignCond}) and both conditions in  (\ref{Eq:StndCnd1b}),
\begin{align*}
\int_{x<-1}\left|e^{-\varphi(x)}-1+\varphi(x)\right||x|^{-Y-1}dx\leq C_{1}\int_{x<-1}|\varphi(x)||x|^{-Y-1}dx\leq C_{2}\int_{x<-1}|x|^{-Y}dx<\infty.
\end{align*}
Finally, from (\ref{NdCndTSMrt}-i) and (\ref{Eq:StndCnd1b}), for some constant $0<C_{1}<\infty$,
\begin{align*}
\int_{x>1}\left|e^{-\varphi(x)}-1+\varphi(x)\right||x|^{-Y-1}dx\leq C_{1}\int_{x>1}e^{\ln q(x)+x}x^{-Y-1}dx+\int_{x>1}\left|-1+\varphi(x)\right|x^{-Y-1}dx<\infty.
\end{align*}
\end{rem}

The following decomposition of the process $X$ in terms of the compensated measure {$\bar{N}$} is also of use throughout:
\begin{equation}\label{RX}
X_{t}=\sigma W_{t}^{*}+Z_{t}+t\tilde{\gamma}=\sigma W_{t}^{*}+\int_{0}^{t}\int_{\bbr_{0}}x{\bar{N}}(ds,dx)+t\tilde{\gamma},\quad {t\geq 0}.
\end{equation}
The representation (\ref{RX}) can be deduced from the L\'{e}vy-It\^{o} decomposition of the process $(X_{t})_{t\geq 0}$ (cf.~\cite[Theorem 13.4 and Corollary 13.7]{Kallenberg} or~\cite[Theorem 19.2]{Sato:1999}) with the stated L\'{e}vy triplet $(0,\tilde{b},\tilde{\nu})$ of $X$ under $\widetilde{\bbp}$ and where, by construction, $\tilde{b}$ must be such that {$\widetilde{\bbe}(X_{1})=\tilde{\gamma}$}.

\section{{The Pure-Jump Model}}\label{Sec:PureJump}

{This section finds} the second-order asymptotic behavior for the at-the-money call option prices (\ref{CallPriceDfn}) in the pure-jump model (i.e., $\sigma=0$ {in (\ref{CGMY1stDfn}) and, thus, $X_{t}=L_{t}$, for $t\geq 0$}). The proofs of all the results of this section are deferred to Appendix \ref{proofA}. Before stating our {results}, {the call option price (\ref{CallPriceDfn}) needs to be rewritten} in a suitable form.
\begin{lem}\label{Lm:NRATMPJ}
With the probability measure $\widetilde{\bbp}$ defined in (\ref{EMMN}) and the parameter {$\widetilde{\gamma}:=\widetilde{\bbe}\left(X_{1}\right)=\widetilde{\bbe}\left(L_{1}^{*}\right)$},
\begin{align}\label{LOP}
t^{-\frac{1}{Y}}\frac{1}{S_{0}}{\mathbb{E}\left[\left(S_{t}-S_{0}\right)^{+}\right]}=e^{-(\tilde{\gamma}+\eta)t}\int_{-\tilde{\gamma}t^{1-\frac{1}{Y}}}^{\infty}e^{-t^{\frac{1}{Y}}v}\,
\widetilde{\bbe}\left(e^{-\widetilde{U}_{t}}{\bf 1}_{\left\{t^{-\frac{1}{Y}}Z_{t}\geq v\right\}}\right)dv.
\end{align}
\end{lem}
The next two lemmas are crucial to obtain the main result of the section:
\begin{lem}\label{Lm:AsymBhvrToSt}
In the setting and under the assumptions of Section \ref{Sec:TmpStble}, for any $\xi\geq 0$,
\begin{align}\label{LmtMGFTldU}
{\rm (i)}\,\,\,\lim_{t\to 0}\widetilde{\bbe}\left(e^{-\xi t^{-\frac{1}{Y}}\left(\widetilde{U}_{t}+Z_{t}\right)}\right)=e^{\bar{\eta}\xi^{Y}},\quad
{\rm (ii)}\,\,\,\lim_{t\to 0}\widetilde{\bbe}\left(e^{-\xi t^{-\frac{1}{Y}}\widetilde{U}_{t}}\right)=e^{\bar{\eta}^{*}}\xi^{Y},
\end{align}
where $\bar{\eta}:=\Gamma(-Y)\left(C_{+}\,M^{Y}+C_{-}\,G^{Y}\right)$ and $\bar{\eta}^{*}=\Gamma(-Y)\left(C_{+}(M^{*})^{Y}+C_{-}(G^{*})^{Y}\right)$. In particular, both $t^{-1/Y}(\widetilde{U}_{t}+Z_{t})$ and $t^{-1/Y}\widetilde{U}_{t}$ converge to a $Y$-stable distribution as $t\to 0$.
\end{lem}
\begin{lem}\label{KLTCTBR}
In the setting and under the assumptions of Section \ref{Sec:TmpStble}, the following two assertions hold true:
\begin{enumerate}
\item For any $v>0$,
    \begin{align}\label{SmplLmtSLJ1}
    \lim_{t\to 0}t^{-1}\widetilde{\bbp}\left(Z_{t}^{+}+\widetilde{U}_{t}\geq v\right)=\int_{\bbr_{0}}{\bf 1}_{\{x^{-}-\ln\bar{q}(x)\geq v\}}\tilde{\nu}(dx).
    \end{align}
\item There exist constants $0<\tilde{\kappa}<\infty$ and $t_{0}>0$ such that
    \begin{equation}\label{TIITI}
    {\rm (i)}\,\,\,t^{-1}\widetilde{\bbp}\left(\widetilde{U}_{t}\geq v\right)\leq\tilde{\kappa}v^{-Y},\quad {\rm (ii)}\,\,\,t^{-1}\widetilde{\bbp}\left(Z_{t}^{+}+\widetilde{U}_{t}\geq v\right)\leq\tilde{\kappa}v^{-Y},
    \end{equation}
    for any $0<t\leq t_{0}$ and $v>0$.
\end{enumerate}
\end{lem}

{The following theorem} gives the second-order asymptotic behavior of {ATM} call option prices under an exponential {tempered-stable-like} L\'{e}vy model. {It is useful to recall from Section \ref{Sec:TmpStble} that, under $\widetilde{\bbp}$, $\{Z_{t}\}_{t\geq{}0}$ is a strictly stable L\'evy process with L\'evy measure $\tilde{\nu}(dx):=|x|^{-Y-1}\left(C_{+}{\bf 1}_{\{x>0\}}+C_{-}{\bf 1}_{\{x<0\}}\right)dx$.}
\begin{thm}\label{2ndASY}
For the exponential L\'{e}vy model (\ref{ExpLvMdl}) without Brownian component {({i.e., $\sigma=0$ in (\ref{CGMY1stDfn})})} and under the conditions of Section \ref{Sec:TmpStble},
\begin{equation}\label{CL2}		
\lim_{t\to 0}t^{\frac{1}{Y}-1}\left(t^{-\frac{1}{Y}}\frac{1}{S_{0}}{\mathbb{E}\left[\left(S_{t}-S_{0}\right)^{+}\right]}-\widetilde{\bbe}\left(Z_{1}^{+}\right)\right)=\tilde{\vartheta}+\tilde{\gamma}\,\widetilde{\bbp}\left(Z_{1}\geq 0\right),
\end{equation}
where, in terms of the function $\bar{q}$ introduced in (\ref{DfnBarq0}),
\begin{align}\label{vartheta}
\tilde{\vartheta}&:=C_{+}\int_{0}^{\infty}\left(e^{x}\bar{q}(x)-\bar{q}(x)-x\right)x^{-Y-1}dx,\\
\label{Deftildegamma} \tilde{\gamma}&:=\widetilde{\bbe}\left(L^{*}_{1}\right)={\Blue b+\frac{C_{+}-C_{-}}{Y-1}+C_{+}\int_{0}^{1}x^{-Y}\left(1-\bar{q}(x)\right)dx-C_{-}\int_{-1}^{0}|x|^{-Y}\left(1-\bar{q}(x)\right)dx}.
\end{align}
\end{thm}
\begin{rem}\label{Rem:LedingTermPJ}
The leading term $\widetilde{\bbe}\left(Z_{1}^{+}\right)$ can be explicitly computed via the absolute first moment of $Z_{1}$ since $\widetilde{\bbe}\left|Z_{1}\right|=2\widetilde{\bbe}\left(Z_{1}^{+}\right)-\widetilde{\bbe}\left(Z_{1}\right)=2\widetilde{\bbe}\left(Z_{1}^{+}\right)$. It turns out that
\begin{align}
\widetilde{\bbe}\left(Z_{1}^{+}\right)=\frac{1}{2}\widetilde{\bbe}|Z_{1}|&=\frac{1}{\pi}\left(C_{+}+C_{-}\right)^{\frac{1}{Y}}\Gamma(-Y)^{\frac{1}{Y}}\left|\cos\left(\frac{Y\pi}{2}\right)\right|^{\frac{1}{Y}}\Gamma\left(1-\frac{1}{Y}\right)\left(1+\left(\frac{C_{+}-C_{-}}{C_{+}+C_{-}}\right)^{2}\tan^{2}\left(\frac{Y\pi}{2}\right)\right)^{\frac{1}{2Y}}\nonumber\\
\label{1stOrdFAGen} &\quad\,\,\,\times\cos\left(\frac{1}{Y}\arctan\left(\frac{C_{+}-C_{-}}{C_{+}+C_{-}}\tan\left(\frac{Y\pi}{2}\right)\right)\right),
\end{align}
(see the proof of~\cite[(1.2.13)]{SamoTaqqu}). In the symmetric case (i.e. $C_{-}=C_{+}=C$), (\ref{1stOrdFAGen}) simplifies to (\ref{1stOrdFA}). The following {closed form} formula for the probability $\widetilde{\bbp}\left(Z_{1}\geq{}0\right)$ is also known:
\begin{equation}\label{PstStb}
	\widetilde{\bbp}\left(Z_{1}\geq{}0\right)=\frac{1}{2}+\frac{1}{\pi Y}\arctan\left(\frac{C_{+}-C_{-}}{C_{+}+C_{-}}
	\tan\left(\frac{Y\pi }{2}\right)\right);
\end{equation}
see \cite[Section VIII.1]{Bertoin} and \cite[Section 2.2]{Zolotarev}. {\Blue Furthermore, after plugging the value $b$, as determined by the condition (\ref{NdCndTSMrt}-ii), in (\ref{Deftildegamma}), we get the following expression for the limit in (\ref{CL2}):}
\begin{align*}
	{\Blue \tilde{\vartheta}+\tilde{\gamma}\widetilde{\bbp}\left(Z_{1}\geq 0\right)}&=
	{\Blue C_{+}\widetilde{\bbp}\left(Z_{1}< 0\right)\int_{0}^{\infty}\left(e^{x}\bar{q}(x)-\bar{q}(x)-x\right)x^{-Y-1}dx}\\
	&\quad - {\Blue C_{-}\widetilde{\bbp}\left(Z_{1}\geq{} 0\right)\int_{-\infty}^{0}\left(e^{x}\bar{q}(x)-\bar{q}(x)-x\right)|x|^{-Y-1}dx.}
\end{align*}
\end{rem}
\begin{rem}\label{Rem:ApproxPJ}
From (\ref{CL2}), it follows that the short-time second-order asymptotic behavior of the ATM call option price (\ref{CallPriceDfn}) has the form:
\begin{equation}\label{ExpAsymBehCGMY}
\frac{1}{S_{0}}{\mathbb{E}\left[\left(S_{t}-S_{0}\right)^{+}\right]}=d_{1}t^{{\frac{1}{Y}}}+d_{2}t +o(t),\quad t\to 0,
\end{equation}
with $d_{1}:=\widetilde{\bbe}\left(Z_{1}^{+}\right)$ given in (\ref{1stOrdFAGen}) and $d_{2}:=\tilde{\vartheta}+\tilde{\gamma}\widetilde{\bbp}\left(Z_{1}\geq 0\right)$. {In particular,} the first-order term only synthesizes information on the degree of jump activity as measured by $Y$ and the intensity of small jumps as measured by $C_{+}$ and $C_{-}$. However, the second-order term also incorporates some information about the tempering function $\bar{q}$. Let us also point out that even though the parameters $M$ and $G$ introduced in (\ref{KyCondValAllRsIntro}) do not explicitly show up in the second-order expansion, their existence guarantees that the last two terms in the right-hand side of (\ref{Deftildegamma}) are well-defined. In particular, the result is expected to be true if $M=0$ or $G=0$, but not if either $M$ or $G$  {is} $\infty$.
\end{rem}

{Let us} proceed to study the asymptotic behavior of the corresponding Black-Scholes implied {volatility}. Throughout, let $\hat{\sigma}(t)$ denote the ATM Black-Scholes implied volatility at maturity $t$ with zero interest rates and {zero} dividend yield. The following result {gives an easy derivation of the second-order asymptotic expansion of $\hat{\sigma}(t)$ as $t\rightarrow 0$. {For a more cumbersome, yet methodical,} procedure to derive an asymptotic expansion for $\hat{\sigma}(t)$ of arbitrary order, we refer the reader to the recent manuscript \cite{GaoLee:2013}.}
\begin{cor}\label{AsyIVPCGMY}
For the exponential L\'{e}vy model (\ref{ExpLvMdl}) without Brownian component, the implied volatility $\hat{\sigma}$ has the following small-time behavior:
\begin{align}\label{AsyIVPureCGMY}
\hat{\sigma}(t)=\sigma_{1}t^{\frac{1}{Y}-\frac{1}{2}}+\sigma_{2}t^{{\frac{1}{2}}}+o(t^{{\frac{1}{2}}}),\quad t\rightarrow 0,
\end{align}
where
\begin{align}\label{1stCoefIVPureCGMY}
\sigma_{1}:=\sqrt{2\pi}\,\widetilde{\bbe}\left(Z_{1}^{+}\right),\quad
\sigma_{2}:=\sqrt{2\pi}\left(\tilde{\vartheta}+\tilde{\gamma}\widetilde{\bbp}\left(Z_{1}\geq 0\right)\right),
\end{align}
and where $\tilde{\vartheta}$ and $\tilde{\gamma}$ are respectively given in (\ref{vartheta}) and (\ref{Deftildegamma}).
\end{cor}

\section{{The Pure-Jump Model With A Nonzero Brownian Component}}\label{Sect:NonZeroBrwn}

In this part, {the case of a nonzero independent Brownian component is considered}. Concretely, throughout, $(X_{t})_{t\geq 0}$ is a L\'{e}vy process with triplet $(\sigma^{2},b,\nu)$ as introduced in Section \ref{Sec:TmpStble} and $\sigma\neq 0$. The first-order asymptotic behavior for the ATM European call options in this mixed model is obtained in~\cite{Tankov} using Fourier methods. {The second-order correction term for the at-the-money European call option price is given next. As previously done}, we change the probability measure $\bbp^{*}$ to $\widetilde{\bbp}$ so that $X_{t}=t\tilde{\gamma}+\sigma W_{t}^{*}+Z_{t}$, {$t\geq 0$,} with $(Z_{t})_{t\geq 0}$ a strictly $Y$-stable L\'{e}vy process under $\widetilde{\bbp}$ (see (\ref{SSSP})). Recall also that, under both $\bbp^{*}$ and $\widetilde{\bbp}$, $W^{*}$ is still a standard Brownian motion, independent of $(Z_{t})_{t\geq 0}$. We will also make use of the decompositions (\ref{DcmLL}) and (\ref{RX}). The {proofs} of the {results} below {are} presented in Appendix \ref{ProofsSectGenCse}.
\begin{thm}\label{2ndOAsyCGMYB}
For the exponential L\'{e}vy model (\ref{ExpLvMdl}) with a nonzero {independent} Brownian component and under the conditions of Section \ref{Sec:TmpStble}, the ATM European call option price is such that
\begin{align}\label{2ndAsyCGMYB}
\lim_{t\rightarrow 0}t^{\frac{Y}{2}-1}\left(t^{-\frac{1}{2}}\frac{1}{S_{0}}{\bbe\left[\left(S_{t}-S_{0}\right)^{+}\right]-\sigma\bbe^{*}\left(\left(W_{1}^{*}\right)^{+}\right)}\right)=\frac{C_{-}+C_{+}}{2Y(Y-1)}\,\sigma^{1-Y}\,\bbe^{*}\left(\left|W_{1}^{*}\right|^{1-Y}\right).
\end{align}
\end{thm}
\begin{rem}\label{Rem:ApproxGen}
The $(1-Y)$-centered moment of a standard normal distribution is given by (see, e.g.,~\cite[(25.6)]{Sato:1999}):
\begin{align*}
\bbe^{*}\left(\left|W_{1}^{*}\right|^{1-Y}\right)=\frac{2^{\frac{1-Y}{2}}}{\sqrt{\pi}}\Gamma\left(1-\frac{Y}{2}\right).
\end{align*}
Thus, the second-order asymptotic behavior of the ATM call option price (\ref{CallPriceDfn}) in short-time takes the form
\begin{equation}\label{ExpAsymBehCGMYBM}
\frac{1}{S_{0}}{\mathbb{E}\left[\left(S_{t}-S_{0}\right)^{+}\right]}=d_{1}t^{\frac{1}{2}}+d_{2}t^{\frac{3-Y}{2}}+o\left(t^{\frac{3-Y}{2}}\right),\quad t\to 0,
\end{equation}
with
\begin{equation}\label{PrcDefn2ndTrm}
d_{1}:=\frac{\sigma}{\sqrt{2\pi}},\quad d_{2}:=\frac{2^{\frac{1-Y}{2}}}{\sqrt{\pi}}\Gamma\left(1-\frac{Y}{2}\right)\frac{\left(C_{-}+C_{+}\right)\sigma^{1-Y}}{2Y(Y-1)}.
\end{equation}
The first-order term only synthesizes the information {on the volatility parameter $\sigma$ of the continuous component}. In fact, the first-order term of the ATM call option price under the mixed tempered-stable-like model is the same as the one under the Black-Scholes model. The second-order term further incorporates the information on the degree of jump activity $Y$ and the net intensity of small jumps as measured by the parameter $\bar{C}:=C_{-}+C_{+}$. However, these two-terms do not reflect the individual intensities of the small negative or positive jumps as measured by the values of $C_{-}$ and $C_{+}$, {respectively}.  This fact suggests that it could be useful to consider a third-order approximation. For more information on the latter in the CGMY model, we refer the reader to~\cite{FigGH:2013}.
\end{rem}

The {forthcoming} proposition provides the small-time asymptotic behavior for the ATM Black-Scholes implied volatility under the tempered-stable-like model with {a nonzero} independent Brownian component. Unlike the pure-jump case, {the first-order asymptotics can only be derived from} Theorem \ref{2ndOAsyCGMYB}. The second-order term for the implied volatility requires {the} third-order asymptotics for the ATM call option price.
\begin{cor}\label{AsyIVCGMYB}
For the exponential L\'{e}vy model (\ref{ExpLvMdl}) with {a nonzero} independent Brownian component, the implied volatility $\hat{\sigma}$ has the following small-time behavior:
\begin{align}\label{AsyIVGerCGMY}
\hat{\sigma}(t)=\sigma+\frac{(C_{+}+C_{-})2^{-\frac{Y}{2}}}{Y(Y-1)}\Gamma\left(1-\frac{Y}{2}\right)\sigma^{1-Y}t^{1-\frac{Y}{2}}+o\left(t^{1-\frac{Y}{2}}\right),\quad t\rightarrow 0.
\end{align}
\end{cor}
\begin{rem}\label{CompMedSca}
It is {of interest} to compare (\ref{AsyIVGerCGMY}) with the corresponding asymptotics under the presence of a jump component $L$ of finite activity (namely, a compound Poisson process). Indeed, \cite{MedSca07} {argued}, under a relatively strong technical condition (see Proposition 5 therein for details), {that} the implied volatility is asymptotically $\sigma+I_{1}\sqrt{t}+O(t)$, {as $t\to 0$,} for a certain constant $I_{1}$. As {seen from} (\ref{AsyIVGerCGMY}), $\hat{\sigma}(t)$ converges {at a slower rate} under a tempered stable-like infinite jump activity component. In fact, the higher the degree of jump activity, the slower the rate of convergence is {and, thus}, the stronger is the effect of the intensity of small jumps in the asymptotic behavior of the implied volatility in short-time.
\end{rem}
\begin{rem}\label{MoreOnCalibration}
As mentioned in the introduction, short-time asymptotics of option prices are relevant in numerical calibration. Here, we illustrate an approach for this purpose inspired by a method proposed in~\cite{MedSca07}. As outlined in the introduction, it is expected that (\ref{AsyIVGerCGMY}) is valid under a stochastic volatility model with $\sigma$ replaced with the spot volatility at time $0$, say $\sigma_{0}$, which itself can be random depending on an additional risky factor (see~\cite{FigOlaf} for further information about this type of results). Now, two major interrelated issues arise. {The} resulting asymptotic formula would involve the unobserved spot volatility, which is changing in time. Thus, apparently the calibration will have to be carried out day by day, but unfortunately, a limited number of short-term options are available in the market at {any} given day. To overcome these issues, {argue} as follows: First, it is reasonable that observed prices of options ``sufficiently close-to-the-money" can be considered as being at-the-money {(see~\cite{FigOlaf} for a more formal justification of this step)}. Hence, if $\kappa$ denotes the log-moneyness $\ln\left(K/S_{0}\right)$ of the option and $\hat\sigma(t;\kappa)$ denotes the corresponding {Black-Scholes} implied volatility under the model, {it is expected} that
\begin{equation}\label{AsyIVGerCGMY2}
\hat\sigma(t;\kappa)\approx\sigma_{0}+\frac{(C_{+}+C_{-})2^{-Y}}{Y(Y-1)}\Gamma\left(1-\frac{Y}{2}\right)\sigma_{0}^{1-Y}t^{1-\frac{Y}{2}},
\end{equation}
when $\kappa\approx 0$. Next, consider the observed implied volatility corresponding to the {closest}-to-the-money option, hereafter denoted by $\hat\sigma^{*}$, and solve the equation
\begin{equation}\label{ImpliedObsEq}
\hat{\sigma}^{*}=\sigma_{0}+\frac{(C_{+}+C_{-})2^{-Y}}{Y(Y-1)}\Gamma\left(1-\frac{Y}{2}\right)\sigma_{0}^{1-Y}t^{1-\frac{Y}{2}},
\end{equation}
for the spot volatility $\sigma_{0}$. {Denote} the {smallest of such solutions} by $\sigma_{0}\left(t,\hat{\sigma}^{*};C_{+},C_{-},Y\right)$. Then, {substitute it into} (\ref{AsyIVGerCGMY2}) to get the equation:
\begin{equation}\label{AsyIVGerCGMY3}
\hat\sigma(t;\kappa)\approx\sigma_{0}\left(t,\hat{\sigma}^{*};C_{+},C_{-},Y\right)+\frac{(C_{+}+C_{-})2^{-Y}}{Y(Y-1)}\Gamma\left(1-\frac{Y}{2}\right)\sigma_{0}\left(t,\hat{\sigma}^{*};C_{+},C_{-},Y\right)^{1-Y}t^{1-\frac{Y}{2}},
\end{equation}
which is free of the unobserved spot volatility $\sigma_{0}$. Finally, {calibrate} (\ref{AsyIVGerCGMY3}) to other sufficiently close-to-the-money implied volatilities by minimizing a weighted sum of squared errors. As in~\cite{MedSca07}, this approach has the advantage of {enabling} to calibrate the parameters across calendar dates simultaneously and not day by day. Note that, alternatively, one can plug $\sigma_{0}(t,\hat{\sigma}^{*};C_{+},C_{-},Y)$ into other type of asymptotics such as the out-of-the-money asymptotics obtained in~\cite{FigGH:2011}.
\end{rem}

\section{Asymptotics For ATM {Option Prices} Under A CGMY {Model}}\label{Sec:CGMY}

In this section, {the second-order asymptotic expansions of the two previous sections is specialized} to the CGMY model. The result presented here were first reported in~\cite{FigGH:2012b}. Recall that under the CGMY model, the L\'{e}vy measure of the pure-jump component $(L_{t})_{t\geq 0}$ is given by
\begin{equation}\label{LevyMeasCGMY2}
\nu(dx)=|x|^{-Y-1}q(x)dx=|x|^{-Y-1}\left(Ce^{-Mx}\,{\bf 1}_{\{x>0\}}+Ce^{Gx}\,{\bf 1}_{\{x<0\}}\right)dx,
\end{equation}
with corresponding parameters $C,\,G,\,M>0$ and $Y\in(1,2)$. Then, the characteristic function of the log-return process $X_{t}=\sigma W_{t}+L_{t}$ takes the form
\begin{equation}\label{CFVCGMY}
\varphi_{t}(u)=\bbe\left(e^{iuX_{t}}\right)=\exp\left(t\left[icu-\frac{\sigma^{2}u^{2}}{2}+C\Gamma(-Y)\left((M-iu)^{Y}+(G+iu)^{Y}-M^{Y}-G^{Y}\right)\right]\right),
\end{equation}
for a constant $c\in\bbr$. The martingale condition (\ref{NdCndTSMrt}) implies that $M>1$ and
\begin{align}\label{cVValMartM}
c&=-C\Gamma(-Y)\left((M-1)^{Y}+(G+1)^{Y}-M^{Y}-G^{Y}\right)-\frac{\sigma^{2}}{2},
\end{align}
(see, e.g.,~\cite[Proposition 4.2]{Tankov}). In particular, the center {$\gamma:=\bbe\left(X_{1}\right)=\bbe\left(L_{1}\right)$} of $X$ and the parameter $b$ of $X$ (relative to the truncation function $x{\bf 1}_{\{|x|\leq 1\}}$) are given by
\begin{align*}
\gamma=c-CY\Gamma(-Y)\left(M^{Y-1}-G^{Y-1}\right),\quad b=c-\int_{|x|>1}x\nu(dx)-CY\Gamma(-Y)\left(M^{Y-1}-G^{Y-1}\right).
\end{align*}

Under the share measure $\bbp^{*}$ introduced in Section \ref{Sec:TmpStble}, $(X_{t})_{t\geq 0}$ has L\'{e}vy triplet $(b^{*},(\sigma^{*})^{2},\nu^{*})$ given by
\begin{equation}\label{b*TripletCGMY}
\sigma^{*}:=\sigma,\quad b^{*}:=c+\sigma^{2}-\int_{|x|>1}x\nu^{*}(dx)-CY\Gamma(-Y)\left(\left(M^{*}\right)^{Y-1}-\left(G^{*}\right)^{Y-1}\right),\quad\nu^{*}(dx):=e^{x}\nu(dx),
\end{equation}
with $M^{*}=M-1$ and $G^{*}=G+1$. Under the probability measure $\widetilde{\bbp}$, the centered process $(Z_{t})_{t\geq 0}$ is symmetric, and its center {$\tilde{\gamma}:=\widetilde{\bbe}\left(X_{1}\right)=\widetilde{\bbe}\left(L_{1}^{*}\right)$} is given by (see~\cite{FigGH:2012b} for the detailed computation)
\begin{align}
\tilde{\gamma}&=\tilde{b}+\int_{\{|x|> 1\}}x\tilde{\nu}(dx)=b^{*}+\int_{|x|\leq 1}x(\tilde{\nu}-\nu^{*})(dx)+\int_{\{|x|>1\}}x\tilde{\nu}(dx)\nonumber\\
\label{Cent} &=-C\Gamma(-Y)\left((M-1)^{Y}+(G+1)^{Y}-M^{Y}-G^{Y}\right)+\frac{\sigma^{2}}{2}.
\end{align}
{The value of $\eta$ defined in (\ref{Uplusminuseta}) is also needed. Under the CGMY model, it} is now given by
\begin{align}\label{eta}
\eta=C\int_{0^{+}}^{\infty}\left(e^{-M^{*}x}\!-1+M^{*}x\right)x^{-Y-1}dx+C\int_{-\infty}^{0^{-}}\left(e^{G^{*}x}\!-1-G^{*}x\right)|x|^{-Y-1}dx=C\Gamma(-Y)\left(\left(M^{*}\right)^{Y}\!\!+\!\left(G^{*}\right)^{Y}\right),
\end{align}
where, {to obtain} the last equality above, we used the analytic continuation {of the representation (14.19) given in \cite{Sato:1999}}.

{Let us now} explicitly write the second-order expansions. First, {compute} the term $\vartheta$ of (\ref{vartheta}). To this end, it is convenient to use the representation given in (\ref{D12Domi}) below noting that in the CGMY case, $\bar{q}(x):=e^{-Mx}{\bf 1}_{\{x>0\}}+e^{Gx}{\bf 1}_{\{x<0\}}$. Hence,
\begin{align}
\vartheta&=C\int_{0}^{\infty}\left(e^{-v}-1\right)\int_{\bbr_{0}}{\bf 1}_{\{x^{-}-\ln\bar{q}(x)\geq v\}}|x|^{-Y-1}dxdv\nonumber\\
&=C\int_{0}^{\infty}\left(e^{-v}-1\right)\int_{0}^{\infty}{\bf 1}_{\{Mx\geq v\}}x^{-Y-1}dxdv+C\int_{0}^{\infty}\left(e^{-v}-1\right)\int_{-\infty}^{0}{\bf 1}_{\{-x-Gx\geq v\}}(-x)^{-Y-1}dxdv\nonumber\\
&=\frac{C}{Y}\left(M^{Y}+(G+1)^{Y}\right)\int_{0}^{\infty}\left(e^{-v}-1\right)v^{-Y}dv\nonumber\\
\label{varthetaCGMY} &=-C\Gamma(-Y)\left(\left(M^{*}+1\right)^{Y}+\left(G^{*}\right)^{Y}\right),
\end{align}
where {for} the last equality $\int_{0}^{\infty}\left(e^{-v}-1\right)v^{-Y}dv=\Gamma(1-Y)=-Y\Gamma(-Y)$ is used (see~\cite[(14.18)]{Sato:1999}). {Next, from} (\ref{Cent})-(\ref{varthetaCGMY}) and recalling that $\tilde{\vartheta}=\vartheta+\eta$ (see (\ref{DExLmN})), it follows that in the pure-jump CGMY model,
\begin{align*}
d_{2}:=\vartheta+\eta+\frac{\tilde{\gamma}}{2}=\frac{C\Gamma(-Y)}{2}\left((M-1)^{Y}-M^{Y}-(G+1)^{Y}+G^{Y}\right),
\end{align*}
while in the general CGMY model with non-zero Brownian component, $C_{+}=C_{-}=C$ and, thus,
\begin{align*}
d_{2}:=\frac{C\sigma^{1-Y}2^{\frac{1-Y}{2}}}{Y(Y-1)\sqrt{\pi}}\Gamma\left(1-\frac{Y}{2}\right).
\end{align*}

\section{Numerical {Examples}}\label{Numerics}

In this part, {the performance of the previous approximations is assessed} through a detailed numerical analysis for the CGMY model.

\subsection{The {Numerical Methods}}

Let us first select a suitable numerical scheme to compute the ATM option prices by considering two methods: Inverse Fourier Transform (IFT) and Monte Carlo (MC). Before introducing the IFT, let us set some notations. The characteristic function corresponding to the Black-Scholes model with volatility $\Sigma$ is given by
\begin{align*}
\varphi_{t}^{BS,\Sigma}(v)=\exp\left(-\frac{\Sigma^{2}t}{2}\left(v^{2}+iv\right)\right).
\end{align*}
The corresponding call option price at the log-moneyness {$k=\log(S_{0}/K)$,} under the Black-Scholes model with volatility $\Sigma$ is denoted by $C_{BS}^{\Sigma}(k)$; that is,
\begin{align*}
C_{BS}^{\Sigma}(k)=S_{0}e^{-rt}{\mathbb{E}\left[\left(e^{(r-\Sigma^{2}/2)t+\Sigma W_{t}}-e^{k}\right)^{+}\right]}.
\end{align*}
{Recall also} that the characteristic function, under the mixed CGMY model with a Brownian component, is denoted by $\varphi_{t}$ (see (\ref{CFVCGMY})) and {denote} the corresponding call option price at log-moneyness $k$ by $C(k)$. The IFT method is based on the following inversion formula (see~\cite[Section 11.1.3]{CT04}):
\begin{equation}\label{InvFormOptP}
z_{_{T}}(k):=C(k)-C_{BS}^{\Sigma}(k)=\frac{1}{2\pi}\int_{-\infty}^{\infty}e^{-ivk}\zeta_{_{T}}(v)dv,
\end{equation}
where
\begin{align}\label{DfnZeta}
\zeta_{_{T}}(v)&:=e^{ivr}\frac{\varphi_{_{T}}(v-i)-\varphi_{_{T}}^{BS,\Sigma}(v-i)}{iv(1+iv)}.
\end{align}
{Fix $r=0$ and, since our interest is only} in ATM option prices, set $k=0$. {The integral in (\ref{InvFormOptP}) is numerically computed using} {the} Simpson's rule:
\begin{align*}
z_{_{T}}(0):=\frac{1}{2\pi}\int_{-\infty}^{\infty}\zeta_{_{T}}(v)dv=\Delta\sum_{m=0}^{P-1}w_{M}^{*}\zeta_{_{T}}\left(v_{M}^{*}\right),
\end{align*}
with $\Delta=Q/(P-1)$, $v_{M}^{*}=-Q/2+m\Delta$, and $w_{0}=1/2$, $w_{2\ell-1}=4/3$, and $w_{2\ell}=2/3$, for $\ell=1,\ldots,P/2$.

{Let us introduce} a Monte Carlo method based on the risk-neutral option price representation under the probability measure $\widetilde{\bbp}$. Under this probability measure and using the notation (\ref{Uplusminuseta}) as well as the relations (\ref{DcmLL}) and (\ref{RX}), we have:
\begin{align*}
{\bbe\left[\left(e^{X_{T}}\!-\!1\right)^{+}\right]}\!=\!\bbe^{*}\!\left[e^{-X_{T}}\left(e^{X_{T}}\!-\!1\right)^{+}\right]\!=\!\widetilde{\bbe}\left[e^{-U_{T}}\left(1\!-\!e^{-X_{T}}\right)^{+}\right]\!=\!\widetilde{\bbe}\left[e^{-M^{*}\bar{U}^{+}_{T}+G^{*}\bar{U}^{-}_{T}-\eta T}\left(1\!-\!e^{-\bar{U}^{+}_{T}-\bar{U}^{-}_{T}-T\tilde{\gamma}-\sigma W_{T}}\right)^{+}\right],
\end{align*}
which can be easily computed by Monte Carlo method using that, under $\widetilde{\bbp}$, the variables $\bar{U}^{+}_{T}$ and $-\bar{U}^{-}_{T}$ are independent $Y$-stable random variables with scale, skewness, and location parameters $TC|\cos(\pi Y/2)|\Gamma(-Y)$, $1$ and $0$, respectively. Standard simulation methods are available to generate stable random variables.

\begin{figure}[htp]
{\par\centering
\includegraphics[width=8.0cm,height=9.0cm]{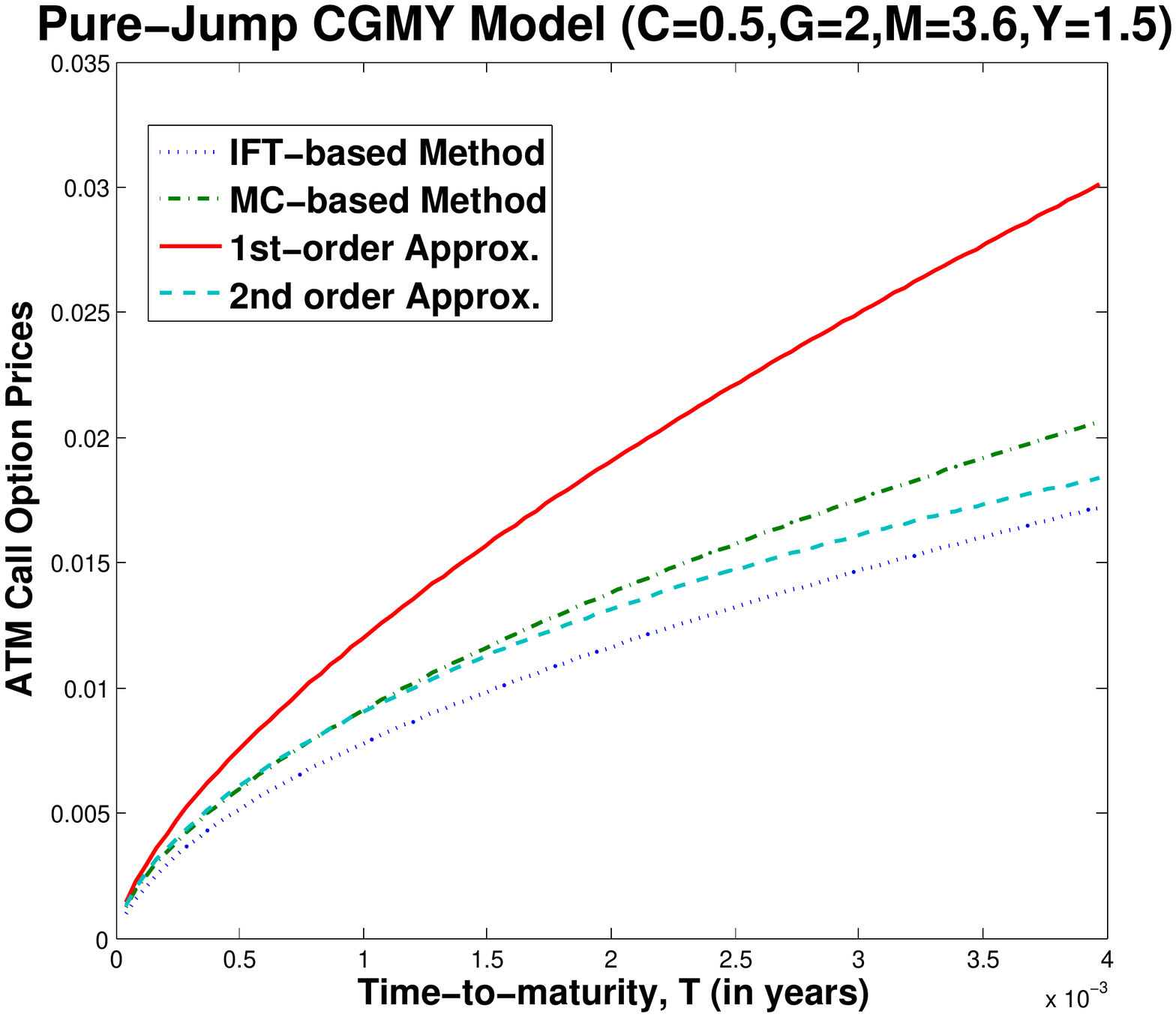}
\includegraphics[width=8.0cm,height=9.0cm]{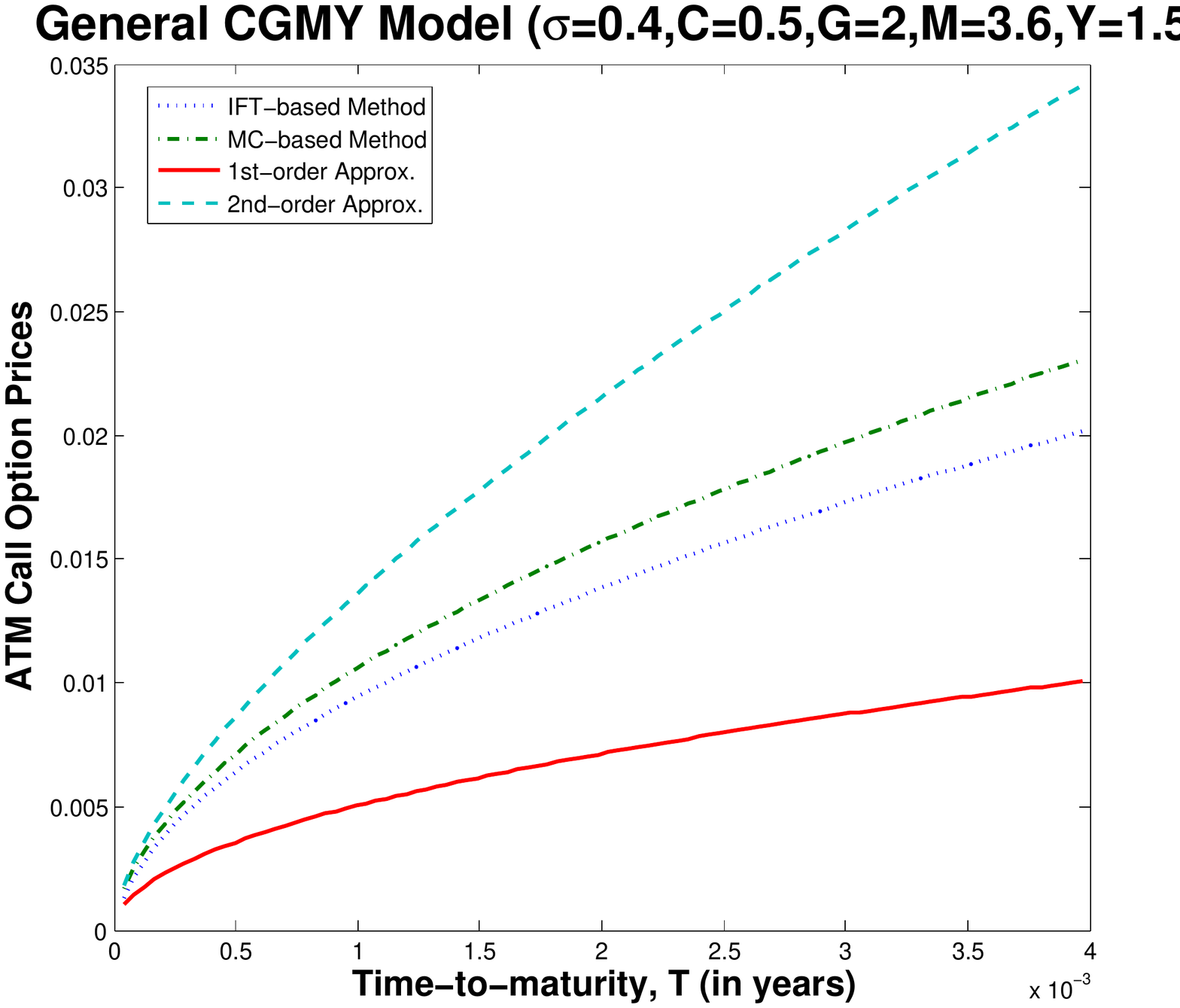}
\par}\vspace{-1.2 cm}
\caption{Comparisons of ATM call option prices for two methods (Inverse Fourier Transform and Monte-Carlo method) with the first- and second-order approximations. MC-based price is based on $100,000$ simulations while the IFT-based method is based on the parameter values {$P=2^{14}$} and {$Q=800$}. The parameter $\sigma$ in the {mixed} CGMY model is set to be $0.1$.}\label{Figure1}
\end{figure}

Next, {take} the following set of parameters
\begin{equation*}
C=0.5,\quad G=2,\quad M=3.6,\quad Y=1.5.
\end{equation*}
Figure \ref{Figure1} compares the first- and second-order approximations as given in Remarks \ref{Rem:ApproxPJ} and \ref{Rem:ApproxGen} to the prices based on the Inverse Fourier Transform (IFT-based price) and the Monte Carlo method (MC-based price) under both the pure-jump case and the mixed CGMY case with $\sigma=0.4$. For the MC-based price, {$100,000$ simulations are used, while for the IFT-based method, $P=2^{14}$ and $Q=800$}. As it can be seen, it is not easy to integrate numerically the characteristic function (\ref{DfnZeta}) since $T$ is quite small and, therefore, the characteristic functions $\varphi_{T}$ and $\varphi_{T}^{BS,\Sigma}$ are quite flat. The Monte Carlo method turns out to be much more accurate and faster. It is also interesting to note that the second-order approximation is in general much more accurate in the pure-jump model than in the mixed model with nonzero continuous component. This observation is consistent with the last comment of Remark \ref{Rem:ApproxGen}.

\subsection{Results {For} {Different Parameter Settings}}

\begin{figure}[hbt]
{\par\centering
\includegraphics[width=8.0cm,height=9.0cm]{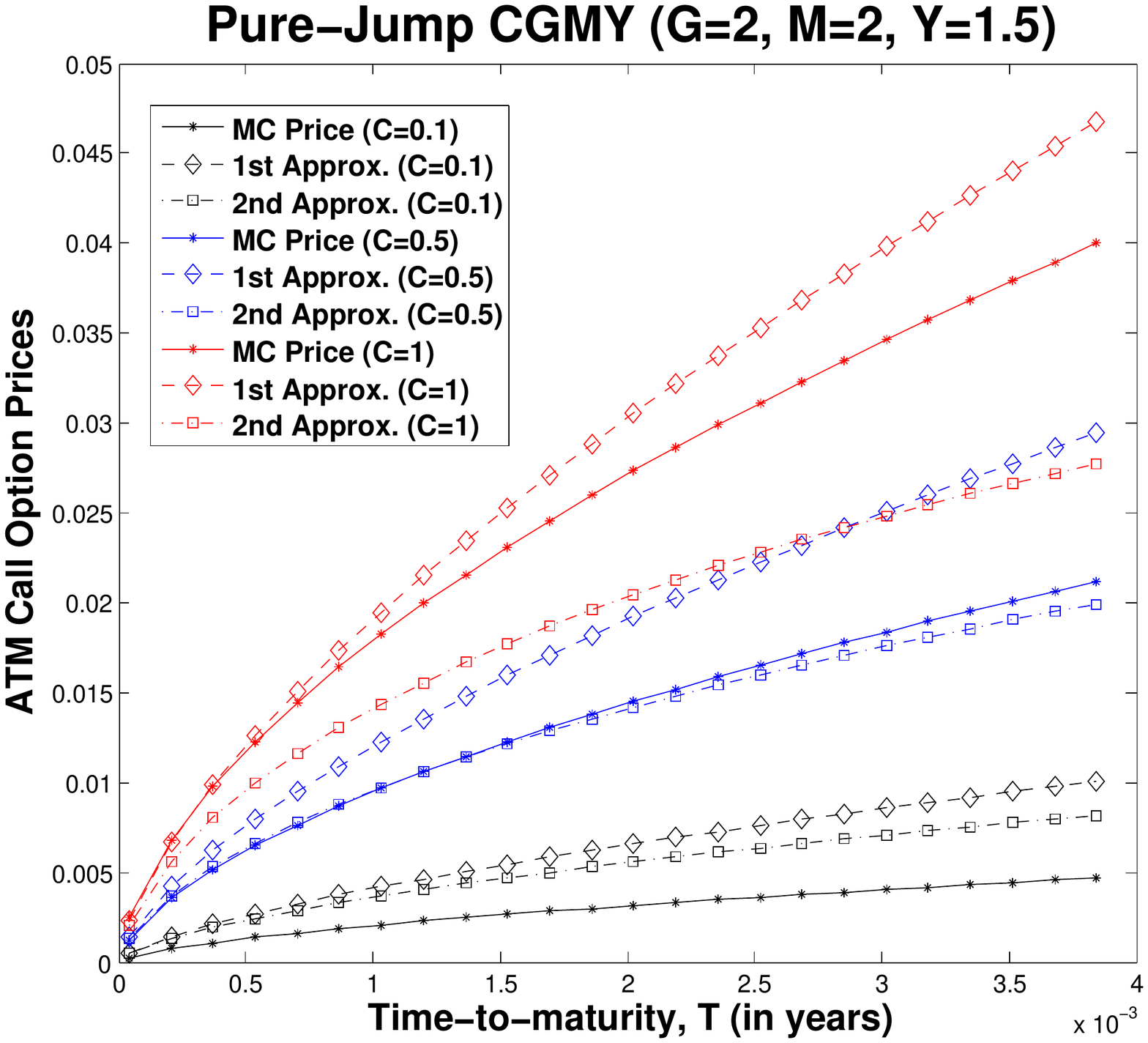}
\includegraphics[width=8.0cm,height=9.0cm]{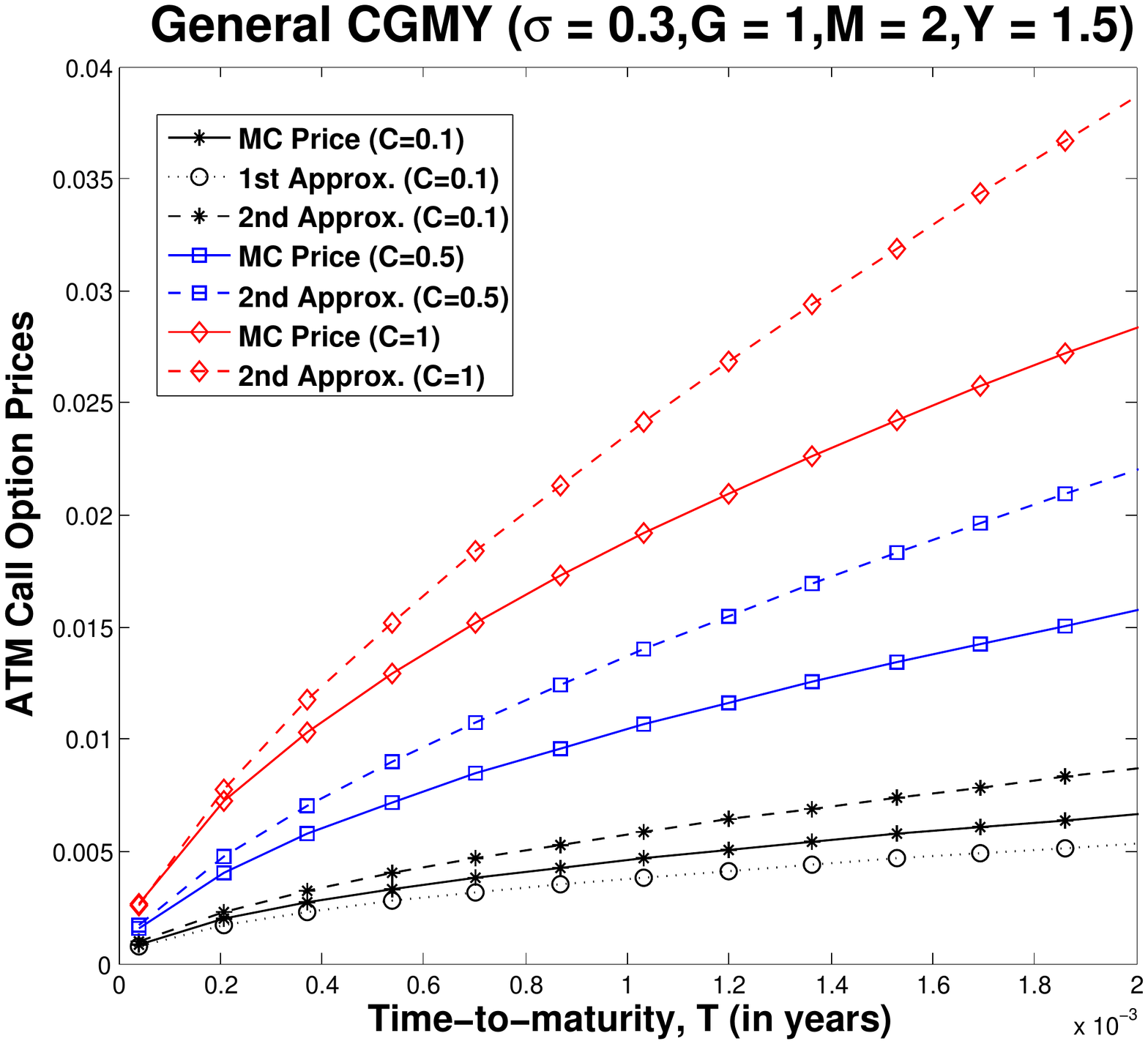}
\par}\vspace{-1.2 cm}
\caption{Comparisons of ATM call option prices {with} short-time approximations for different values of the jump intensity parameter $C$.}\label{Figure2}
\end{figure}

\begin{figure}[htp]
{\par\centering
\includegraphics[width=8.0cm,height=9.0cm]{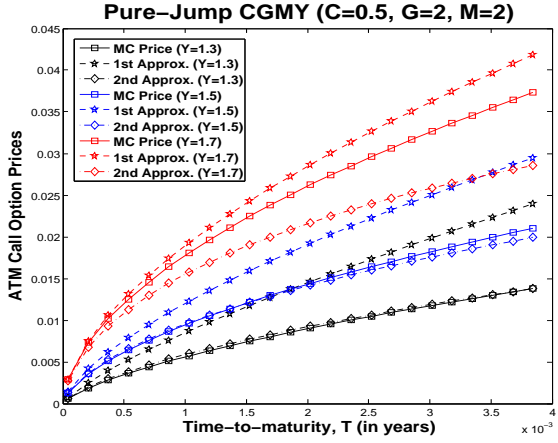}
\includegraphics[width=8.0cm,height=9.0cm]{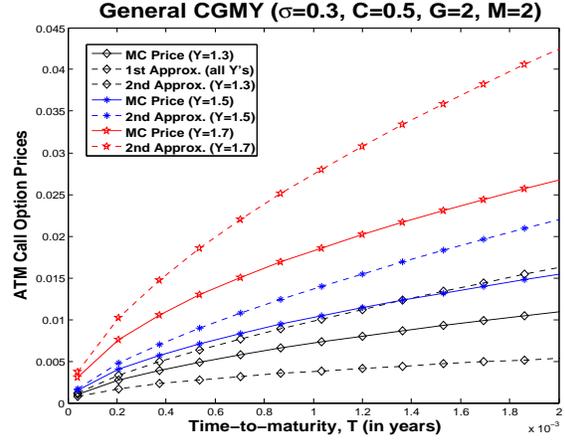}
\par}\vspace{-1.2 cm}
\caption{Comparisons of ATM call option prices with short-time approximations for different values of the tail-heaviness parameter $Y$.}\label{Figure3}
\end{figure}

\begin{figure}[htp]
{\par\centering
\includegraphics[width=8.0cm,height=9.0cm]{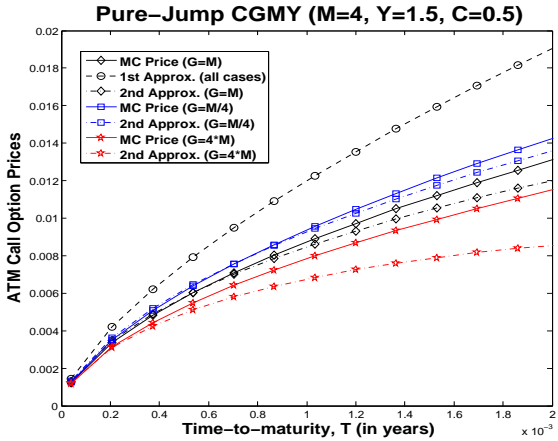}
\includegraphics[width=8.0cm,height=9.0cm]{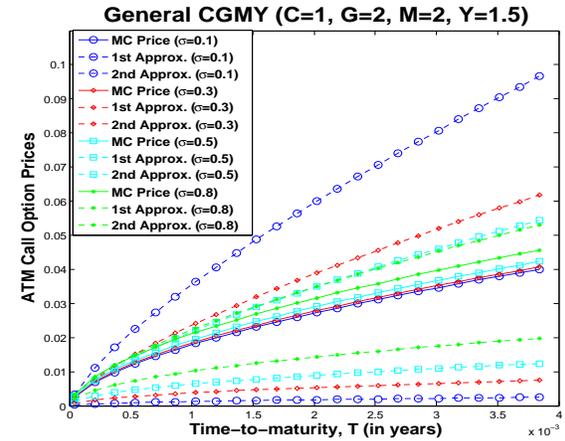}
\par}\vspace{-1.2 cm}
\caption{Comparisons of ATM call option prices with short-time approximations for different values of $G$ or $M$ and different values of the volatility parameter $\sigma$.}\label{Figure4}
\end{figure}

{The performances of the approximations for different settings of parameters are investigated below}:
\begin{enumerate}
\item Figure \ref{Figure2} compares the first- and second-order approximations with the MC prices for different values of $C$, fixing the values of all the other parameters. In the pure-jump case, the second-order approximation is significantly better for moderately small values of $C$, but for larger values of $C$, this is not the case unless $T$ is extremely small. For a nonzero continuous component, the first-order approximation is extremely poor as it only takes into account the parameter $\sigma$.
\item Figure \ref{Figure3} compares the first- and second-order approximations with the MC prices for different values of $Y$, fixing the values of all the other parameters. In both cases, the second-order approximation is significantly better for values of $Y$ around $1.5$, which is consistent with the observation that $|d_{2}|\to\infty$ as $Y\to 1$ or $Y\to 2$. For a nonzero continuous component, the first-order approximation is again extremely poor when compared to the 2nd order approximation.
\item {The left panel of Figure \ref{Figure4} analyzes} the effect of the relative intensities of the negative jumps compared to the positive jumps in the pure-jump CGMY case. That is, {the value $M$ is fixed} to be $4$ and consider different values for $G$.  As expected, since the first-order approximation does not take into account this information, the second-order approximation performs significantly better.
\item In the right panel of Figure \ref{Figure4}, we analyze the effect of the volatility of the continuous component in the mixed CGMY case. The second-order approximation is, in general, much better than the {first-order} approximation and, interestingly enough, the quality of the second-order approximations improves as the values of $\sigma$ increases. In fact, it seems that the second-order approximation and the MC prices collapse to a steady curve as $\sigma$ increases.
\end{enumerate}

\appendix
\section{Proofs of Section \ref{Sec:PureJump}: {The Pure-Jump Model}}\label{proofA}

For simplicity, unless otherwise stated, throughout this section, {fix} $S_{0}=1$.

\medskip
\noindent
\textbf{Proof of Lemma \ref{Lm:NRATMPJ}.}

\noindent
From (\ref{CMR}),
\begin{align}	
{t^{-\frac{1}{Y}}\mathbb{E}\left[\left(S_{t}-S_{0}\right)^{+}\right]}&=t^{-\frac{1}{Y}}\mathbb{P}^{*}\left(X_{t}\geq E\right)=t^{-\frac{1}{Y}}\int_{0}^{\infty}e^{-x}\mathbb{P}^{*}\left(X_{t}\geq x\right)dx\nonumber\\
\label{LOP1} &=\int_{0}^{\infty}e^{-t^{\frac{1}{Y}}u}\mathbb{P}^{*}\left(t^{-\frac{1}{Y}}X_{t}\geq u\right)du.
\end{align}	
Next, using the change of probability measure (\ref{EMMN}),
\begin{align*}
\mathbb{P}^{*}\left(t^{-\frac{1}{Y}}X_{t}\geq u\right)&=\bbe^{*}\left({\bf 1}_{\left\{t^{-\frac{1}{Y}}X_{t}\geq u\right\}}\right)=\widetilde{\bbe}\left(e^{-U_{t}}{\bf 1}_{\left\{t^{-\frac{1}{Y}}X_{t}\geq u\right\}}\right),
\end{align*}
and, moreover, (\ref{DcmLL}) and (\ref{RX}),
\begin{align*}
\mathbb{P}^{*}\left(t^{-\frac{1}{Y}}X_{t}\geq u\right)&=e^{-\eta t}\widetilde{\bbe}\left(e^{-\widetilde{U}_{t}}{\bf 1}_{\left\{t^{-\frac{1}{Y}}Z_{t}\geq u-\tilde{\gamma}t^{1-\frac{1}{Y}}\right\}}\right).
\end{align*}
Then,
\begin{align}
{t^{-\frac{1}{Y}}\mathbb{E}\left[\left(S_{t}-S_{0}\right)^{+}\right]}&=\int_{0}^{\infty}e^{-t^{\frac{1}{Y}}u-\eta t}\,\widetilde{\bbe}\left(e^{-\widetilde{U}_{t}}{\bf 1}_{\left\{t^{-\frac{1}{Y}}Z_{t}\geq u-\tilde{\gamma}t^{1-\frac{1}{Y}}\right\}}\right)du,
\end{align}
and, changing variables $\left(v=u-\tilde{\gamma}t^{1-1/Y}\right)$, the result follows.\hfill\qed

\medskip
\noindent
\textbf{Proof of Lemma \ref{Lm:AsymBhvrToSt}.}

\noindent
Fix $Q_{t}:=\widetilde{U}_{t}+Z_{t}$. From the usual moment generating formula for Poisson integrals, the change of variables $y=t^{-1/Y}x$, and recalling that $\varphi(x)+x=-\ln \bar{q}(x)$,
\begin{align}
\widetilde{\bbe}\left(e^{-\xi t^{-\frac{1}{Y}}Q_{t}}\right)&=\widetilde{\bbe}\left(\exp\left\{-\xi t^{-\frac{1}{Y}}\int_{0}^{t}\int\left(\varphi(x)+x\right){\bar{N}}(ds,dx)\right\}\right)\nonumber\\
&=\exp\left\{t\int\left(e^{\xi t^{-\frac{1}{Y}}\ln\bar{q}(x)}-1-\xi t^{-\frac{1}{Y}}\ln\bar{q}(x)\right)\tilde{\nu}(dx)\right\}\nonumber\\
&=\exp\left\{C_{+}\int_{0}^{\infty}\left(e^{\xi t^{-\frac{1}{Y}}\ln\bar{q}\left(t^{\frac{1}{Y}}y\right)}-1-\xi t^{-\frac{1}{Y}}\ln\bar{q}\left(t^{\frac{1}{Y}}y\right)\right)y^{-Y-1}dy\right.\nonumber\\
\label{LIIntc1} &\qquad\quad\,\,\left.+\,C_{-}\int_{-\infty}^{0}\left(e^{\xi t^{-\frac{1}{Y}}\ln\bar{q}\left(t^{\frac{1}{Y}}y\right)}-1-\xi t^{-\frac{1}{Y}}\ln\bar{q}\left(t^{\frac{1}{Y}}y\right)\right)|y|^{-Y-1}dy\right\}.
\end{align}	
Let $\Xi(u):=e^{u}-1-u$ and $B:=\sup_{w}\left|\ln\bar{q}(w)\right|/|w|$, which is necessarily finite in light of of (\ref{Eq:StndCnd1b}) and (\ref{UsflLmt}-ii). Now, using that $\ln\bar{q}(u)\leq 0$, for any $u\in\bbr_{0}$ (from both conditions in (\ref{Eq:StndCnd1a})), and that $\left|\Xi(u)\right|\leq K\left(u^{2}\wedge|u|\right)$, for $u\leq 0$ and some constant $0<K<\infty$,
\begin{align*}
\Xi\left(\xi t^{-\frac{1}{Y}}\ln\bar{q}\left(t^{\frac{1}{Y}}y\right)\right)|y|^{-Y-1}&\leq {K\left[\left(\xi t^{-\frac{1}{Y}}\ln\bar{q}\left(t^{\frac{1}{Y}}y\right)\right)^{2}\wedge\left|\xi t^{-\frac{1}{Y}}\ln\bar{q}\left(t^{\frac{1}{Y}}y\right)\right|\right]}\,|y|^{-Y-1}\\
&\leq K\left(\left(\xi yB\right)^{2}\wedge\left|\xi yB\right|\right)\,|y|^{-Y-1}\\
&\leq K\left(\left(\xi B\right)^{2}\vee\left|\xi B\right|\right)\,\left(y^{2}\wedge|y|\right)\,|y|^{-Y-1},
\end{align*}
which is integrable. Therefore, one can pass the limit inside the integrals in (\ref{LIIntc1}) and, using (\ref{UsflLmt}),
\begin{align*}
\lim_{t\to 0}\widetilde{\bbe}\left(e^{-\xi t^{-\frac{1}{Y}}Q_{t}}\right)&=\exp\left\{C_{+}\int_{0}^{\infty}\left(e^{-y\xi M}-1+y\xi M\right)y^{-Y-1}dy+C_{-}\int_{-\infty}^{0}\left(e^{-|y|\xi G}-1+|y|\xi G\right)|y|^{-Y-1}dy\right\}\\
&=\exp\left\{\left(C_{+}\int_{0}^{\infty}\left(e^{-uM}-1+uM\right)u^{-Y-1}du+C_{-}\int_{-\infty}^{0}\left(e^{-|u|G}-1+|u|G\right)|u|^{-Y-1}du\right)\xi^{Y}\right\}.
\end{align*}
Finally, {the analytic continuation of the representation (14.19) given in \cite{Sato:1999} shows} that the last expression is of the form $\exp(\bar{\eta}\xi^{Y})$, with $\bar{\eta}$ given as in the statement of the lemma.
	
For (\ref{LmtMGFTldU}-ii), {proceed} as above to get
\begin{align}\label{LIIntc1OTldU}
\widetilde{\bbe}\left(e^{-\xi t^{-\frac{1}{Y}}\widetilde{U}_{t}}\right)\!=\!\exp\left\{C_{+}\!\!\int_{0}^{\infty}\!\!\Xi\!\left(\xi t^{-\frac{1}{Y}}\ln\bar{q}^{*}\!\!\left(t^{\frac{1}{Y}}y\right)\right)y^{-Y-1}dy+C_{-}\!\!\int_{-\infty}^{0}\!\!\Xi\!\left(\xi t^{-\frac{1}{Y}}\ln\bar{q}^{*}\!\!\left(t^{\frac{1}{Y}}y\right)\right)|y|^{-Y-1}dy\right\}.
\end{align}
Now, for all $y\in\bbr_{0}$,
\begin{equation}\label{EsEstNN}
\xi t^{-\frac{1}{Y}}\ln\bar{q}^{*}\left(t^{\frac{1}{Y}}y\right)=\xi t^{-\frac{1}{Y}}\ln\bar{q}\left(t^{\frac{1}{Y}}y\right)+\xi y=\xi y\left(\frac{1}{yt^{\frac{1}{Y}}}\ln\bar{q}\left(t^{\frac{1}{Y}}y\right)+1\right)\leq 0,
\end{equation}
from both conditions in (\ref{Eq:StndCnd1a}). Using (\ref{EsEstNN}), {proceed as above to justify passing the limit into the integrals in (\ref{LIIntc1OTldU}). Next, using (\ref{UsflLmtq*}),} conclude that
\begin{align*}
\lim_{t\to 0}\widetilde{\bbe}\left(e^{-\xi t^{-\frac{1}{Y}}\widetilde{U}_{t}}\right)&=\exp\left\{C_{+}\int_{0}^{\infty}\left(e^{-y\xi M^{*}}\!\!-\!1\!+\!y\xi M^{*}\right)y^{-Y-1}dy+C_{-}\int_{-\infty}^{0}\left(e^{-|y|\xi G^{*}}\!\!-\!1\!+\!|y|\xi G^{*}\right)|y|^{-Y-1}dy\right\}\\
&=\exp\left\{\left(C_{+}\!\int_{0}^{\infty}\!\left(e^{-uM^{*}}\!\!-\!1\!+\!uM^{*}\right)u^{-Y-1}du+C_{-}\!\int_{-\infty}^{0}\!\left(e^{-|u|G^{*}}\!\!-\!1\!+\!|u|G^{*}\right)|u|^{-Y-1}du\right)\xi^{Y}\right\}.
\end{align*}
Finally, (\ref{LmtMGFTldU}-ii) follows once more from the analytic continuation of~\cite[(14.19)]{Sato:1999}.\hfill\qed

\medskip
{\noindent\textbf{Proof of Lemma \ref{KLTCTBR}.}}

\smallskip
\noindent
{{\rm (1)} {The proof of the first assertion of this lemma makes use of the following ``small/large jumps decomposition" of $Z$:}}
\begin{equation}\label{DcmpSmLr1}
{\bar{Z}_{t}^{(\varepsilon)}}:=\int_{0}^{t}\int_{|x|\geq\varepsilon}xN(ds,dx),\qquad Z_{t}^{(\varepsilon)}:=Z_{t}-{\bar{Z}_{t}^{(\varepsilon)}},\qquad {t\geq 0},
\end{equation}
{for a suitably chosen $\varepsilon>0$.} Under $\widetilde{\bbp}$, {$(\bar{Z}_{t}^{(\varepsilon)})_{t\geq 0}$} is a drift-less L\'{e}vy process with finite L\'{e}vy measure {${\bf 1}_{\{|x|\geq \varepsilon\}}\tilde{\nu}(dx)$} {and, thus}, is a compound Poisson process. Denote respectively by $(N_{t}^{(\varepsilon)})_{t\geq 0}$ and $(\xi_{i}^{(\varepsilon)})_{i\geq 1}$ the counting process and the sizes of the jumps of {$(\overline{Z}_{t}^{(\varepsilon)})_{t\geq 0}$}, so that ${\bar{Z}_{t}^{(\varepsilon)}}=\sum_{i=1}^{N_{t}^{(\varepsilon)}}\xi_{i}^{(\varepsilon)}$, {for any $t\geq 0$}. In particular, {$(N_{t}^{(\varepsilon)})_{t\geq 0}$} is a Poisson process with intensity $\lambda_{\varepsilon}:={\widetilde\bbe N_{1}^{(\varepsilon)}=\tilde{\nu}(|x|\geq{}\varepsilon)}$ and $(\xi_{i}^{(\varepsilon)})_{i\geq 1}$ are i.i.d, random variables with distribution ${\bf 1}_{\{|x|\geq{}\varepsilon\}}\tilde{\nu}(dx)/\lambda_{\varepsilon}$. Next, {define} the corresponding processes for $\widetilde{U}$:
\begin{equation}\label{DcmpSmLr2}
{\bar{\widetilde{U}}_{t}^{(\varepsilon)}}:=\int_{0}^{t}\int_{|x|\geq\varepsilon}\varphi(x)N(ds,dx)=-\sum_{i=1}^{N_{t}^{(\varepsilon)}}\ln\bar{q}^{*}\left(\xi_{i}^{(\varepsilon)}\right),\qquad\widetilde{U}_{t}^{(\varepsilon)}:=\widetilde{U}_{t}-{\bar{\widetilde{U}}_{t}^{(\varepsilon)}},\qquad {t\geq 0}.
\end{equation}
 {For any $t\geq 0$, recalling} that $N_{t}^{(\varepsilon)}$ is Poisson distributed with mean $\lambda_{\varepsilon}t$ and using (\ref{DcmpSmLr1}) and (\ref{DcmpSmLr2}), {conditioning on $N_{t}^{(\varepsilon)}$ gives}
\begin{align*}
t^{-1}\widetilde{\bbp}\left(Z_{t}^{+}+\widetilde{U}_{t}\geq v\right)&=t^{-1}\widetilde{\bbp}\left(\left(Z_{t}^{(\varepsilon)}+\sum_{i=1}^{N_{t}^{(\varepsilon)}}\xi_{i}^{(\varepsilon)}\right)^{+}+\widetilde{U}_{t}^{(\varepsilon)}-\sum_{i=1}^{N_{t}^{(\varepsilon)}}\ln\bar{q}^{*}\left(\xi_{i}^{(\varepsilon)}\right)\geq v\right)\\
&=t^{-1}\widetilde{\bbp}\left(\!\left(Z_{t}^{(\varepsilon)}\right)^{+}\!\!\!+\widetilde{U}_{t}^{(\varepsilon)}\geq v\!\right)e^{-\lambda_{\varepsilon}t}\!+\!e^{-\lambda_{\varepsilon}t}\lambda_{\varepsilon}\widetilde{\bbp}\left(\!\left(Z_{t}^{(\varepsilon)}\!+\xi_{1}^{(\varepsilon)}\right)^{+}\!\!\!+\widetilde{U}_{t}^{(\varepsilon)}\!-\ln\bar{q}^{*}\left(\xi_{1}^{(\varepsilon)}\right)\geq v\!\right)\!+\!O(t).
\end{align*}
The first term above can be made $O(t)$ by taking $\varepsilon\in(0,\varepsilon_{0})$, for some small enough $\varepsilon_{0}>0$ (see, e.g.,~\cite[Section 26]{Sato:1999} and~\cite[Lemma 3.2]{Ruschendorf}). Indeed, {first note that the supports} of the L\'{e}vy measures of {$(Z^{(\varepsilon)}_{t})_{t\geq 0}$} and {$(\widetilde{U}^{(\varepsilon)}_{t})_{t\geq 0}$} are respectively $\{x:|x|\leq{}\varepsilon\}$ and $\{\varphi(x):|x|\leq \varepsilon\}=\{-\ln\bar{q}(x)-x:|x|\leq \varepsilon\}$. Next, since $\lim_{x\to 0}\bar{q}(x)=1$, one can choose $\varepsilon$ small enough so that the supports are contained in a ball of arbitrarily small radius $\delta$, which in turn implies that $\widetilde{\bbp}\left({Z}_{t}^{(\varepsilon)}\geq v/2\right)=O(t^{2})$ and $\widetilde{\bbp}\left({\widetilde{U}}_{t}^{(\varepsilon)}\geq v/2\right)=O(t^{2})$, by taking $\delta>0$ small enough. For the second term, first from (\ref{KyCondValAllRsIntro}), there exists $\varepsilon_{0}>0$ such that, for all $0<\varepsilon<\varepsilon_{0}$,
\begin{align*}
a_{0}(v)&:=\lambda_{\varepsilon}\widetilde{\bbp}\left(\left(\xi_{1}^{(\varepsilon)}\right)^{+}-{\ln\bar{q}^{*}}\left(\xi_{1}^{(\varepsilon)}\right)\geq v\right)=\int_{\bbr_{0}}{\bf 1}_{\{{x^{+}}-\ln\bar{q}^{*}(x)\geq v\}}\tilde{\nu}(dx)=\int_{\bbr_{0}}{\bf 1}_{\{{x^{-}}-\ln\bar{q}(x)\geq v\}}\tilde{\nu}(dx)\\
&\,=C_{+}\int_{0}^{\infty}{\bf 1}_{\{-\ln\bar{q}(x)\geq v\}}x^{-Y-1}dx+C_{-}\int_{-\infty}^{0}{\bf 1}_{\{-x-\ln\bar{q}(x)\geq v\}}|x|^{-Y-1}dx,
\end{align*}
where for the second {equality recall} that $\xi_{i}^{(\varepsilon)}$ has distribution ${\bf 1}_{\{|x|\geq \varepsilon\}}\tilde{\nu}(dx)/\lambda_{\varepsilon}$. Also, since
\begin{align*}		 F_{v,\varepsilon}(z,u):=\widetilde{\bbp}\left(\left(z+\xi_{1}^{(\varepsilon)}\right)^{+}+u-{\ln\bar{q}^{*}}\left(\xi_{1}^{(\varepsilon)}\right)\geq v\right){,}
\end{align*}
is continuous at $(0,0)$, for any fixed $0<\varepsilon<\varepsilon_{0}$, the function
\begin{align*}
A_{t}(v)&:=t^{-1}\widetilde{\bbp}\left(Z_{t}^{+}+\widetilde{U}_{t}\geq v\right)-a_{0}(v){,}
\end{align*}
is such that
\begin{align*}		
\lim_{t\to 0}A_{t}(v)&=\lambda_{\varepsilon}\lim_{t\to 0}{\left[\widetilde{\bbp}\left(\left({Z}_{t}^{(\varepsilon)}+\xi_{1}^{(\varepsilon)}\right)^{+}+\widetilde{U}_{t}^{(\varepsilon)}-\ln\bar{q}^{*}\left(\xi_{1}^{(\varepsilon)}\right)\geq v\right)-\lambda_{\varepsilon}\widetilde{\bbp}\left(\left(\xi_{1}^{(\varepsilon)}\right)^{+}-\ln\bar{q}^{*}\left(\xi_{1}^{(\varepsilon)}\right)\geq v\right)\right]}\\
&=\lambda_{\varepsilon}\lim_{t\to 0}{\widetilde{\bbe}\left[F_{v,\varepsilon}\left(Z_{t}^{(\varepsilon)},\widetilde{U}_{t}^{(\varepsilon)}\right)-F_{v,\varepsilon}(0,0)\right]}\\
&=\lambda_{\varepsilon}{\widetilde{\bbe}\left[\lim_{t\to 0}F_{v,\varepsilon}\left(Z_{t}^{(\varepsilon)},\widetilde{U}_{t}^{(\varepsilon)}\right)-F_{v,\varepsilon}(0,0)\right]}=0,
\end{align*}
where dominated convergence is used to obtain the last equality.

\smallskip
\noindent
{\rm (2)}
Throughout this part, $\tilde{\kappa}>0$ denotes a generic finite constant that may vary from line to line. First, note that (\ref{KIn}) implies that
\begin{equation}\label{TIIStabl}
t^{-1}\widetilde{\bbp}\left(Z_{t}^{+}\geq v\right)\leq\tilde{\kappa}v^{-Y},
\end{equation}
for any $0<t\leq 1$ and $v>0$. So, it suffices to show the analog inequality for {$(\widetilde{U}_{t})_{t\geq 0}$}.
{To this end, {use the following ``small/large jumps decomposition" of $\widetilde{U}$}:
	\begin{equation}\label{DcmpSmLr3}
{\bar{\widetilde{U}}_{t}^{(\varepsilon)}}:=\int_{0}^{t}\int_{|\varphi(x)|\geq\varepsilon}\varphi(x)N(ds,dx),\qquad\widetilde{U}_{t}^{(\varepsilon)}:=\widetilde{U}_{t}-{\bar{\widetilde{U}}_{t}^{(\varepsilon)}},\qquad t\geq 0,
\end{equation}
with $\varepsilon:=v/4$.
Note that, in view of both (\ref{Eq:StndCnd1b}) and (\ref{UsflLmtq*}-ii), there exists a constant $r<\infty$ such that
\begin{equation}\label{ERelVarphi}
	|\varphi(x)|=|\ln \bar{q}^{*}(x)|\leq r |x|,
\end{equation}
for all $x\neq{}0$. In particular, $\bar{\widetilde{U}}^{(\varepsilon)}$ in (\ref{DcmpSmLr3}) is a compound Poisson process with intensity of jumps $\lambda_{\varepsilon}:=\tilde\nu\left(|\varphi(x)|\geq{}\varepsilon\right)$ and, thus, denoting the counting process of the jumps of $\bar{\widetilde{U}}^{(\varepsilon)}$ by $(N_{t}^{(\varepsilon)})_{t\geq{}0}$,
\begin{align*}
	t^{-1}\widetilde{\bbp}\left({\bar{\widetilde{U}}_{t}^{(\varepsilon)}}\geq\frac{v}{2}\right)\leq t^{-1}\widetilde{\bbp}\left(N_{t}^{(\varepsilon)}\neq 0\right)=t^{-1}\left(1-e^{-\lambda_{\varepsilon}t}\right)\leq\lambda_{\varepsilon}\leq\bar{C}\int_{|\varphi(x)|\geq\varepsilon}|x|^{-Y-1}dx,
\end{align*}
where $\bar{C}:=C_{+}+C_{-}$. Next, using (\ref{ERelVarphi}) and $\varepsilon=v/4$,
\begin{equation}\label{NEsH1}
	t^{-1}\widetilde{\bbp}\left({\bar{\widetilde{U}}_{t}^{(\varepsilon)}}\geq\frac{v}{2}\right)\leq
	\bar{C}\int_{|x|\geq \frac{v}{4r}}|x|^{-Y-1}dx\leq{}\tilde{\kappa} v^{-Y},
\end{equation}
for some constant $\tilde\kappa$.
Also,
\begin{align*}
\widetilde{\bbe}\left(\widetilde{U}_{t}^{(\varepsilon)}\right)&
=-t\int_{|\varphi(x)|\geq\varepsilon}\varphi(x)\tilde{\nu}(dx)\\
&=t\int_{|\varphi(x)|\geq\varepsilon}\ln\bar{q}(x)\tilde{\nu}(dx)+t\,C_{-}\int_{|\varphi(x)|\geq\varepsilon,x<0}x|x|^{-Y-1}dx+t\,C_{+}\int_{|\varphi(x)|\geq\varepsilon,x>0}x|x|^{-Y-1}dx,
\end{align*}
and, since $\bar{q}(x)\leq 1$ for all $x\in\bbr\backslash\{0\}$ (from (\ref{Eq:StndCnd1a})),
\[
	\widetilde{\bbe}\left(\widetilde{U}_{t}^{(\varepsilon)}\right)\leq tC_{+}\int_{x\geq \frac{\varepsilon}{r}}x^{-Y}dx=C_{+}(Y-1)^{-1}{(4r)^{Y-1}}tv^{1-Y}.
\]
Thus, whenever $t$ and $v$ are such that $C_{+}(Y-1)^{-1}{(4r)^{Y-1}}tv^{1-Y}\leq v/4$, we have
\begin{align*}
\widetilde{\bbp}\left(\widetilde{U}_{t}^{(\varepsilon)}\geq\frac{v}{2}\right)=\widetilde{\bbp}\left(\widetilde{U}_{t}^{(\varepsilon)}-\widetilde{\bbe}\left(\widetilde{U}_{t}^{(\varepsilon)}\right)+\widetilde{\bbe}\left(\widetilde{U}_{t}^{(\varepsilon)}\right)\geq\frac{v}{2}\right)\leq\widetilde{\bbp}\left(\widetilde{U}_{t}^{(\varepsilon)}-\widetilde{\bbe}\left(\widetilde{U}_{t}^{(\varepsilon)}\right)\geq\frac{v}{4}\right).
\end{align*}
Next, {by} a concentration inequality for centered random variables ({see \cite[Corollary 1]{Hou:2002})} together with the identity $e^{-x\log(1+x)}{\bf 1}_{\{x>0\}}\leq x^{-1}{\bf 1}_{\{x>0\}}$,
\begin{align*}
\widetilde{\bbp}\left(\widetilde{U}_{t}^{(\varepsilon)}\geq\frac{v}{2}\right)\leq\widetilde{\bbp}\left(\widetilde{U}_{t}^{(\varepsilon)}-\widetilde{\bbe}\left(\widetilde{U}_{t}^{(\varepsilon)}\right)\geq\frac{v}{4}\right)\leq e^{\frac{v}{4\varepsilon}-\left(\frac{v}{4\varepsilon}+\frac{tV_{\varepsilon}^2}{\varepsilon^2}\right)
\log\left(1+\frac{\varepsilon v}{4tV_{\varepsilon}^2}\right)}\leq \left(\frac{4eV_{\varepsilon}^2}{\varepsilon v}\right)^{\frac{v}{4\varepsilon}}t^{\frac{v}{4\varepsilon}}\leq \frac{{16}eV_{v/4}^2}{v^{2}}t,
\end{align*}
where $V_{\varepsilon}^2:={\rm Var}\left(\widetilde{U}_{1}^{(\varepsilon)}\right)$ and, in the last inequality, $\varepsilon=v/4$. Now, using again (\ref{ERelVarphi}) and setting {$\bar{C}:=C_{+}+C_{-}$,
\begin{align*}
V_{v/4}^{2}&=\int_{|\varphi(x)|<\frac{v}{4}}\left(\varphi(x)\right)^{2}\tilde{\nu}(dx)\\
&\leq \bar{C}\int_{|\varphi(x)|<\frac{v}{4},|x|<\frac{v}{4}}(\varphi(x))^{2}|x|^{-Y-1}dx+
\bar{C}\int_{|\varphi(x)|<\frac{v}{4},|x|\geq\frac{v}{4}}(\varphi(x))^{2}|x|^{-Y-1}dx\\
&\leq \bar{C}r^{2}\int_{|x|<\frac{v}{4}}|x|^{1-Y}dx+
\frac{v^{2}}{16}\bar{C}\int_{|x|\geq\frac{v}{4}}|x|^{-Y-1}dx\\
&\leq\tilde{\kappa}v^{2-Y},
\end{align*}
for} some $0<\tilde{\kappa}<\infty$. Therefore, whenever $C_{+}(Y-1)^{-1}(4r)^{Y-1}tv^{1-Y}\leq v/4$ (or equivalently, $C_{+}(Y-1)^{-1}r^{Y-1}4^{Y}tv^{-Y}\leq 1$),
\begin{align*}
t^{-1}\widetilde{\bbp}\left(\widetilde{U}_{t}^{(\varepsilon)}\geq\frac{v}{2}\right)\leq\frac{eV_{v/4}^{2}}{v^{2}}\leq\tilde{\kappa}v^{-Y},
\end{align*}
for some $0<\tilde{\kappa}<\infty$. {Moreover,} for any $t>0$ and $v>0$,
\begin{align}\nonumber
t^{-1}\widetilde{\bbp}\left(\widetilde{U}_{t}^{(\varepsilon)}\geq\frac{v}{2}\right)&=t^{-1}\widetilde{\bbp}\left(\widetilde{U}_{t}^{(\varepsilon)}\geq\frac{v}{2}\right){\bf 1}_{\left\{C_{+}(Y-1)^{-1}r^{Y-1}4^{Y}tv^{-Y}\leq 1\right\}}+t^{-1}\widetilde{\bbp}\left(\widetilde{U}_{t}^{(\varepsilon)}\geq\frac{v}{2}\right){\bf 1}_{\left\{C_{+}(Y-1)^{-1}r^{Y-1}4^{Y}tv^{-Y}>1\right\}}\\
&\leq\tilde{\kappa}v^{-Y}+\frac{C_{+}}{t}(Y-1)^{-1}r^{Y-1}4^{Y}tv^{-Y}\leq\tilde{\kappa}'v^{-Y},\label{NEsH2}
\end{align}
for some constant $0<\tilde{\kappa}'<\infty$. Combining (\ref{NEsH1}) and (\ref{NEsH2}), {finally leads to}
\begin{equation} \label{UsflTlEstTldU}
t^{-1}\widetilde{\bbp}\left(\widetilde{U}_{t}\geq v\right)\leq t^{-1}\widetilde{\bbp}\left({\bar{\widetilde{U}}_{t}^{(\varepsilon)}}\geq\frac{v}{2}\right)+t^{-1}\widetilde{\bbp}\left(\widetilde{U}_{t}^{(\varepsilon)}\geq\frac{v}{2}\right)\leq\tilde{\kappa}v^{-Y},
\end{equation}
for all $v>0$ and $t>0$ and some constant $0<\tilde{\kappa}<\infty$}.\hfill\qed

\medskip
\noindent
\textbf{Proof of Theorem \ref{2ndASY}.}

\noindent
For simplicity, {the case $\tilde{\gamma}=0$ is treated first,} so that, in light of Lemma \ref{Lm:NRATMPJ},
\begin{align}\label{SmplOP}
{t^{-\frac{1}{Y}}\mathbb{E}\left[\left(S_{t}-S_{0}\right)^{+}\right]}=e^{-\eta t}\int_{0}^{\infty}e^{-t^{\frac{1}{Y}}v}\widetilde{\bbe}\left(e^{-\widetilde{U}_{t}}{\bf 1}_{\left\{t^{-\frac{1}{Y}}Z_{t}\geq v\right\}}\right)dv.
\end{align}
The general case is resolved in Lemma \ref{GenCseGam} below. Let
\begin{align*}
D(t):={t^{-\frac{1}{Y}}\mathbb{E}\left[\left(S_{t}-S_{0}\right)^{+}\right]}-\widetilde{\bbe}\left(Z_{1}^{+}\right),
\end{align*}
which can be written as
\begin{align}
D(t)&:=\left[\int_{0}^{\infty}e^{-t^{\frac{1}{Y}}v}\widetilde{\bbe}\left(e^{-\widetilde{U}_{t}}{\bf 1}_{\left\{t^{-\frac{1}{Y}}Z_{t}\geq v\right\}}\right)dv-\widetilde{\bbe}\left(Z_{1}^{+}\right)\right]+\left(e^{-\eta t}-1\right)\widetilde{\bbe}\left(Z_{1}^{+}\right)\nonumber\\
&\quad\,\,+(e^{-\eta t}-1){\left[\int_{0}^{\infty}e^{-t^{\frac{1}{Y}}v}\widetilde{\bbe}\left(e^{-\widetilde{U}_{t}}{\bf 1}_{\left\{t^{-\frac{1}{Y}}Z_{t}\geq v\right\}}\right)dv-{\widetilde{\bbe}}\left(Z_{1}^{+}\right)\right]}\nonumber\\
\label{DecomD} &=:D_{1}(t)+D_{2}(t)+D_{3}(t).
\end{align}
{Let us show that}
\begin{equation}\label{KNLFMP}
t^{\frac{1}{Y}-1}D_{1}(t)\rightarrow\vartheta+\eta=:\tilde{\vartheta},\quad\text{as }\,t\to 0,
\end{equation}
for a certain constant $\vartheta$, while it is clear that $D_{3}(t)=o(D_{1}(t))$ and that $t^{1/Y-1}D_{2}(t)=o(1)$, as $t\to 0$. First, note that
\begin{align}
D_{1}(t)&=\widetilde{\bbe}\left(e^{-\widetilde{U}_{t}}\int_{0}^{t^{-\frac{1}{Y}}Z_{t}^{+}}e^{-t^{\frac{1}{Y}}v}dv\right)-\widetilde{\bbe}\left(Z_{1}^{+}\right)\nonumber\\
&=t^{-\frac{1}{Y}}{\left[\widetilde{\bbe}\left(e^{-\widetilde{U}_{t}}\right)-\widetilde{\bbe}\left(e^{-\left(\widetilde{U}_{t}+Z_{t}^{+}\right)}\right)\right]}-\widetilde{\bbe}\left(Z_{1}^{+}\right)\nonumber\\
\label{D1} &=t^{-\frac{1}{Y}}{\left[e^{\eta t}-\widetilde{\bbe}\left(e^{-\left(\widetilde{U}_{t}+Z_{t}^{+}\right)}\right)\right]}-\widetilde{\bbe}\left(Z_{1}^{+}\right),
\end{align}
where to obtain the last equality {use (\ref{ExpMmntTildeU})}. Next, by the self-similarity of $(Z_{t})_{t\geq 0}$ under $\widetilde{\bbp}$ (see (\ref{SSCN})), $\widetilde{\bbe}\left(Z_{1}^{+}\right)=t^{-1/Y}\widetilde{\bbe}\left(Z_{t}^{+}\right)$ and, since {$\widetilde{\bbe}\left(\widetilde{U}_{t}\right)=0$},
\begin{align}
t^{\frac{1}{Y}-1}D_{1}(t)&=t^{\frac{1}{Y}-1}{\left[\frac{e^{\eta t}-1}{t^{\frac{1}{Y}}}+\frac{1-\widetilde{\bbe}\left(e^{-(Z_{t}^{+}+\widetilde{U}_{t})}\right)-\widetilde{\bbe}\left(Z_{t}^{+}+\widetilde{U}_{t}\right)}{t^{\frac{1}{Y}}}\right]}\nonumber\\
&=\frac{e^{\eta t}-1}{t}+t^{-1}\widetilde{\bbe}\left(\int_{0}^{Z_{t}^{+}+\widetilde{U}_{t}}\left(e^{-v}-1\right)dv{\bf 1}_{\{Z_{t}^{+}+\widetilde{U}_{t}\geq 0\}}\right)-t^{-1}\widetilde{\bbe}\left(\int_{Z_{t}^{+}+\widetilde{U}_{t}}^{0}\left(e^{-v}-1\right)dv{\bf 1}_{\{Z_{t}^{+}+\widetilde{U}_{t}\leq 0\}}\right)\nonumber\\
&=\frac{e^{\eta t}-1}{t}+t^{-1}\int_{0}^{\infty}\left(e^{-v}-1\right)\widetilde{\bbp}\left(Z_{t}^{+}+\widetilde{U}_{t}\geq v\right)dv-t^{-1}\int_{0}^{\infty}\left(e^{v}-1\right)\widetilde{\bbp}\left(Z_{t}^{+}+\widetilde{U}_{t}\leq -v\right)dv\nonumber\\
\label{DecomD1} &=: D_{11}(t)+D_{12}(t)+D_{13}(t).
\end{align}
Clearly,
\begin{align}\label{limD11}
D_{11}(t)\rightarrow\eta,\quad\text{as }\,t\to 0.
\end{align}
{For any $t\geq 0$, let} $Q_{t}:=Z_{t}+\widetilde{U}_{t}$ and note that $Q_{t}\leq Z_{t}^{+}+\widetilde{U}_{t}$. Then, for $D_{13}(t)$, using {respectively} $e^{y}-1\leq y e^{y}$, $y>0$, Markov's inequality, and since $Y<2$,
\begin{align}
0\leq {D_{13}(t)}&\leq t^{\frac{1}{Y}-1}\int_{0}^{\infty}\left(e^{t^{\frac{1}{Y}}u}-1\right)\widetilde{\bbp}\left(t^{-\frac{1}{Y}}Q_{t}\leq-u\right)du\nonumber\\
&\leq t^{\frac{2}{Y}-1}\int_{0}^{\infty}e^{t^{\frac{1}{Y}}u}u\widetilde{\bbp}\left(t^{-\frac{1}{Y}}Q_{t}\leq -u\right)du\nonumber\\	
\label{limD13} &\leq t^{\frac{2}{Y}-1}\int_{0}^{\infty}e^{(t^{\frac{1}{Y}}-1)u}u\,du\cdot\widetilde{\bbe}\left(e^{-t^{-\frac{1}{Y}}Q_{t}}\right){\to 0,\quad t\to 0,}
\end{align}
where in the last step we applied (\ref{LmtMGFTldU}-i).

{For $D_{12}$, from (\ref{TIITI}-ii), by dominated convergence, one passes} the limit inside the integrals so that
\begin{align*}
\lim_{t\to 0}D_{12}(t)=\int_{0}^{\infty}\left(e^{-v}-1\right)\lim_{t\to 0}{\left[t^{-1}\widetilde{\bbp}\left(Z_{t}^{+}+\widetilde{U}_{t}\geq v\right)\right]}dv.
\end{align*}
{Then, from (\ref{SmplLmtSLJ1}),}
\begin{align}\label{D12Domi}
\lim_{t\to 0}D_{12}(t)=\int_{0}^{\infty}\left(e^{-v}-1\right)\int_{\bbr_{0}}{\bf 1}_{\left\{x^{-}-\ln\bar{q}(x)\geq v\right\}}\tilde{\nu}(dx)dv=:\vartheta.
\end{align}
Combining (\ref{DecomD1}), (\ref{limD11}), (\ref{limD13}), and (\ref{D12Domi}), it follows that
\begin{equation}\label{DExLmN}	
\lim_{t\to 0}t^{\frac{1}{Y}-1}\left({t^{-\frac{1}{Y}}\mathbb{E}\left[\left(S_{t}-S_{0}\right)^{+}\right]}-\widetilde{\bbe}\left(Z_{1}^{+}\right)\right)=\lim_{t\to 0}t^{\frac{1}{Y}-1}D_{1}(t)=\vartheta+\eta.
\end{equation}
Finally, to get the expression in (\ref{vartheta}), recall that $\tilde{\nu}(dx)=|x|^{-Y-1}\left(C_{+}{\bf 1}_{\{x>0\}}+C_{-}{\bf 1}_{\{x<0\}}\right)dx$ and, thus, applying Fubini's theorem to the right-hand side of (\ref{D12Domi}) gives
\begin{align*}
\vartheta &=C_{+}\int_{0}^{\infty}\left(e^{-v}-1\right){\left(\int_{0}^{\infty}{\bf 1}_{\left\{-\ln\bar{q}(x)\geq v\right\}}x^{-Y-1}dx\right)}dv+C_{-}\int_{0}^{\infty}\left(e^{-v}-1\right){\left(\int_{-\infty}^{0}{\bf 1}_{\left\{-x-\ln\bar{q}(x)\geq v\right\}}|x|^{-Y-1}dx\right)}dv\\
&=C_{+}\int_{0}^{\infty}{\left(\int_{0}^{-\ln\bar{q}(x)}\left(e^{-v}-1\right)dv\right)}x^{-Y-1}dx+C_{-}\int_{-\infty}^{0}{\left(\int_{0}^{-x-\ln\bar{q}(x)}\left(e^{-v}-1\right)dv\right)}|x|^{-Y-1}dx\\
&=C_{+}\int_{0}^{\infty}\left(1-e^{\ln\bar{q}(x)}+\ln\bar{q}(x)\right)x^{-Y-1}dx+C_{-}\int_{-\infty}^{0}\left(1-e^{x+\ln\bar{q}(x)}+x+\ln\bar{q}(x)\right)|x|^{-Y-1}dx.
\end{align*}
One can similarly show that the constant $\eta$ defined in (\ref{Uplusminuseta}) can be written as:
\begin{align*}
\eta=C_{+}\int_{0}^{\infty}\left(e^{x+\ln\bar{q}(x)}-1-\ln\bar{q}(x)-x\right)x^{-Y-1}dx+C_{-}\int_{-\infty}^{0}\left(e^{x+\ln\bar{q}(x)}-1-\ln\bar{q}(x)-x\right)|x|^{-Y-1}dx.
\end{align*}
Combining {the} expressions for $\vartheta$ and $\eta$ yields (\ref{vartheta}). The expression for $\tilde{\gamma}$ in (\ref{Deftildegamma}) follows from
\begin{align*}
\tilde{\gamma}={\widetilde{\bbe}\left(L^{*}_{1}\right)}=\tilde{b}+\int_{\{|x|> 1\}}x\,\tilde{\nu}(dx)=b^{*}+\int_{|x|\leq 1}x\left(\tilde{\nu}-\nu^{*}\right)(dx)+\int_{\{|x|>1\}}x\,\tilde{\nu}(dx),
\end{align*}
and standard simplifications. This concludes the proof.\hfill\qed

\begin{lem}\label{GenCseGam}
If $\tilde{\gamma}\neq 0$ in (\ref{LOP}), then
\begin{align}\label{2ndASP0}
\lim_{t\to 0}t^{\frac{1}{Y}-1}\left(t^{-\frac{1}{Y}}\frac{1}{S_{0}}{\bbe\left[\left(S_{t}-S_{0}\right)^{+}\right]}-S_{0}\,\widetilde{\bbe}\left(Z_{1}^{+}\right)\right)=\tilde{\vartheta}+\tilde{\gamma}\widetilde{\bbp}\left(Z_{1}\geq 0\right).
\end{align}
\end{lem}

\noindent
\textbf{Proof.}
Without loss of generality, fix $S_{0}=1$, and also assume that $\tilde{\gamma}>0$ (the case $\tilde\gamma<0$ being similar). Using (\ref{LOP}),
\begin{align}
t^{\frac{1}{Y}-1}\left({t^{-\frac{1}{Y}}\bbe\left[\left(S_{t}-S_{0}\right)^{+}\right]}-\widetilde{\bbe}\left(Z_{1}^{+}\right)\right)&=t^{\frac{1}{Y}-1}{\left[e^{-(\tilde{\gamma}+\eta)t}\int_{0}^{\infty} e^{-t^{\frac{1}{Y}}v}\widetilde{\bbe}\left(e^{-\widetilde{U}_{t}}{\bf 1}_{\left\{t^{-\frac{1}{Y}}Z_{t}\geq v\right\}}\right)dv-\widetilde{\bbe}\left(Z_{1}^{+}\right)\right]}\nonumber\\
&\quad+t^{\frac{1}{Y}-1}e^{-(\tilde{\gamma}+\eta)t}\int_{-\tilde{\gamma}t^{1-\frac{1}{Y}}}^{0}e^{-t^{\frac{1}{Y}}v}\widetilde{\bbe}\left(e^{-\widetilde{U}_{t}}{\bf 1}_{\left\{t^{-\frac{1}{Y}}Z_{t}\geq v\right\}}\right)dv\nonumber\\
&=:\widetilde{D}_{11}(t)+\widetilde{D}_{12}(t).\nonumber
\end{align}
As in the proof of (\ref{KNLFMP}), it can be shown that
\begin{align}\label{limtdD11}
\lim_{t\to 0}\widetilde{D}_{11}(t)=\tilde{\vartheta}.
\end{align}
For $\widetilde{D}_{12}(t)$, changing variables to $u=t^{1/Y-1}v$ and probability measure to $\bbp^{*}$, we have
\begin{align*}
\widetilde{D}_{12}(t)=e^{-\tilde{\gamma}t}\int_{-\tilde{\gamma}}^{0}e^{-tu}\widetilde{\bbe}\left(e^{-(\widetilde{U}_{t}+\eta t)}{\bf 1}_{\left\{Z_{t}\geq tu\right\}}\right)du=e^{-\tilde{\gamma}t}\int_{-\tilde{\gamma}}^{0}e^{-tu}\bbp^{*}\left(t^{-\frac{1}{Y}}Z_{t}\geq t^{1-\frac{1}{Y}}u\right)du.
\end{align*}
Next, recall from Section \ref{Sec:TmpStble} that, under $\mathbb{P}^{*}$, $(L^{*}_{t})_{t\geq 0}$ is a L\'{e}vy process with L\'{e}vy measure $\nu^{*}$ given by
\begin{align*}
\nu^{*}(dx):=e^{x}\nu(dx)=e^{x}s(x)dx=q^{*}(x)|x|^{-Y-1}dx.
\end{align*}
In particular, $\lim_{x\searrow 0}q^{*}(x)=C_{+}$ and $\lim_{x\nearrow 0} q^{*}(x)=C_{-}$ and, since $1<Y<2$, the assumptions of~\cite[Proposition 1]{rosenbaum.tankov.10} are satisfied. Therefore, {both $t^{-1/Y}L^{*}_{t}$ and $t^{-1/Y}Z_{t}$} converge in distribution to a $Y$-stable random variable $\widetilde{Z}$ under $\bbp^{*}$ with center (or mean) $0$ and L\'{e}vy measure $|x|^{-Y-1}\left(C_{+}{\bf 1}_{\{x>0\}}+C_{-}{\bf 1}_{\{x<0\}}\right)dx$. Hence, the distribution of $\widetilde{Z}$ (under $\bbp^{*}$) is the same as the distribution of $Z_{1}$ under $\widetilde{\bbp}$. Thus, Slutsky's lemma implies that $t^{-1/Y}Z_{t}-t^{1-1/Y}u\ld\widetilde{Z}$ and, thus,
\begin{align*}
\lim_{t\to 0}\bbp^{*}\left(t^{-\frac{1}{Y}}Z_{t}-t^{1-\frac{1}{Y}}u\geq 0\right)=\bbp^{*}\left(\widetilde{Z}\geq 0\right)=\widetilde{\bbp}\left({Z}_{1}\geq 0\right).
\end{align*}
Finally, by the dominated convergence theorem,
\begin{align}\label{NDLGNZ}
\lim_{t\to 0}\widetilde{D}_{12}(t)=\tilde{\gamma}\,\widetilde{\bbp}\left(Z_{1}\geq 0\right).
\end{align}
Combining (\ref{NDLGNZ}) with (\ref{limtdD11}) leads to (\ref{2ndASP0}).\hfill\qed

\medskip
\noindent
\textbf{Proof of Corollary \ref{AsyIVPCGMY}.}

\noindent
The small-time asymptotic behavior of the ATM call option price $C_{BS}(t,\sigma)$ at maturity $t$ under the Black-Scholes model with volatility $\sigma$ and zero interest rates is given by (e.g., see~\cite[Corollary 3.4]{FordeJacLee:2010} and recall also that $S_{0}=1$)
\begin{align}\label{AsyIVATMBS}
C_{BS}(t,\sigma)=\frac{\sigma}{\sqrt{2\pi}}t^{\frac{1}{2}}-\frac{\sigma^{3}}{24\sqrt{2\pi}}t^{\frac{3}{2}}+O\left(t^{\frac{5}{2}}\right),\quad t\to 0.
\end{align}
To derive the small-time asymptotics for the implied volatility, {a result analogous to (\ref{AsyIVATMBS}) is needed} when $\sigma$ is replaced by $\hat{\sigma}(t)$. The following representation taken from~\cite[Lemma 3.1]{RoperRut:2007} {is} useful,
\begin{align*}
C_{BS}(t,\sigma)=F(\sigma\sqrt{t})\quad\text{with}\quad F(\theta):=\int_{0}^{\theta}\Phi'\left(\frac{v}{2}\right)dv=\frac{1}{\sqrt{2\pi}}\int_{0}^{\theta}\exp\left(-\frac{v^{2}}{8}\right)dv,
\end{align*}
together with the Taylor expansion for $F$ at $\theta=0$ (see~\cite[Lemma 5.1]{RoperRut:2007}), i.e.,
\begin{align*}
F(\theta)=\frac{1}{\sqrt{2\pi}}\theta-\frac{1}{24\sqrt{2\pi}}\theta^{3}+O\left(\theta^{5}\right),\quad\theta\rightarrow 0.
\end{align*}
Then, since $\hat{\sigma}(t)\to 0$ as $t\to 0$ (see, e.g., \cite[Proposition 5]{Tankov}),
\begin{align}\label{eq:BSOptIV}
C_{BS}(t,\hat{\sigma}(t))=\frac{\hat\sigma(t)}{\sqrt{2\pi}}t^{\frac{1}{2}}-\frac{\hat\sigma(t)^{3}}{24\sqrt{2\pi}}t^{\frac{3}{2}}+O\left(\left(\hat\sigma(t)t^{\frac{1}{2}}\right)^{5}\right),\quad \text{as }\,t\rightarrow 0.
\end{align}
Returning to the proof of Proposition \ref{AsyIVPCGMY}, {equating (\ref{ExpAsymBehCGMY}) and (\ref{eq:BSOptIV}) {gives}
\[
	 \frac{d_{1}t^{{\frac{1}{Y}}}}{\frac{\hat\sigma(t)}{\sqrt{2\pi}}t^{\frac{1}{2}}}=\frac{1+O(t^{1-1/Y})}{1+O\left(\left(\hat\sigma(t)t^{\frac{1}{2}}\right)^{2}\right)},
\]
{showing that}
\begin{align}\label{1stAsyIVPureCGMY}
\hat{\sigma}(t)\sim\sqrt{2\pi}\,d_{1}t^{\frac{1}{Y}-\frac{1}{2}},\qquad t\to 0.
\end{align}
Next, using (\ref{1stAsyIVPureCGMY}) and setting $\tilde{\sigma}(t):=\hat{\sigma}(t)-\sqrt{2\pi}\,d_{1}t^{\frac{1}{Y}-\frac{1}{2}}$, {rewrite} (\ref{eq:BSOptIV}) as follows:
\[
	C_{BS}=d_{1}t^{\frac{1}{Y}}+\frac{\tilde\sigma(t)}{\sqrt{2\pi}}t^{\frac{1}{2}}+O\left(t^{\frac{3}{Y}}\right).
\]	
{Equating  (\ref{ExpAsymBehCGMY}) and the previous expression leads to}
\begin{equation}\label{NFCH}
	d_{2} t +o(t)=\frac{\tilde\sigma(t)}{\sqrt{2\pi}}t^{\frac{1}{2}}+O\left(t^{\frac{3}{Y}}\right),
\end{equation}
which, {together with the fact that the second-term on the right of (\ref{NFCH}) is $o(t)$,} implies that
\begin{align*}
	d_{2}t\sim\frac{\tilde{\sigma}(t)}{\sqrt{2\pi}}\sqrt{t},\qquad t\to 0.
\end{align*}
{Therefore,} $\tilde{\sigma}(t)\rightarrow 0$ as $t\rightarrow 0$ and, moreover, recalling the definition of $d_{2}$,}
\begin{align}\label{2ndAsyIVPureCGMY}
\tilde{\sigma}(t)\sim\sqrt{2\pi}\left(\tilde{\vartheta}+\tilde{\gamma}\widetilde{\bbp}\left(Z_{1}\geq 0\right)\right)\sqrt{t},\qquad t\to 0.
\end{align}
Combining (\ref{1stAsyIVPureCGMY}) and (\ref{2ndAsyIVPureCGMY}) finishes the proof.\hfill\qed

\section{Proofs of Section \ref{Sect:NonZeroBrwn}: {The Pure-Jump Model} With A {Nonzero Brownian Component}}\label{ProofsSectGenCse}

\medskip
\noindent
\textbf{Proof of Theorem \ref{2ndOAsyCGMYB}.}

\noindent
For simplicity, fix $S_{0}=1$. Recalling that $X_{t}=\sigma W_{t}^{*}+L_{t}^{*}$ under $\bbp^{*}$, and using (\ref{CMR}), the self-similarity of $W^{*}$, and the change of variable $u=t^{-1/2}x$,
\begin{align*}
R_{t}&:={t^{-\frac{1}{2}}\bbe\left[\left(S_{t}-S_{0}\right)^{+}\right]}-\sigma\bbe^{*}\left(W_{1}^{*}\right)^{+}=\int_{0}^{\infty}e^{-\sqrt{t}u}\bbp^{*}\left(\sigma W_{1}^{*}\geq u-t^{-\frac{1}{2}}L_{t}^{*}\right)du-\int_{0}^{\infty}\bbp^{*}\left(\sigma W_{1}^{*}\geq u\right)du.
\end{align*}
Next, changing the probability measure to $\widetilde{\bbp}$, using that $L_{t}^{*}=Z_{t}+\tilde{\gamma}t$, $U_{t}=\widetilde{U}_{t}+\eta t$, and the change of variable $y=u-t^{1/2}\tilde{\gamma}$ in the first integral above, lead to
\begin{align}
R_{t}&=\int_{-\sqrt{t}\tilde{\gamma}}^{\infty}e^{-\sqrt{t}y-\tilde{\gamma}t}\,\widetilde{\bbe}\left(e^{-\widetilde{U}_{t}-\eta t}{\bf 1}_{\left\{\sigma W_{1}^{*}\geq y-t^{-\frac{1}{2}}Z_{t}\right\}}\right)dy-\int_{0}^{\infty}\widetilde\bbe\left(e^{-\widetilde{U}_{t}-\eta t}{\bf 1}_{\{\sigma W_{1}^{*}\geq u\}}\right)du\nonumber\\
&=e^{-(\eta+\tilde{\gamma})t}\int_{0}^{\infty}e^{-\sqrt{t}y}{\left[\widetilde{\bbe}\left(e^{-\widetilde{U}_{t}}{\bf 1}_{\left\{\sigma W_{1}^{*}\geq y-t^{-\frac{1}{2}}Z_{t}\right\}}\right)-\widetilde\bbe\left(e^{-\widetilde{U}_{t}}{\bf 1}_{\{\sigma W_{1}^{*}\geq y\}}\right)\right]}dy\nonumber\\
\label{2ndOADecCGMYB} &\quad+e^{-(\eta+\tilde{\gamma})t}\int_{-\sqrt{t}\tilde{\gamma}}^{0}e^{-\sqrt{t}y}\widetilde{\bbe}\left(e^{-\widetilde{U}_{t}}{\bf 1}_{\left\{\sigma W_{1}^{*}\geq y-t^{-\frac{1}{2}}Z_{t}\right\}}\right)dy+\int_{0}^{\infty}\left(e^{-\tilde{\gamma}t-\sqrt{t}y}-1\right)\bbp^{*}\left(\sigma W_{1}^{*}\geq y\right)dy.
\end{align}
Above, the last term is clearly $O(t^{1/2})$ as $t\to 0$, while the middle term can be shown to be asymptotically equivalent to a term that is also $O(t^{1/2})$ by arguments analogous to those of (\ref{NDLGNZ}). Thus, only the first term in (\ref{2ndOADecCGMYB}), which we hereafter denote by $A_{t}$, needs to be studied. Setting $\tilde{\eta}:=\eta+\tilde{\gamma}$, this term can further be expressed as:
\begin{align*}
A_{t}&=e^{-\tilde{\eta}t}\widetilde{\bbe}{\left[e^{-\widetilde{U}_{t}}\int_{0}^{\infty}e^{-\sqrt{t}y}\left({\bf 1}_{\left\{\sigma W_{1}^{*}\geq y-t^{-\frac{1}{2}}Z_{t}\right\}}-{\bf 1}_{\{\sigma W_{1}^{*}\geq y\}}\right)dy\right]},
\end{align*}
To study the asymptotic behavior of $A_{t}$, decompose it into the following three parts:
\begin{align}
A_{t}&=e^{-\tilde{\eta}t}\widetilde{\bbe}\left(e^{-\widetilde{U}_{t}}{\bf 1}_{\left\{W_{1}^{*}\geq 0,\sigma W_{1}^{*}+t^{-\frac{1}{2}}Z_{t}\geq 0\right\}}\!\int_{\sigma W_{1}^{*}}^{\sigma W_{1}^{*}+t^{-\frac{1}{2}}Z_{t}}\!\!\!e^{-\sqrt{t}y}\,dy\right)-e^{-\tilde{\eta}t}\widetilde{\bbe}\left(e^{-\widetilde{U}_{t}}{\bf 1}_{\left\{0\leq\sigma W_{1}^{*}\leq-t^{-\frac{1}{2}}Z_{t}\right\}}\!\int_{0}^{\sigma W_{1}^{*}}\!\!\!e^{-\sqrt{t}y}\,dy\right)\nonumber\\
&\quad+e^{-\tilde{\eta}t}\widetilde{\bbe}\left(e^{-\widetilde{U}_{t}}{\bf 1}_{\left\{0\leq-\sigma W_{1}^{*}\leq t^{-\frac{1}{2}}Z_{t}\right\}}\int_{0}^{\sigma W_{1}^{*}+t^{-\frac{1}{2}}Z_{t}}e^{-\sqrt{t}y}\,dy\right)\nonumber\\
\label{DecomAt}&=: I_{1}(t)-I_{2}(t)+I_{3}(t).
\end{align}
Each of these terms is analyzed in the following three steps:

\smallskip
\noindent
\textbf{Step 1.}
Since $(Z_{t})_{t\geq 0}$ and $(W_{t}^{*})_{t\geq 0}$ are independent,
\begin{align}
I_{1}(t)&=e^{-\tilde{\eta}t}\widetilde{\bbe}\left({\bf 1}_{\left\{W_{1}^{*}\geq 0,\sigma W_{1}^{*}+t^{-\frac{1}{2}}Z_{t}\geq 0\right\}}\frac{e^{-\widetilde{U}_{t}}-e^{-(\widetilde{U}_{t}+Z_{t})}}{\sqrt{t}}e^{-\sqrt{t}\sigma W_{1}^{*}}\right)\nonumber\\
&=e^{-\tilde{\eta}t}\int_{0}^{\infty}\widetilde{\bbe}\left({\bf 1}_{\left\{Z_{t}\geq -t^{\frac{1}{2}}y\right\}}\frac{e^{-\widetilde{U}_{t}}\left(1-e^{-Z_{t}}\right)}{\sqrt{t}}\right)e^{-\sqrt{t}y}\frac{e^{-\frac{y^{2}}{2\sigma^{2}}}}{\sqrt{2\pi\sigma^{2}}}\,dy\nonumber\\
\label{IntJ1} &=:e^{-\tilde{\eta}t}\int_{0}^{\infty}J_{1}(t,y)\,e^{-\sqrt{t}y}\frac{e^{-\frac{y^{2}}{2\sigma^{2}}}}{\sqrt{2\pi\sigma^{2}}}\,dy.
\end{align}
Using the self-similarity of $(Z_{t})_{t\geq 0}$ and since $\widetilde{\bbe}Z_{t}=0$, $J_{1}(t,y)$ is then decomposed as:
\begin{align}
J_{1}(t,y)&=\widetilde{\bbe}{\left[{\bf 1}_{\left\{Z_{t}\geq -t^{\frac{1}{2}}y\right\}}\left(\frac{e^{-\widetilde{U}_{t}}-e^{-(\widetilde{U}_{t}+Z_{t})}}{\sqrt{t}}-t^{-\frac{1}{2}}Z_{t}\right)\right]}+t^{\frac{1}{Y}-\frac{1}{2}}\widetilde{\bbe}{\left[\left(-Z_{1}\right){\bf 1}_{\left\{-Z_{1}\geq t^{\frac{1}{2}-\frac{1}{Y}}y\right\}}\right]}\nonumber\\
\label{DecomJ1} &=:\!J_{11}(t,y)\!+\!J_{12}(t,y).
\end{align}
Let us first consider $J_{12}(t,y)$. From (\ref{Asydenpz00})-(\ref{AsytailZ100b}), there exists a constant $\lambda>0$ such that
\begin{equation}\label{Eq:NdBnd1}
t^{\frac{Y}{2}-1}J_{12}(t,y)\leq\lambda y^{1-Y},
\end{equation}
for any $0<t\leq 1$ and $y>0$ (see Appendix \ref{AddtnProofs} for proof of this claim). Moreover, for any fixed $y>0$,
\begin{align*}
t^{\frac{Y}{2}-1}J_{12}(t,y)=t^{\frac{Y}{2}+\frac{1}{Y}-\frac{3}{2}}\int_{t^{\frac{1}{2}-\frac{1}{Y}}y}^{\infty}up_{Z}(-u)\,du
=t^{\frac{Y}{2}-\frac{1}{Y}-\frac{1}{2}}\int_{y}^{\infty}wp_{Z}\left(-t^{\frac{1}{2}-\frac{1}{Y}}w\right)dw.
\end{align*}
Using (\ref{Asydenpz00}), there exists $0<t_{0}<1$ such that
\begin{align*}
t^{\frac{Y}{2}-\frac{1}{Y}-\frac{1}{2}}wp_{Z}\left(-t^{\frac{1}{2}-\frac{1}{Y}}w\right)\leq 2\left(C_{+}\vee C_{-}\right)w^{-Y},
\end{align*}
for any $0<t<t_{0}$ and $w\geq y$. Therefore, by the dominated convergence theorem, and in light of (\ref{Asydenpz00}),
\begin{align}
\lim_{t\rightarrow 0}t^{\frac{Y}{2}-1}e^{-\tilde{\eta}t}\int_{0}^{\infty}J_{12}(t,y)e^{-\sqrt{t}y}\frac{e^{-\frac{y^{2}}{2\sigma^{2}}}}{\sqrt{2\pi\sigma^{2}}}\,dy&=\int_{0}^{\infty}\left(\lim_{t\rightarrow 0}t^{\frac{Y}{2}-1}J_{12}(t,y)\right)\frac{e^{-\frac{y^{2}}{2\sigma^{2}}}}{\sqrt{2\pi\sigma^{2}}}\,dy\nonumber\\
&=\int_{0}^{\infty}{\left[\int_{y}^{\infty}w\left(\lim_{t\rightarrow 0}t^{\frac{Y}{2}-\frac{1}{Y}-\frac{1}{2}}p_{Z}\left(-t^{\frac{1}{2}-\frac{1}{Y}}w\right)\right)dw\right]}\frac{e^{-\frac{y^{2}}{2\sigma^{2}}}}{\sqrt{2\pi\sigma^{2}}}\,dy\nonumber\\ &=C_{-}\int_{0}^{\infty}\left(\int_{y}^{\infty}{w^{-Y}}dw\right)\frac{e^{-\frac{y^{2}}{2\sigma^{2}}}}{\sqrt{2\pi\sigma^{2}}}\,dy\nonumber\\
\label{AsyInJ12} &=\frac{C_{-}}{Y-1}\int_{0}^{\infty}y^{1-Y}\frac{e^{-\frac{y^{2}}{2\sigma^{2}}}}{\sqrt{2\pi\sigma^{2}}}\,dy.
\end{align}
For $J_{11}(t,y)$,
\begin{align}
J_{11}(t,y)&=t^{-\frac{1}{2}}\widetilde{\bbe}\left({\bf 1}_{\left\{Z_{t}\geq 0\right\}}\int_{\widetilde{U}_{t}}^{\widetilde{U}_{t}+Z_{t}}\left(e^{-x}-1\right)dx\right)-t^{-\frac{1}{2}}\widetilde{\bbe}\left({\bf 1}_{\left\{-t^{\frac{1}{2}}y\leq Z_{t}\leq 0\right\}}\int_{\widetilde{U}_{t}+Z_{t}}^{\widetilde{U}_{t}}\left(e^{-x}-1\right)dx\right)\nonumber\\
&=t^{-\frac{1}{2}}\int_{\bbr}\left(e^{-x}-1\right)T_{1}(t,x,y)\,dx-t^{-\frac{1}{2}}\int_{\bbr}\left(e^{-x}-1\right)T_{2}(t,x,y)\,dx,
\end{align}
where, for $t>0$ and $y>0$, we set
\begin{align*}
T_{1}(t,x,y):=\widetilde{\bbp}\left(Z_{t}\geq 0,\,\widetilde{U}_{t}\leq x\leq\widetilde{U}_{t}+Z_{t}\right),\quad T_{2}(t,x,y):=\widetilde{\bbp}\left(-t^{\frac{1}{2}}y\leq Z_{t}\leq 0,\,\widetilde{U}_{t}+Z_{t}\leq x\leq\widetilde{U}_{t}\right).
\end{align*}
By (\ref{TIITI}-ii), there exists $0<\tilde{\kappa}<\infty$ such that, for any $x>0$ and $0<t\leq 1$,
\begin{align}\label{NEFS0}
T_{1}(t,x,y)\leq\widetilde{\bbp}\left(x\leq\widetilde{U}_{t}+Z_{t}\right)\leq\widetilde{\bbp}\left(x\leq\widetilde{U}_{t}+Z_{t}^{+}\right)\leq\tilde{\kappa}tx^{-Y}.
\end{align}
Hence,
\begin{align}
0&\leq e^{-\tilde{\eta}t}t^{\frac{Y}{2}-1}\int_{0}^{\infty}{\left(\int_{0}^{\infty}\frac{\left(1-e^{-x}\right)}{\sqrt{t}}T_{1}(t,x,y)\,dx\right)}\frac{e^{-\sqrt{t}y}e^{-\frac{y^{2}}{2\sigma^{2}}}}{\sqrt{2\pi\sigma^{2}}}\,dy\nonumber\\
\label{AsyDecomJ111} &\leq e^{-\tilde{\eta}t}\tilde{\kappa}t^{\frac{Y-1}{2}}\int_{0}^{\infty}{\left(\int_{0}^{\infty}\left(1-e^{-x}\right)x^{-Y}\,dx\right)}\frac{e^{-\frac{y^{2}}{2\sigma^{2}}}}{\sqrt{2\pi\sigma^{2}}}\,dy\rightarrow 0,\quad\text{as }\,t\rightarrow 0,
\end{align}
since $Y>1$. Similarly, using (\ref{TIITI}-i), there exists a constant $0<\tilde{\kappa}<\infty$ such that
\begin{align}\label{EstT2post}
T_{2}(t,x,y)&\leq\widetilde{\bbp}\left(\widetilde{U}_{t}\geq x\right)\leq\tilde{\kappa}tx^{-Y},
\end{align}
for any $x>0$ and $0<t\leq 1$, {and thus}, as in (\ref{AsyDecomJ111}),
\begin{align}\label{AsyDecomJ112}
\lim_{t\rightarrow 0}{t^{\frac{Y}{2}-1}e^{-\tilde{\eta}t}}\int_{0}^{\infty}{\left(\int_{0}^{\infty}\frac{\left(1-e^{-x}\right)}{\sqrt{t}}T_{2}(t,x,y)\,dx\right)}\frac{e^{-\sqrt{t}y}e^{-\frac{y^{2}}{2\sigma^{2}}}}{\sqrt{2\pi\sigma^{2}}}\,dy=0.
\end{align}
For $x<0$, using (\ref{LmtMGFTldU}-ii) and Markov's inequality, there exist $0<t_{0}<1$ and $0<\tilde{\kappa}<\infty$ such that
\begin{align}\label{EstT1neq}
T_{1}(t,x,y)&\leq\widetilde{\bbp}\left(t^{-\frac{1}{Y}}\widetilde{U}_{t}\leq t^{-\frac{1}{Y}}x\right)\leq\widetilde{\bbe}\left(e^{-t^{-\frac{1}{Y}}\widetilde{U}_{t}}\right)e^{t^{-\frac{1}{Y}}x}\leq\tilde{\kappa}e^{t^{-\frac{1}{Y}}x},
\end{align}
for any $0<t\leq t_{0}$. Therefore,
\begin{align*}
0\leq t^{\frac{Y-3}{2}}\int_{-\infty}^{0}\left(e^{-x}-1\right)T_{1}(t,x,y)\,dx\leq\tilde{\kappa}t^{\frac{Y-3}{2}}\int_{-\infty}^{0}\left(e^{\left(t^{-\frac{1}{Y}}-1\right)x}-e^{t^{-\frac{1}{Y}}x}\right)dx=\tilde{\kappa}t^{\frac{Y-3}{2}}\frac{t^{\frac{2}{Y}}}{1-t^{\frac{1}{Y}}}.
\end{align*}
Hence, by the dominated convergence theorem, {as $t\to 0$,}
\begin{align}\label{AsyDecomJ113}
0\leq e^{-\tilde{\eta}t} t^{\frac{Y}{2}-1}\int_{0}^{\infty}{\left(\int_{-\infty}^{0}\frac{e^{-x}-1}{\sqrt{t}}T_{1}(t,x,y)\,dx\right)}\frac{e^{-\sqrt{t}y}e^{-\frac{y^{2}}{2\sigma^{2}}}}{\sqrt{2\pi\sigma^{2}}}\,dy\leq\tilde{\kappa}t^{\frac{Y-3}{2}}\frac{t^{\frac{2}{Y}}}{1-t^{\frac{1}{Y}}}\int_{0}^{\infty}\frac{e^{-\frac{y^{2}}{2\sigma^{2}}}}{\sqrt{2\pi\sigma^{2}}}\,dy\rightarrow 0,
\end{align}
since $2/Y>1>(3-Y)/2$, for $1<Y<2$. Similarly, using  (\ref{LmtMGFTldU}-i), for $x<0$,
\begin{align}\label{IneqForT2}
T_{2}(t,x,y)\leq\widetilde{\bbp}\left(\widetilde{U}_{t}+Z_{t}\leq x\right)\leq\widetilde{\bbe}\left(e^{-t^{-\frac{1}{Y}}\left(\widetilde{U}_{t}+Z_{t}\right)}\right)e^{t^{-\frac{1}{Y}}x}\leq\tilde{\kappa}e^{t^{-\frac{1}{Y}}x},
\end{align}
for any $0<t\leq t_{0}$ and some constant $0<\tilde{\kappa}<\infty$. Therefore, as in (\ref{AsyDecomJ113}),
\begin{align}\label{AsyDecomJ114}
\lim_{t\rightarrow 0}e^{-\eta t}t^{\frac{Y}{2}-1}\int_{0}^{\infty}{\left(\int_{-\infty}^{0}\frac{\left(1-e^{-x}\right)}{\sqrt{t}}T_{2}(t,x,y)\,dx\right)}\frac{e^{-\sqrt{t}y}e^{-\frac{y^{2}}{2\sigma^{2}}}}{\sqrt{2\pi\sigma^{2}}}\,dy=0.
\end{align}
Combining (\ref{AsyInJ12}), (\ref{AsyDecomJ111}), (\ref{AsyDecomJ112}), (\ref{AsyDecomJ113}) and (\ref{AsyDecomJ114}) finally gives
\begin{align}\label{AsyI1}
\lim_{t\rightarrow 0}t^{\frac{Y}{2}-1}I_{1}(t)={\frac{C_{-}}{Y-1}}\int_{0}^{\infty}y^{1-Y}\frac{e^{-\frac{y^{2}}{2\sigma^{2}}}}{\sqrt{2\pi\sigma^{2}}}\,dy.
\end{align}

\smallskip
\noindent
\textbf{Step 2.}
The asymptotic behavior of $I_{2}(t)$ is now studied. Using the independence of $(Z_{t})_{t\geq 0}$ and $(W_{t}^{*})_{t\geq 0}$,
\begin{align}
I_{2}(t)&=e^{-\tilde{\eta}t}\widetilde{\bbe}\left(e^{-\widetilde{U}_{t}}{\bf 1}_{\left\{0\leq\sigma W_{1}^{*}\leq -t^{-\frac{1}{2}}Z_{t}\right\}}\frac{1-e^{-\sqrt{t}\sigma W_{1}^{*}}}{\sqrt{t}}\right)\nonumber\\
&=e^{-\tilde{\eta}t}\int_{0}^{\infty}\widetilde{\bbe}\left(e^{-\widetilde{U}_{t}}{\bf 1}_{\left\{Z_{t}\leq-t^{\frac{1}{2}}y\right\}}\right)\frac{1-e^{-\sqrt{t}y}}{\sqrt{t}}\frac{e^{-\frac{y^{2}}{2\sigma^{2}}}}{\sqrt{2\pi\sigma^{2}}}\,dy\nonumber\\
\label{DecomI2} &=e^{-\tilde{\eta}t}\!\!\!\int_{0}^{\infty}\!\!\widetilde{\bbe}\!{\left[\!\left(e^{-\widetilde{U}_{t}}\!-\!1\right)\!{\bf 1}_{\left\{Z_{t}\leq -t^{\frac{1}{2}}y\right\}}\!\right]}\!\frac{1\!-\!e^{-\sqrt{t}y}}{\sqrt{t}}\frac{e^{-\frac{y^{2}}{2\sigma^{2}}}}{\sqrt{2\pi\sigma^{2}}}dy\!+\!e^{-\tilde{\eta}t}\!\!\!\int_{0}^{\infty}\!\!\widetilde{\bbp}\left(Z_{t}\!\leq\!-t^{\frac{1}{2}}y\right)\!\frac{1\!-\!e^{-\sqrt{t}y}}{\sqrt{t}}\frac{e^{-\frac{y^{2}}{2\sigma^{2}}}}{\sqrt{2\pi\sigma^{2}}}dy.
\end{align}
By (\ref{KIn}) and the self-similarity of $(Z_{t})_{t\geq 0}$, for $y>0$,
\begin{align*}
\widetilde{\bbp}\left(Z_{t}\leq -t^{\frac{1}{2}}y\right)\frac{1-e^{-\sqrt{t}y}}{\sqrt{t}}=\widetilde{\bbp}\left(Z_{1}\leq -t^{\frac{1}{2}-\frac{1}{Y}}y\right)\frac{1-e^{-\sqrt{t}y}}{\sqrt{t}}\leq\kappa t^{1-\frac{Y}{2}}y^{1-Y}\leq\kappa y^{1-Y},
\end{align*}
which, when multiplied by $\exp(-y^{2}/2\sigma^{2})$, becomes integrable on $[0,\infty)$. Hence, by (\ref{AsytailZ100b}) and the dominated convergence theorem,
\begin{align}\label{AsyI2second}
\lim_{t\rightarrow 0}t^{\frac{Y}{2}-1}e^{-\tilde{\eta}t}\int_{0}^{\infty}\widetilde{\bbp}\left(Z_{t}\leq -t^{\frac{1}{2}}y\right)\frac{1-e^{-\sqrt{t}y}}{\sqrt{t}}\frac{e^{-\frac{y^{2}}{2\sigma^{2}}}}{\sqrt{2\pi\sigma^{2}}}\,dy
=\frac{C_{-}}{Y}\int_{0}^{\infty}y^{1-Y}\frac{e^{-\frac{y^{2}}{2\sigma^{2}}}}{\sqrt{2\pi\sigma^{2}}}\,dy.
\end{align}
To find the asymptotic behavior of the first integral in (\ref{DecomI2}), decompose it as:
\begin{align}
{\widetilde{\bbe}\left[\left(e^{-\widetilde{U}_{t}}-1\right){\bf 1}_{\left\{Z_{t}\leq -t^{\frac{1}{2}}y\right\}}\right]}&=\widetilde{\bbe}\left({\bf 1}_{\left\{Z_{t}\leq -t^{\frac{1}{2}}y,\widetilde{U}_{t}<0\right\}}\int_{\widetilde{U}_{t}}^{0}e^{-u}\,du\right)-\widetilde{\bbe}\left({\bf 1}_{\left\{Z_{t}\leq -t^{\frac{1}{2}}y,\widetilde{U}_{t}\geq 0\right\}}\int_{0}^{\widetilde{U}_{t}}e^{-u}\,du\right)\nonumber\\
\label{DocomFirIntI2} &=:J_{21}(t,y)+J_{22}(t,y).
\end{align}
For $J_{21}(t,y)$, using Markov's inequality and (\ref{LmtMGFTldU}-ii), there exist $0<\tilde{\kappa}<\infty$ and $0<t_{0}<1$ such that
\begin{align*}
J_{21}(t,y)&=\int_{-\infty}^{0}e^{-x}\widetilde{\mathbb{P}}\left(Z_{t}\leq -t^{\frac{1}{2}}y,\,t^{-\frac{1}{Y}}\widetilde{U}_{t}\leq t^{-\frac{1}{Y}}x\right)dx\leq\widetilde{\mathbb{E}}\left(e^{-t^{-1/Y}\bar{U}_{t}}\right)\int_{-\infty}^{0}e^{-x}\exp{\left(t^{-\frac{1}{Y}}x\right)}dx\leq\frac{\tilde{\kappa}t^{\frac{1}{Y}}}{1-t^{\frac{1}{Y}}},
\end{align*}
for any $0<t<t_{0}$ and $y\geq 0$. Since $1-Y/2<1/2<1/Y$, for $1<Y<2$, by the dominated convergence theorem,
\begin{align}\label{AsyInJ21}
0\leq t^{\frac{Y}{2}-1}e^{-\tilde{\eta}t}\int_{0}^{\infty}J_{21}(t,y)\frac{1-e^{-\sqrt{t}y}}{\sqrt{t}}\frac{e^{-\frac{y^{2}}{2\sigma^{2}}}}{\sqrt{2\pi\sigma^{2}}}\,dy\leq\tilde{\kappa}t^{\frac{Y}{2}-1}\frac{t^{\frac{1}{Y}}}{1-t^{\frac{1}{Y}}}\int_{0}^{\infty}y\frac{e^{-\frac{y^{2}}{2\sigma^{2}}}}{\sqrt{2\pi\sigma^{2}}}\,dy\rightarrow 0,\quad\text{as }\,t\rightarrow 0.
\end{align}
Next, further decompose the second term $J_{22}(t,y)$ in (\ref{DocomFirIntI2}) as:
\begin{align}
J_{22}(t,y)&={\widetilde{\bbe}\left[\left(e^{-\widetilde{U}_{t}}-1+\widetilde{U}_{t}\right){\bf 1}_{\left\{Z_{t}\leq -t^{\frac{1}{2}}y,\widetilde{U}_{t}\geq 0\right\}}\right]}-\widetilde{\bbe}\left(\widetilde{U}_{t}{\bf 1}_{\left\{Z_{t}\leq -t^{\frac{1}{2}}y,\widetilde{U}_{t}\geq 0\right\}}\right)\nonumber\\
\label{DecomJ22} &=:J_{22}^{(1)}(t,y)-J_{22}^{(2)}(t,y).
\end{align}
Using (\ref{TIITI}-i) and the fact that $1-Y/2<1/2<1/Y$, for $1<Y<2$,
\begin{align}
0\leq J_{22}^{(2)}(t,y)&\leq t^{\frac{1}{Y}}{\widetilde{\bbe}\left[\left(t^{-\frac{1}{Y}}\widetilde{U}_{t}\right)^{+}\right]}=t^{\frac{1}{Y}}\int_{0}^{\infty}\bbp\left(t^{-\frac{1}{Y}}\widetilde{U}_{t}\geq u\right)du\leq t^{\frac{1}{Y}}\left(1+\int_{1}^{\infty}\bbp\left(t^{-\frac{1}{Y}}\widetilde{U}_{t}\geq u\right)du\right)\nonumber\\
\label{AsyInJ222} &\leq t^{\frac{1}{Y}}\left(1+\int_{1}^{\infty}\tilde{\kappa}t(t^{\frac{1}{Y}}u)^{-Y}du\right),
\end{align}
which is clearly $o(t^{1-\frac{Y}{2}})$. Moreover,
\begin{align*}
J_{22}^{(1)}(t,y)\!=\!\widetilde{\bbe}\!\left(\int_{0}^{\widetilde{U}_{t}}\!\!\!\!\left(1\!-\!e^{-w}\right)\!dw{\bf 1}_{\left\{Z_{t}\leq -t^{\frac{1}{2}}y,\widetilde{U}_{t}\geq 0\right\}}\!\right)\!=\!\!\int_{0}^{\infty}\!\!\!\!\left(1\!-\!e^{-w}\right)\!\widetilde{\bbp}\!\left(\widetilde{U}_{t}\!\geq\!w,Z_{t}\!\leq\!-t^{\frac{1}{2}}y\right)\!dw\!\leq\!\!\int_{0}^{\infty}\!\!\!\!\left(1\!-\!e^{-w}\right)\!\widetilde{\bbp}\!\left(\widetilde{U}_{t}\!\geq\!w\right)\!dw.
\end{align*}
Using (\ref{EstT2post}), {$t^{-1/2}\left(1-e^{-\sqrt{t}y}\right)\leq y$}, $y>0$, and by the dominated convergence theorem,
\begin{align}\label{AsyInJ221}
0\leq t^{\frac{Y}{2}-1}e^{-\tilde{\eta}t}\!\int_{0}^{\infty}\!J_{22}^{(1)}(t,y)\frac{1-e^{-\sqrt{t}y}}{\sqrt{t}}\frac{e^{-\frac{y^{2}}{2\sigma^{2}}}}{\sqrt{2\pi\sigma^{2}}}\,dy\leq t^{\frac{Y}{2}}\int_{0}^{\infty}\!\left(1-e^{-w}\right)w^{-Y}dw\cdot\int_{0}^{\infty}y\frac{e^{-\frac{y^{2}}{2\sigma^{2}}}}{\sqrt{2\pi\sigma^{2}}}\,dy\rightarrow 0,
\end{align}
 {as $t\to 0$.} Combining (\ref{AsyI2second}), (\ref{AsyInJ21}), (\ref{AsyInJ222}) and (\ref{AsyInJ221}) lead to
\begin{align}\label{AsyI2}
\lim_{t\rightarrow 0}t^{\frac{Y}{2}-1}I_{2}(t)=\frac{C_{-}}{Y}\int_{0}^{\infty}y^{1-Y}\frac{e^{-\frac{y^{2}}{2\sigma^{2}}}}{\sqrt{2\pi\sigma^{2}}}\,dy.
\end{align}

\smallskip
\noindent
\textbf{Step 3.}
To finish, let us study the behavior of $I_{3}(t)$. Note that
\begin{align}
I_{3}(t)&=e^{-\tilde{\eta}t}\int_{0}^{\infty}\widetilde{\bbe}\left(e^{-\widetilde{U}_{t}}{\bf 1}_{\{Z_{t}\geq\sqrt{t}y\}}\frac{1-e^{\sqrt{t}y}e^{-Z_{t}}}{\sqrt{t}}\right)\frac{e^{-\frac{y^{2}}{2\sigma^{2}}}}{\sqrt{2\pi\sigma^{2}}}\,dy\nonumber\\
\label{DefnJ31J32} &=e^{-\tilde{\eta}t}\!\!\!\int_{0}^{\infty}\!\!\widetilde{\bbe}\left(e^{-\widetilde{U}_{t}}{\bf 1}_{\left\{Z_{t}\geq\sqrt{t}y\right\}}\right)\!\frac{1\!-\!e^{\sqrt{t}y}}{\sqrt{t}}\!\frac{e^{-\frac{y^{2}}{2\sigma^{2}}}}{\sqrt{2\pi\sigma^{2}}}dy\!+\!e^{-\tilde{\eta}t}\!\!\!\int_{0}^{\infty}\!\!\widetilde{\bbe}\left(e^{-\widetilde{U}_{t}}{\bf 1}_{\left\{Z_{t}\geq\sqrt{t}y\right\}}\!\frac{1\!-\!e^{-Z_{t}}}{\sqrt{t}}\right)e^{\sqrt{t}y}\frac{e^{-\frac{y^{2}}{2\sigma^{2}}}}{\sqrt{2\pi\sigma^{2}}}dy.
\end{align}
First, decompose $J_{31}(t,y):=\widetilde{\bbe}\left(e^{-\widetilde{U}_{t}}{\bf 1}_{\left\{Z_{t}\geq\sqrt{t}y\right\}}\right)$ as:
\begin{align}\label{DecomJ31}
J_{31}(t,y)={\widetilde{\bbe}\left[\left(e^{-\widetilde{U}_{t}}-1\right){\bf 1}_{\{Z_{t}\geq\sqrt{t}y\}}\right]}+\widetilde{\bbp}\left(Z_{t}\geq\sqrt{t}y\right)=:J_{31}^{(1)}(t,y)+J_{31}^{(2)}(t,y).
\end{align}
By (\ref{KIn}), it is easy to see that $J_{31}^{(2)}(t,y)\leq \kappa t^{1-\frac{Y}{2}}y^{-Y}$, for any $0<t\leq 1$ and $y\geq 0$.
Hence, {the} dominated convergence theorem together with (\ref{AsytailZ100}) lead to
\begin{align}\label{AsyInJ312}
\lim_{t\rightarrow 0}e^{-\tilde{\eta}t}t^{\frac{Y}{2}-1}\int_{0}^{\infty}J_{31}^{(2)}(t,y)\frac{1-e^{\sqrt{t}y}}{\sqrt{t}}\frac{e^{-\frac{y^{2}}{2\sigma^{2}}}}{\sqrt{2\pi\sigma^{2}}}\,dy=-\frac{C_{+}}{Y}\int_{0}^{\infty}y^{1-Y}\frac{e^{-\frac{y^{2}}{2\sigma^{2}}}}{\sqrt{2\pi\sigma^{2}}}\,dy.
\end{align}
As before, $J_{31}^{(1)}(t,y)$ can be further decomposed as
\begin{align*}
J_{31}^{(1)}(t,y)={\widetilde{\bbe}\left[\left(e^{-\widetilde{U}_{t}}-1\right){\bf 1}_{\left\{Z_{t}\geq\sqrt{t}y,\widetilde{U}_{t}\geq 0\right\}}\right]}+\int_{0}^{\infty}e^{u}\widetilde{\bbp}\left(Z_{t}\geq\sqrt{t}y,\,\widetilde{U}_{t}\leq -u\right)du.
\end{align*}
Now, for $u>0$, $y>0$ and $t>0$, by Markov's inequality and (\ref{LmtMGFTldU}-ii), there exist $0<\tilde{\kappa}<\infty$ and $0<t_{0}<1$, such that
\begin{align}\label{EstJ31first}
\widetilde{\bbp}\left(Z_{t}\geq\sqrt{t}y,\,\widetilde{U}_{t}\leq -u\right)\leq\widetilde{\bbp}\left(\widetilde{U}_{t}\leq -u\right)\leq\widetilde{\bbe}\left(e^{-t^{-\frac{1}{Y}}\bar{U}_{t}}\right)e^{-t^{-\frac{1}{Y}}u}\leq\tilde{\kappa}e^{-t^{-\frac{1}{Y}}u},
\end{align}
for all $0<t<t_{0}$. Also, as done in (\ref{AsyInJ222}), there exist $0<K<\infty$ and $0<t_{0}<1$ such that
\begin{align}\label{EstJ31sec}
0\leq\widetilde{\bbe}{\left[\left(1-e^{-\widetilde{U}_{t}}\right){\bf 1}_{\left\{Z_{t}\geq\sqrt{t}y,\widetilde{U}_{t}\geq 0\right\}}\right]\leq\widetilde{\bbe}\left(\widetilde{U}_{t}^{+}\right)}\leq Kt^{\frac{1}{Y}},
\end{align}
for all $0<t<t_{0}$. Hence, for any $y\geq 0$ and $0<t<t_{0}<1$, and since $1-Y/2<1/2<1/Y$, for $1<Y<2$,
\begin{align}\label{AsyJ311}
t^{\frac{Y}{2}-1}\left|J_{31}^{(1)}(t,y)\right|&\leq\tilde{\kappa}t^{\frac{Y}{2}-1}\int_{0}^{\infty}e^{-u\left(t^{-\frac{1}{Y}}-1\right)}du+Kt^{\frac{1}{Y}+\frac{Y}{2}-1}=\tilde{\kappa}t^{\frac{Y}{2}-1}\frac{t^{\frac{1}{Y}}}{1-t^{\frac{1}{Y}}}+Kt^{\frac{1}{Y}+\frac{Y}{2}-1}\rightarrow 0,\quad\text{as }\,t\to 0.
\end{align}
Since both control functions in (\ref{EstJ31first}) and (\ref{EstJ31sec}) are independent of $y$, combining (\ref{AsyInJ312}) and (\ref{AsyJ311}), and by the dominated convergence theorem,
\begin{align}\label{AsyInJ31}
\lim_{t\rightarrow 0}e^{-\tilde{\eta}t}t^{\frac{Y}{2}-2}\int_{0}^{\infty}J_{31}(t,y)\frac{1-e^{\sqrt{t}y}}{\sqrt{t}}\frac{e^{-\frac{y^{2}}{2\sigma^{2}}}}{\sqrt{2\pi\sigma^{2}}}\,dy=-\frac{C_{+}}{Y}\int_{0}^{\infty}y^{1-Y}\frac{e^{-\frac{y^{2}}{2\sigma^{2}}}}{\sqrt{2\pi\sigma^{2}}}\,dy.
\end{align}
Next, by the self-similarity of $(Z_{t})_{t\geq 0}$, the term $J_{32}(t,y):=\widetilde{\bbe}\left(t^{-1/2}e^{-\widetilde{U}_{t}}{\bf 1}_{\{Z_{t}\geq t^{\frac{1}{2}}y\}}\left(1-e^{-Z_{t}}\right)\right)$ appearing in (\ref{DefnJ31J32}) is decomposed as:
\begin{align}
J_{32}(t,y)&={\widetilde{\bbe}\left[{\bf 1}_{\left\{Z_{t}\geq\sqrt{t}y\right\}}\left(\frac{e^{-\widetilde{U}_{t}}-e^{-(Z_{t}+\widetilde{U}_{t})}}{\sqrt{t}}-t^{-\frac{1}{2}}Z_{t}\right)\right]}+\widetilde{\bbe}\left(t^{\frac{1}{Y}-\frac{1}{2}}Z_{1}{\bf 1}_{\left\{Z_{1}\geq t^{\frac{1}{2}-\frac{1}{Y}}y\right\}}\right)\nonumber\\
\label{DecomJ32} &=:\!J_{32}^{(1)}(t,y)+J_{32}^{(2)}(t,y).
\end{align}
Note that $J_{32}^{(2)}(t,y)$ is quite similar to $J_{12}(t,y)$ in (\ref{DecomJ1}) and, thus, the corresponding integral has an asymptotic behavior similar to (\ref{AsyInJ12}). Concretely,
\begin{align}
\lim_{t\rightarrow 0}t^{\frac{Y}{2}-1}e^{-\tilde{\eta}t}\int_{0}^{\infty}\widetilde{\bbe}\left(t^{\frac{1}{Y}-\frac{1}{2}}Z_{1}{\bf 1}_{\left\{Z_{1}\geq t^{\frac{1}{2}-\frac{1}{Y}}y\right\}}\right)e^{\sqrt{t}y}\frac{e^{-\frac{y^{2}}{2\sigma^{2}}}}{\sqrt{2\pi\sigma^{2}}}\,dy=\frac{C_{+}}{Y-1}\int_{0}^{\infty}y^{1-Y}\frac{e^{-\frac{y^{2}}{2\sigma^{2}}}}{\sqrt{2\pi\sigma^{2}}}\,dy.
\label{AsyInJ32sec}
\end{align}
Next, decompose $J_{32}^{(1)}(t,y)$ as:
\begin{align*}
J_{32}^{(1)}(t,y)&=t^{-\frac{1}{2}}\widetilde{\bbe}\left({\bf 1}_{\left\{Z_{t}\geq t^{\frac{1}{2}}y\right\}}\int_{\widetilde{U}_{t}}^{Z_{t}+\widetilde{U}_{t}}\left(e^{-x}-1\right)dx\right)\\
&=t^{-\frac{1}{2}}\!\!\!\int_{-\infty}^{0}\!\!\left(e^{-x}\!-\!1\right)\widetilde{\bbp}\!\left(Z_{t}\!\geq\!\sqrt{t}y,\widetilde{U}_{t}\leq x\leq Z_{t}+\widetilde{U}_{t}\right)\!dx+t^{-\frac{1}{2}}\!\!\int_{0}^{\infty}\!\!\!\left(e^{-x}\!-\!1\right)\widetilde{\bbp}\!\left(Z_{t}\!\geq\!\sqrt{t}y,\widetilde{U}_{t}\leq x\leq Z_{t}+\widetilde{U}_{t}\right)\!dx.
\end{align*}
Note that for $x>0$,
\begin{align*}
\widetilde{\bbp}\left(Z_{t}\geq\sqrt{t}y,\,\widetilde{U}_{t}\leq x\leq Z_{t}+\widetilde{U}_{t}\right)\leq\widetilde{\bbp}\left(x\leq Z_{t}+\widetilde{U}_{t}\right),
\end{align*}
while for $x<0$,
\begin{align*}
\widetilde{\bbp}\left(Z_{t}\geq\sqrt{t}y,\,\widetilde{U}_{t}\leq x\leq Z_{t}+\widetilde{U}_{t}\right)\leq\widetilde{\bbp}\left(\widetilde{U}_{t}\leq x\right).
\end{align*}
Using the estimates (\ref{NEFS0}) and (\ref{EstT1neq}), arguments as in getting (\ref{AsyDecomJ111}) and (\ref{AsyDecomJ113}) give
\begin{align}\label{AsyInJ321}
\lim_{t\rightarrow 0}e^{-\tilde{\eta}t}t^{\frac{Y}{2}-1}\int_{0}^{\infty}J_{32}^{(1)}(t,y)e^{\sqrt{t}y}\frac{e^{-\frac{y^{2}}{2\sigma^{2}}}}{\sqrt{2\pi\sigma^{2}}}\,dy=0.
\end{align}
Combining (\ref{AsyInJ31}), (\ref{AsyInJ32sec}) and (\ref{AsyInJ321}) leads to
\begin{align}\label{AsyI3}
\lim_{t\rightarrow 0}t^{\frac{Y}{2}-1}I_{3}(t)=\frac{C_{+}}{Y(Y-1)}\int_{0}^{\infty}y^{1-Y}\frac{e^{-\frac{y^{2}}{2\sigma^{2}}}}{\sqrt{2\pi\sigma^{2}}}\,dy.
\end{align}
Finally, from (\ref{DecomAt}), (\ref{AsyI1}), (\ref{AsyI2}) and (\ref{AsyI3}), and since $1-Y/2<1/2$, for $1<y<2$, (\ref{2ndAsyCGMYB}) is obtained.\hfill\qed

\medskip
\noindent
\textbf{Proof of Corollary \ref{AsyIVCGMYB}.}\\
\noindent
When the diffusion component is nonzero, \cite[Proposition 5]{Tankov} implies that $\hat{\sigma}(t)\rightarrow\sigma$ as $t\rightarrow 0$. In particular, ${\hat{\sigma}(t)}t^{1/2}\rightarrow 0$ as $t\rightarrow 0$ and, thus, (\ref{eq:BSOptIV}) above remains true. Let $\tilde{\sigma}(t):={\hat{\sigma}(t)}-\sigma$, then $\tilde{\sigma}(t)\rightarrow 0$ as $t\rightarrow 0$, and (\ref{eq:BSOptIV}) can be written as
\begin{align}
C_{BS}(t,\hat{\sigma}(t))&=\frac{\sigma}{\sqrt{2\pi}}t^{\frac{1}{2}}+\frac{\tilde{\sigma}(t)}{\sqrt{2\pi}}t^{\frac{1}{2}}-\frac{\hat\sigma(t)^{3}}{24\sqrt{2\pi}}t^{\frac{3}{2}}+O\left(\left(\hat\sigma(t)t^{\frac{1}{2}}\right)^{5}\right)\nonumber\\
\label{eq:BSOptIVBC} &=\frac{\sigma}{\sqrt{2\pi}}t^{\frac{1}{2}}+\frac{\tilde{\sigma}(t)}{\sqrt{2\pi}}t^{\frac{1}{2}}+O\left(t^{\frac{3}{2}}\right).
\end{align}
Comparing (\ref{ExpAsymBehCGMYBM})-(\ref{PrcDefn2ndTrm}) and (\ref{eq:BSOptIVBC}) gives
\begin{align*}
\frac{(C_{+}+C_{-})2^{-\frac{1+Y}{2}}\sigma^{1-Y}}{Y(Y-1)\sqrt{\pi}}\Gamma\left(1-\frac{Y}{2}\right)t^{\frac{3-Y}{2}}\sim\frac{\tilde{\sigma}(t)}{\sqrt{2\pi}}\sqrt{t},\quad t\to 0,
\end{align*}
and, therefore,
\begin{align*}
\tilde{\sigma}(t)\sim\frac{(C_{+}+C_{-})2^{-\frac{Y}{2}}\sigma^{1-Y}}{Y(Y-1)}\Gamma\left(1-\frac{Y}{2}\right)t^{1-\frac{Y}{2}},\quad t\rightarrow 0.
\end{align*}
The proof is now complete.\hfill\qed

\section{Additional Proofs}\label{AddtnProofs}

\medskip
\noindent
\textbf{Verification of (\ref{KIn}).}

\noindent
It suffices to show that, for some constant $0<\kappa<\infty$ and any $v>0$ and $0<t\leq 1$,
\begin{align*}
{\widetilde{\bbp}\left(Z_{1}\geq t^{-\frac{1}{Y}}v\right)\leq\kappa t|v|^{-Y}.}
\end{align*}
Indeed, the same argument for $-Z_{1}$ shows an analogous bound for $\widetilde{\bbp}\left(Z_{1}\leq t^{-1/Y}v\right)$, with $v<0$, which in turn will imply (\ref{KIn}). First, in light of (\ref{Asydenpz00}), there exist {$0<R<\infty$} and {$H>0$}, such that for any {$u\geq H$},
\begin{align}\label{AsyInqdenpz}
p_{Z}(u)\leq Ru^{-Y-1}.
\end{align}
Thus, whenever $t^{-1/Y}v\geq H$,
\begin{align*}
\widetilde{\bbp}\left(Z_{1}\geq t^{-\frac{1}{Y}}v\right)\leq R\int_{t^{-\frac{1}{Y}}v}^{\infty}u^{-Y-1}du=\frac{R}{Y}\left(t^{-\frac{1}{Y}}v\right)^{-Y}=\frac{R}{Y}tv^{-Y}.
\end{align*}
Since $\widetilde{\bbp}\left(Z_{1}\geq t^{-1/Y}v\right)\leq 1<H^{Y}t v^{-Y}$, when $t^{-1/Y}v< H$,
\begin{align*}
\widetilde{\bbp}\left(Z_{1}\geq t^{-\frac{1}{Y}}v\right)&\leq {\bf 1}_{\left\{t^{-\frac{1}{Y}}v\geq H\right\}}\widetilde{\bbp}\left(Z_{1}\geq t^{-\frac{1}{Y}}v\right)+{\bf 1}_{\left\{t^{-\frac{1}{Y}}v< H\right\}}\widetilde{\bbp}\left(Z_{1}\geq t^{-\frac{1}{Y}}v\right)\\
&\leq {\bf 1}_{\left\{t^{-\frac{1}{Y}}v\geq H\right\}}\frac{R}{Y}tv^{-Y}+{\bf 1}_{\left\{t^{-\frac{1}{Y}}v<H\right\}}H^{Y}tv^{-Y}\\
&\leq tv^{-Y}\left(\frac{R}{Y}+H^{Y}\right),
\end{align*}
for all $0<t\leq 1$ and $v>0$.\hfill\qed

\medskip
\noindent
\textbf{Verification of (\ref{Eq:NdBnd1})}

\noindent
This is similar to the proof of (\ref{KIn}). Using (\ref{AsyInqdenpz}), for any $t>0$ and $y>0$,
\begin{align*}
t^{\frac{Y}{2}-1}J_{12}(t,y)&=t^{\frac{Y}{2}-1} t^{\frac{1}{Y}-\frac{1}{2}}\int_{t^{\frac{1}{2}-\frac{1}{Y}}y}^{\infty}up_{Z}(u)du\\
&= t^{\left(\frac{1}{Y}-\frac{1}{2}\right)(1-Y)}{\bf 1}_{\left\{t^{\frac{1}{2}-\frac{1}{Y}}y\geq H\right\}}\int_{t^{\frac{1}{2}-\frac{1}{Y}}y}^{\infty}up_{Z}(u)du\\
&\quad +t^{\left(\frac{1}{Y}-\frac{1}{2}\right)(1-Y)}{\bf 1}_{\left\{{t^{\frac{1}{2}-\frac{1}{Y}}y\leq H}\right\}}\left(\int_{H}^{\infty}up_{Z}(u)du+\int_{t^{\frac{1}{2}-\frac{1}{Y}}y}^{{{}H}}up_{Z}(u)du\right)\\
&\leq t^{\left(\frac{1}{Y}-\frac{1}{2}\right)(1-Y)}{\bf 1}_{\left\{t^{\frac{1}{2}-\frac{1}{Y}}y\geq H\right\}}\int_{t^{\frac{1}{2}-\frac{1}{Y}}y}^{\infty}{Ru^{-Y}}du\\
&\quad+t^{\left(\frac{1}{Y}-\frac{1}{2}\right)(1-Y)}{\bf 1}_{\left\{t^{\frac{1}{2}-\frac{1}{Y}}y\leq H\right\}}\left(\int_{H}^{\infty}Ru^{-Y}du+H\widetilde{\bbp}\left(Z_{1}\geq t^{\frac{1}{2}-\frac{1}{Y}}y\right)\right).
\end{align*}
Then, using (\ref{KIn}),
\begin{align*}
t^{\frac{Y}{2}-1}J_{12}(t,y)&\leq t^{\left(\frac{1}{Y}-\frac{1}{2}\right)(1-Y)}{R(Y-1)^{-1}}y^{1-Y}t^{\left(\frac{1}{2}-\frac{1}{Y}\right)(1-Y)}\\
&\quad+t^{\left(\frac{1}{Y}-\frac{1}{2}\right)(1-Y)}{\bf 1}_{\left\{t^{\frac{1}{2}-\frac{1}{Y}}y\leq H\right\}}\left(RH^{1-Y}+H\left(\frac{H}{t^{\frac{1}{2}-\frac{1}{Y}}y}\right)^{Y-1}\right)\\
&\leq R(Y-1)^{-1}y^{1-Y}+t^{\left(\frac{1}{Y}-\frac{1}{2}\right)(1-Y)}Rt^{\left(\frac{1}{2}-\frac{1}{Y}\right)(1-Y)}y^{1-Y}+H^{Y}y^{1-Y},
\end{align*}
and (\ref{Eq:NdBnd1}) follows with $\lambda:=RY/(Y-1)+H^{Y}$. \hfill\qed

\bibliographystyle{jtbnew}

\begin{thebibliography}{37}
\expandafter\ifx\csname natexlab\endcsname\relax\def\natexlab#1{#1}\fi
\expandafter\ifx\csname url\endcsname\relax
  \def\url#1{\texttt{#1}}\fi
\expandafter\ifx\csname urlprefix\endcsname\relax\def\urlprefix{URL }\fi
\providecommand{\selectlanguage}[1]{\relax}

\bibitem[A\"it-Sahalia and Jacod(2009)]{AitJacod09}
A\"it-Sahalia, Y., and J. Jacod (2009):
\newblock Estimating the degree of activity of jumps in high-frequency data.
\newblock \emph{{Annals of Statistics}} {37}(5A), {2202--2244}.

\bibitem[Applebaum(2004)]{Applebaum2004} Applebaum, D. (2004):  \textit{L\'{e}vy Processes and Stochastic Calculus}, Cambridge {University} Press, Cambridge.

\bibitem[Belomestny(2010)]{Belomestny:2010}
Belomestny, D. (2010):
\newblock {Spectral estimation of the fractional order of a L\'evy process},
\newblock \emph{{Annals of Statistics}} {38}(1), {317-351}.

\bibitem[Barndorff-Nielsen(1997)]{BNS:1998}
Barndorff-Nielsen, O. E. (1997):
\newblock {Processes of normal inverse Gaussian type},
\newblock {\em {Finance and Stochastics}}, 2, 41-68.

\bibitem[Berestycki et al.(2002)]{BBF:2002}
Berestycki, H., J. Busca, and I. Florent (2002):
\newblock{Asymptotics and calibration of local volatility models},
\newblock{\em {Quantitative Finance}}, 2, 61-69.

\bibitem[Berestyki(2004)]{BBF2:2004}
Berestyki, H. , J. Busca, and I. Florent (2004):
\newblock{Computing the implied volatility in stochastic volatility models},
\newblock{\em {Communications on Pure and Applied Mathematics}}, Vol LVII, 1352-1373.

{
\bibitem[Bertoin(1996)]{Bertoin} Bertoin, J. (1996):
\newblock {L\'evy Processes}, \newblock{Cambridge University Press},
\newblock{Cambridge}

\bibitem[Boyarchenko  and Levendorksii(2002)]{BL02} Boyarchenko S.I., and S.Z. Levendorksii (2002):
\newblock {Non-Gaussian Merton-Black- Scholes theory}, \newblock {\em {Adv. Ser. Stat. Sci. Appl. Probab.}} 9. World {Scientific} Publishing Co., Inc., River Edge, NJ.
}

\bibitem[Carr et al.(2002)]{CGMY:2002}
Carr, P. , H. Geman, D. Madan, and M. Yor (2002):
\newblock {The fine structure of asset returns: an empirical investigation},
\newblock {\em Journal of Business}, 75, 305-332.

\bibitem[Carr and Madan(2009)]{CM09}
Carr, P., and D. Madan (2009):
\newblock {Saddle point methods for option pricing},
\newblock {\em {The Journal of Computational Finance}}, 13(1), 49-61.

\bibitem[Cont et al.(1997)]{Cont:1997}
Cont, R., J.~Bouchaud, and M.~Potters (1997):
\newblock {Scaling in financial data: stable laws and beyond}.
\newblock {\em In ``Scale Invariance and Beyond", B. Dubrulle, F. Graner and D. Sornette, eds.}.

\bibitem[Cont and Tankov(2004)]{CT04}
Cont, R., and P.~Tankov (2004):
\newblock {\em Financial modelling with jump processes},
\newblock Chapman \& Hall.

\bibitem[Carr and Wu(2003)]{CW03}
Carr, R., and L. Wu (2003):
\newblock {What type of process underlies options? A simple robust test},
\newblock {\em Journal of Finance}, 58(6), 2581-2610.

\bibitem[Eberlein et al.(1998)]{EKP:1998}
Eberlein, E., U. Keller, and K. Prause (1998):
\newblock {New insights into smile, mispricing and value at risk},
\newblock {\em {Journal of {Business}}}, 71, 371-406.

\bibitem[Feng et al.(2010)]{FFF:2010}
Feng, J., M. Forde, and J.P. Fouque (2010):
\newblock {Short maturity asymptotics for a fast mean reverting Heston stochastic volatility model},
\newblock {\em SIAM Journal on Financial Mathematics}, 1, 126-141.

\bibitem[Feng et al.(2012)]{FFK:2012}
Feng, J., J.P. Fouque, and R. Kumar (2012):
\newblock {Small-time asymptotics for fast mean-reverting stochastic volatility models},
\newblock {\em {Annals of Applied Probability}}, 22(4), 1541-1575.

\bibitem[Figueroa-L\'opez and Forde(2012)]{FigForde:2012}
Figueroa-L\'opez, J.E., and M. Forde (2012):
\newblock {The small-maturity smile for exponential L\'evy models},
\newblock{\em SIAM Journal on Financial Mathematics} 3(1), 33-65.

\bibitem[Figueroa-L\'opez et al.(2012)]{FigGH:2011}
Figueroa-L\'opez, J.E., R. Gong, and C. Houdr\'e (2012):
\newblock {Small-time expansions of the distributions, densities, and option prices under stochastic volatility models with L\'{e}vy jumps},
\newblock {\em Stochastic Processes and their Applications}, 122, 1808-1839.

\bibitem[Figueroa-L\'opez et al.(2011)]{FigGH:2012b}
Figueroa-L\'opez, J.E., R. Gong, and C. Houdr\'e (2011).
\newblock {High-order short-time expansions for ATM option prices under the CGMY model},
\newblock {\em Preprint. Available at arXiv:1112.3111v1 [q-fin.CP]}.

\bibitem[Figueroa-L\'opez et al.(2013)]{FigGH:2013}
Figueroa-L\'opez, J.E. , R. Gong, and C. Houdr\'e (2013):
\newblock {A note on high-order short-time expansions for ATM option prices under the CGMY model},
\newblock {\em Preprint. Available at arXiv:1305.4719 [q-fin.CP]}.

{\Blue
\bibitem[Figueroa-L\'opez and \'Olafsson(2013)]{FigOlaf}
Figueroa-L\'opez, J.E., and S. \'Olafsson (2013):
\newblock {Short-time expansions for close-to-the-money options under a L\'evy jump model with stochastic volatility},
\newblock {\em Preprint. Available at arXiv:1404.0601 [q-fin.PR]}.
}

\bibitem[Forde and Jacquier(2009)]{FordeJac:2009}
Forde, M., and A. Jacquier (2009):
\newblock{Small-time asymptotics for implied volatility under the Heston model},
\newblock{\em Int. J. Theor. Appl. Finance}, 12(6), 861-876.

\bibitem[Forde and Jacquier(2011)]{FordeJac:2010}
Forde, M., and A. Jacquier (2011):
\newblock{Small time asymptotics for an uncorrelated local-stochastic volatility model},
\newblock{\em {Applied Mathematical Finance}}, 18(6), 517-535.

\bibitem[Forde et al.(2012)]{FordeJacLee:2010}
Forde, M., A. Jacquier, and R. Lee (2012):
\newblock{The small-time smile and term structure for implied volatility under the Heston model},
\newblock{\em SIAM Journal on Financial Mathematics}, 3:690-708.

{
\bibitem[Gao and Lee(2013)]{GaoLee:2013}
Gao, K., and R. Lee (2013):
\newblock {Asymptotics of implied volatility to arbitrary order},
\newblock {\em{Finance and Stochastics}, Forthcoming}. Available at SSRN: http://ssrn.com/abstract=1768383.
}

\bibitem[Gatheral et al.(2009)]{Gatheral:2009}
Gatheral, J., E. Hsu, P. Laurence, C. Ouyang, and T-H. Wang (2012):
\newblock {Asymptotics of implied volatility in local volatility models},
\newblock {\em{Mathematical Finance}, 22:591-620}.

\bibitem[Henry-Labord\`{e}re(2009)]{Henry:2009}
Henry-Labord\`{e}re, P. (2009):
\newblock{Analysis, geometry, and modeling in finance: advanced methods in option pricing},
\newblock{\em {Chapman \& Hall}}.

\bibitem[Houdr\'e(2002)]{Hou:2002}
Houdr\'{e}, C. (2002):
\newblock {Remarks on deviation inequalities for functions of infinitely divisible random vectors},
\newblock {\em {Annals of Probability}}, 30(3), 1223-1237.

\bibitem[Jacod(2007)]{Jacod05}
Jacod, J. (2007):
\newblock {Asymptotic properties of power variations of L\'evy processes}.
\newblock \emph{ESAIM:P\&S} \textbf{11}, 173--196.

\bibitem[Kallenberg(1997)]{Kallenberg}
Kallenberg, O. (1997):
\newblock {\em {Foundations of Modern Probability}}.
\newblock Springer-Verlag, Berlin, New York, Heidelberg.

\bibitem[Koponen(1995)]{Koponen:1995}
Koponen, I. (1995):
\newblock{Analytic approach to the problem of convergence of truncated L\'{e}vy flights towards the Gaussian stochastic process},
\newblock{\em {Physical Review E}}, 52, 1197-1199.

\bibitem[Kou(2002)]{Kou:2002}
Kou, S. (2002):
\newblock {A jump-diffusion model for option pricing},
\newblock {\em {Management Science}}, 48, 1086-1101.

\bibitem[Kyprianou et al.(2005)]{Kyprianou}
Kyprianou, A., W. Schoutens, and P. Wilmott (2005):
\newblock {Exotic Option Pricing and Advanced L\'evy Models},
\newblock John Wiley \& Sons, West Sussex, England.

\bibitem[Madan et al.(1998)]{MCC:1998}
Madan, D., P. Carr, and E. Chang (1998):
\newblock {The variance gamma process and option pricing},
\newblock {\em {European Finance Review}}, 2, 79-105.

\bibitem[Madan and Milne(1991)]{MadMil:1991}
Madan, D.,  and F. Milne (1991):
\newblock {Option pricing with VG martingale components},
\newblock {\em {Mathematical Finance}}, 1, 39-56.

\bibitem[Madan and Seneta(1990)]{MadSen:1990}
Madan, D., and E. Seneta (1990):
\newblock {The variance gamma (VG) model for share market {returns}},
\newblock {\em {Journal of Business}}, 63, 511-524.

\bibitem[Mandelbrot(1963)]{Mandelbrot}
Mandelbrot, B. (1963):
\newblock {The variation of certain speculative prices}.
\newblock {\em The Journal of Business}, 36:394--419.

\bibitem[Matacz(2000)]{Matacz}
Matacz, A. (2000):
\newblock {Financial modeling and option theory with the truncated L\'evy process},
\newblock {\em Int. J. Theor. Appl. Finance}, 3:143--160.

\bibitem[Medvedev and Scailllet(2007)]{MedSca07}
Medvedev, A., and O.~Scailllet (2007):
\newblock {Approximation and calibration of short-term implied volatility under jump-diffusion stochastic volatility},
\newblock {\em The review of Financial Studies}, 20(2):427-459.

\bibitem[Merton(1976)]{Merton:76}
Merton, R. (1976):
\newblock{Option pricing when underlying stock returns are discontinuous},
\newblock{\em {Journal of Financial Economics}}, 3, 125-144.

\bibitem[Muhle-Karbe and Nutz(2011)]{MuhNut:2009}
Muhle-Karbe, J., and M. Nutz (2011):
\newblock {Small-time asymptotics of option prices and first absolute moments},
\newblock {\em Journal of Applied Probability}, 48(4), 1003-1020.

\bibitem[Paulot(2009)]{Paulot:2009}
Paulot, L, (2009):
\newblock{Asymptotic implied volatility at the second order with application to the SABR model},
\newblock{\em {Preprint}}, 2009.

\bibitem[Press(1967)]{Press}
Press, S.J. (1967):
\newblock {A compound event model for security prices}.
\newblock {\em The Journal of Business}, 40:317--335.

\bibitem[Roper(2009)]{Rop10}
Roper, M. (2009):
\newblock {Implied volatility: small time to expiry asymptotics in exponential L\'{e}vy models},
\newblock {\em Thesis, University of New South Wales}.

\bibitem[Roper and Rutkowski(2007)]{RoperRut:2007}
Roper, M., and M.~Rutkowski (2007):
\newblock {A note on the behaviour of the Black-Scholes implied volatility close to expiry},
\newblock {\em Tech. Report, UNSW}.

\bibitem[Rosenbaum and Tankov(2011)]{rosenbaum.tankov.10}
Rosenbaum, M., and P.~Tankov (2011):
\newblock {Asymptotic results for time-changed L\'evy processes sampled at hitting times}.
\newblock {\em Stochastic processes and their applications}, 121:1607--1633.

\bibitem[R\"uschendorf and Woerner(2002)]{Ruschendorf}
R\"uschendorf, L., and J.~Woerner (2002):
\newblock {Expansion of transition distributions of L\'{e}vy processes in small time}.
\newblock {\em Bernoulli}, 8, 81-96.

\bibitem[Rosi\'nski(2007)]{Rosinski:2007}
Rosi\'nski, J. (2007):
\newblock {Tempering stable processes}.
\newblock {\em {Stochastic processes and their applications}}, 117:\penalty0 677--707.

\bibitem[Samorodnitsky and Taqqu(1994)]{SamoTaqqu}
Samorodnitsky, G., and M. Taqqu (1994):
\newblock{\em {Stable non-Gaussian random processes}},
\newblock Chapman \& Hall, New York.

\bibitem[Sato(1999)]{Sato:1999}
Sato, K. (1999):
\newblock{\em L\'{e}vy processes and infinitely divisible distributions},
\newblock Cambridge University Press.

\bibitem[Tankov(2010)]{Tankov}
Tankov, P. (2010):
\newblock {Pricing and hedging in exponential L\'evy models: review of recent results},
\newblock{\em {Paris-Princeton Lecture Notes in Mathematical Finance}}, Springer.

\bibitem[Zolotarev(1986)]{Zolotarev}
Zolotarev, V.M. (1986):
\newblock{\em One-dimensional stable distributions},
\newblock Amer. Math. Soc., Providence, R.I.

\end{thebibliography}

\end{document}